# Large-Area Patternable Solar-Powered Bistable Organic Crystalline Film for Nonlinear Optical Communication

Ankur Khapre[1,3], Avulu Vinod Kumar[1,3], Haseeba Nasreen Punathil[1], Biswajit Kumar Barman[1], Bedanta Kumar Deka[2], Shilpa Mangalassery[2], Sri Ram G Naraharishetty[2], Rajadurai Chandrasekar*[1]

1. Advanced Photonic Materials and Technology Laboratory, School of Chemistry and Centre for Nanotechnology, University of Hyderabad, Prof. C. R. Rao Road, Gachibowli, Hyderabad 500046, Telangana, India
2. School of Physics, University of Hyderabad, Prof. C. R. Rao Road, Gachibowli, Hyderabad 500046, Telangana, India
3. Equal contribution of authors

*E-mail: r.chandrasekar@uohyd.ac.in

**Abstract:** Reversible control of crystal symmetry offers a powerful route to programmable optical functionality. However, achieving solid-state bistability between centrosymmetric and non-centrosymmetric crystalline phases remains a formidable challenge; examples of materials that enable stable switching of second-order nonlinear optical (NLO) responses are exceptionally rare. Here we report a solar-powered, symmetry-bistable organic material based on the photoisomerizable molecule (*E*/*Z*)-2-(4-(4-bromophenyl)thiazol-2-yl)-3-(4-(dimethylamino)phenyl)acrylonitrile (*E*/*Z*-BTDPA). The crystallizable *E*- and *Z*-isomers adopt distinct molecular packing arrangements that reversibly toggle between these states, controlling second-order NLO activity. The *E*-form exhibits strong second-harmonic generation (SHG), whereas the *Z*-form is SHG-inactive and displays two-photon luminescence. This bistable behavior is retained in flexible thin films, where sunlight-driven photoisomerization enables reversible photoswitching of the second-order electric susceptibility ($\chi^2$), large-area optical patterning, and real-time NLO communication via waveform generation and text-string transcription at telecommunication wavelengths. This sustainable strategy bypasses rigid inorganic architectures, establishing photoinduced symmetry bistability as a scalable paradigm for all-optical computing and advanced communication networks.

**Introduction:**

The demand for universal renewable energy-based technology is rising as the human population surges, and it is projected to rise by 0.5%-1.6% per year until 2050.[1-3] Sunlight reaching the Earth's surface (≈1000 $W/m^2$) represents one of the most accessible renewable energy sources. Certain classes of molecules can harvest solar energy and undergo structural changes, leading to pronounced effects on their mechanical,[4,5] electrical,[6-8] optical[9-11] and magnetic properties.[12,13] In particular, the solar-light-driven photoisomerization of organic molecules involving structural transformations, such as *E*/*Z*,[14,15] keto-enol,[16] and cyclization,[17] are of interest. Channelizing these transformations in a controllable and reversible manner holds great importance for various technological applications including sensors[18,19], optical signal processing[20], THz generation[21], optical communication[22] and computing.[23] Non-centrosymmetric (NCS) organic crystals producing nonlinear optical (NLO) effects are well known in literature.[24-27] However, achieving reversible symmetry-breaking or symmetry bistability in an organic crystalline state whereby molecular arrangements are toggled between centrosymmetric (CS) and NCS phases remains a formidable challenge. Such a transition would enable the reversible switching of the second-order electric susceptibility between $\chi^{(2)}=0$ and $\chi^{(2)}\neq0$, a capability that is imperative for next-generation NLO technologies. Typically, reversible symmetry-breaking phenomenon induced by external stimuli like heat, electric current, mechanical strain, etc., are known only in a few crystals.[28-34] Usually, the inorganic NLO devices are not mechanically compliant as they are patterned on stiff silicon substrates[28-30,35] and whereas, the lack of control over molecular packing of organic materials during processing prohibits the fabrication of organic NLO devices. Therefore, large efforts are devoted towards the development of flexible NLO optical switching devices, especially with reversible symmetry-breaking ability via a sustainable source like solar-radiation.

Photochromic and bistable organic molecules hold great potential for NLO applications, as organic chromophores have very large molecular hyperpolarizabilities, often exceeding those of conventional inorganic crystals when properly aligned in a non-centrosymmetric lattice or periodically poled polymer matrix.[36,37] These molecules can be easily patterned on flexible polymer substrates, often achieved with a careful choice of solvents. The photochromism helps realize the optical switching in structurally bistable molecular crystals using optical stimuli. However, to attain a large macroscopic NLO response, noncentrosymmetrically aligned active molecules must have a large first hyperpolarizability tensor ($\beta$), and excellent thermal and photochemical stabilities. *E*/*Z* photoisomerization is often limited by the low thermal stability of metastable states and the requirement for ultraviolet (UV) light (254 to 365 nm) to induce isomerization. Advancing *E*/*Z* photoisomerization reactions under sunlight in thin-film state is challenging, and research on solid-state isomerization is limited because of the tightly packed molecular arrangements. Recently, Krainova *et al.* achieved large-scale vacuum deposition (approximately 100 mm in area) of a popular NLO molecule, 2-[3-(4-hydroxystyryl)-5,5-dimethylcyclohex-2-enylidene] malononitrile (OH1), on glass and silicon substrates; however, the device lacked NLO switching behavior as OH1 is insensitive to optical switching using light.[38] Therefore, distinct molecular systems capable of performing reversible photochromism under ambient solar conditions must be discovered. Cyanostilbenes represent a rarely explored class of photoisomerizable molecular systems, as it is extremely challenging to obtain crystal structures for both *E/Z* isomers. The extended donor(D)-acceptor(A) π-system enhances intermolecular interactions, and the ability to induce a non-zero dipole orientation collectively imparts photoisomerization and NLO characteristics to cyanostilbene molecules.

In this work, we demonstrate a cyanostilbene based electron donor-π-acceptor system namely, (*E*/*Z*)-2-(4-(4-bromophenyl)thiazol-2-yl)-3-(4-(dimethylamino)phenyl)acrylonitrile, (*E/Z*-BTDPA) that exhibits sun-light-driven SHG switching in the thin film-state useful for NLO communication. Remarkably, both *E*/*Z* photoisomers grew as single crystals suitable for X-ray diffraction studies. The intramolecular hydrogen bonding interactions provide greater stability to the *Z*-isomer than to the *E*-isomer in the solid-state. Under solar light irradiation, the molecule transforms from a thermodynamically stable *E*-isomer (*E*-BTDPA) to a *Z*-isomer (*Z*-BTDPA) in solution, thin films, and single-crystalline states. The SHG characteristics of the *E*-BTDPA crystals, as a result of polar crystallographic molecular packing, is supported by the higher unit cell hyperpolarizability value ($\beta_E$ = 1.2083 x $10^{-28}$ esu) compared to nonpolar packing in *Z*-isomer ($\beta_Z$ = 0.392 x $10^{-30}$ esu). The adaptability of polar and nonpolar isomers in solid states using solar energy enabled the fabrication of previously unknown all-organic NLO switches in the thin-film state. Further, photolithography allows the fabrication of 2.2×2.2 $cm^2$ pixelated square patterns of *E/Z* forms on a flexible polyethylene terephthalate (PET) substrate which when fixed on an oscillating stage, controlled by an embedded program, enables facile NLO communication through SHG signal. This study demonstrates the potential of solar energy-harvesting organic bistable materials for NLO data transfer and communication technologies, opening pathways for molecular-materials-based all-optical switching, all-optical diodes, and optical computing.

**Synthesis, photophysical and photoisomerization dynamics**

BTDPA was synthesized via a Knoevenagel condensation reaction (Figure 1a, Supplementary Figure 1-3). The slow crystallization of BTDPA in a 1:1 $CH_3OH/CH_2Cl_2$ mixture at ambient conditions yielded two visually distinct isomeric solids: *E*-BTDPA (green) and *Z*-BTDPA (orange) (Supplementary Figure 4,5). $^{1}H$ NMR spectroscopy studies of the green solution of *E*-BTDPA (10 mg of *E*-BTDPA in 0.5 mL) in $CDCl_3$ at rt suggested gradual reversible *E*↔*Z* interconversion induced by sunlight and darkness (Figure 1a-d, Supplementary Figure 6-9). *E/Z*-BTDPA exhibited strong positive solvatochromism, with a systematic bathochromic shift in the solution-state UV-Vis absorption and fluorescence (FL) spectra with increasing solvent polarity (Supplementary Figure 10).

Specifically, the absorption maximum shifted from 414 nm in pentane to 446 nm in DMSO, whereas the FL maximum shifted from 469 to 533 nm. These spectral shifts result from an increased dipole moment (polar) upon electronic excitation, where polar solvents preferentially stabilize the excited state, reducing the HOMO–LUMO gap and producing red-shifted transitions characteristic of intramolecular charge transfer systems suitable for improved NLO behavior (Supplementary Figure 11). In the solid state, the *Z*-isomer shows broader absorption up to ≈600 nm than the *E*-isomer. The *E*-isomer showed an FL emission maximum at 535 nm with a shoulder extending to 750 nm (CIE: 0.37, 0.60), whereas the *Z*-isomer emitted at 616 nm with a tail up to 790 nm (CIE: 0.60, 0.39) (Figure 1f).

**Solar-Induced Centro/Noncentrosymmetric SCSC Phase transition:**

The X-ray structure showed that *E*-BTDPA crystallized in a monoclinic noncentrosymmetric *Pn* (7) space group, while *Z*-BTDPA adopted a triclinic centrosymmetric *P*-1 space group (Supplementary Table 1). Both isomers exhibit nearly planar geometries with slight twisting of 12.02° (*E*-isomer) and 8.42° (*Z*-isomer), stabilized by intramolecular hydrogen bonding (Figure 2a). In the *E*-isomer, parallel polar packing stabilized by intermolecular C-H···Br (2.928 Å) hydrogen bonding and C-H··· π (2.752 Å) interactions aligned the molecular dipoles along the crystallographic *c*-axis, generating a macroscopic dipole moment (Figure 2b and Supplementary Figure 12 and 13 a,b). Additionally, intermolecular C-H···π interactions (2.795 Å) organized the molecules into a herringbone arrangement along the *b*-axis (Supplementary Figure 13c). In contrast, the *Z*-isomer is bent and stabilized by an intramolecular C-H(7)···N(3) hydrogen bond (2.207 Å) between the thiazole nitrogen and H atom of the dimethyl amino phenyl ring (Figure 2c). Weak C-H···N (2.545 Å), C–H ···S (2.928 Å), and Br···Br (3.7 Å) interactions organized the molecules into a herringbone arrangement along the *a*-axis (Figure 2d). The head-to-head alignment of nitrile groups and multiple π···π, C-H···π, and C-H···S interactions promoted antiparallel molecular orientation and overall lattice stabilization (Supplementary Figure 14).

Noncovalent interaction (NCI) analysis revealed hydrogen bonding along with van der Waals forces in both isomers (Figure 2e,f), with the *Z*-isomer showing denser packing (1.556 g/cm$^3$) than the *E*-isomer (1.487 g/cm$^3$), indicating enhanced lattice stability, while the higher-energy *E*-isomer was formed under kinetic control. Electrostatic potential (ESP) mapping confirmed a strong dipole moment for *E*-BTDPA (~16.04 D) owing to the asymmetric charge distribution, whereas *Z*-BTDPA exhibited dipole cancellation (~0.002 D) and a nonpolar character (Figure 2g,h). Accordingly, *E*-BTDPA shows a larger hyperpolarizability ($\beta_{trans}$ = 1.2083 x $10^{-28}$ esu) than the *Z*-isomer ($\beta_{cis}$ = 6.39 x $10^{-33}$ esu) supporting the enhanced SHG response.

**Solid-state photoisomerization and nanocrystalline thin-film characterization:**

The solid-state *E* to *Z* reversible interconversion was studied by exposing an *E*-BTDPA single crystal to solar simulator light (450 W) at 10 min intervals (Figure 3a and Supplementary Figure 15). The progressive change in the FL crystal color from green to orange indicated the formation of 100% *Z*-isomer (Figure 3b) within 80 min. In contrast, in the dark or below 4 °C, the *Z*-isomer reverted to the *E*-form within 120 min (Figure 3c), whereas at 77 K, the *Z*-to-*E*-interconversion was instantaneous. Overall, *E*→*Z* photoconversion proceeded faster than the *Z*→*E* (Figure 3b,c and Supplementary Figure 15). Theoretical calculations indicated a higher thermodynamic stability and packing density for the *Z*-isomer (V=854.445 $Å^3$) than for the *E*-isomer (V= 856.681 $Å^3$), restricting molecular motions and slowing *Z*→*E* photoisomerization (Supplementary Table 1). Raman spectroscopy further confirmed conformational changes, showing shifts in C–S stretching (*E*: 1191 $cm^{-1}$; *Z*: 1198 $cm^{-1}$), C–H bending (*E*: 1450 $cm^{-1}$; *Z*: 1484 $cm^{-1}$) and appearance of C≡N stretching band (*Z*: 2205 $cm^{-1}$), consistent with increased steric hindrance and strong intramolecular interactions in the *Z*-isomer (Figure 3d,e).

Photoisomerization was further examined in spin-coated *E*-BTDPA thin films, using solar simulator through a 10 µm photolithography mask to create a 2.2x2.2 $cm^2$ patterned array of *E*- and *Z*-isomers (Supplementary Figure 16-20). FL mapping (excitation: 532 nm CW laser) confirmed the formation of 1D alternative *E*-*Z* microarrays owing to selective *E*→*Z* conversion (Figure 3f, Supplementary Figure 21-22). Field-emission scanning electron microscopy (FE-SEM) images revealed the nanofiber morphology of the thin films (Figure 3f and Supplementary Figure 23). Furthermore, the isomerization rates of *E*→*Z* and vice versa (thickness: 74 to 540 nm) were faster in thinner films owing to improved light penetration (Figure 3g, Supplementary Figure 24-28). Powder X-ray diffraction (PXRD) and PL lifetime measurements further confirmed isomerization, with each phase exhibiting distinct spectral signatures (Supplementary Figure 29-31).

**Nonlinear optical studies of BTDPA:**

Polar crystals lacking inversion symmetry exhibit SHG and two-photon luminescence (TPL) at lower fundamental frequencies. Excitation of ≈10 µm thick *E*-BTDPA crystal with 1100 nm fundamental femtosecond (fs) radiation revealed a sharp SHG signal at 550 nm (Figure 4a and Supplementary Figure 32,33). The pump power dependence showed a quadratic increase in intensity, confirming the second-order nonlinear response (Figure 4a,b, Supplementary Figure 34). Furthermore, excitation across 800–1000 nm produced broad TPL (400–750 nm) along with SHG (Supplementary Figure 34a-c), whereas only SHG signals (SHG ON state) were observed in the 1100–1500 nm region (Figure 4c,d and Supplementary Figure 35b). In contrast, after *E* to *Z* switching under solar irradiation, SHG disappeared (OFF state) and only TPL was detected for 800-1100 nm fundamental excitation (Supplementary Figure 34a,b, and 35d-f). These results highlight the bistable SHG (ON/OFF) state switching between *E*- and *Z*-forms of BTDPA in the telecommunication region (1100-1500 nm).

**Nonlinear optical tuning studies with phase change of BTDPA:**

To realize reversible SHG switching in micropatterns, *E*-BTDPA thin films were spin-coated onto glass coverslips or flexible polyethylene terephthalate (PET) substrates with controlled thicknesses (Figure 5a,b). The pristine *E*-form film exhibited strong SHG, while solar irradiation induced $E\rightarrow Z$ photoisomerization, resulting in TPL emission. Using photomask, 1 mm wide alternating *E* and *Z* arrays were fabricated by selective solar exposure of a 300 nm *E*-BTDPA film, where exposed regions transformed to the *Z*-form and masked regions retained the *E*-form (Figure 5a). Subsequent 90° rotation of the mask generated pixelated 2D square arrays with dimensions of $1\times1$ mm$^2$ on flexible PET substrate (Figure 5b). The $E\rightarrow Z$ conversion time increased with the film thickness (Figures 5c,d and Supplementary Figure 36).

The fabricated 1D arrays comprised alternating photo-masked *E*-isomer regions ($\chi^{(2)}\neq0$; label:1,3,5,7) and exposed *Z*-isomer regions ($\chi^{(2)}=0$; label:2,4,6,8) regions (Figure 5e,g and Supplementary Movie 1). Under 1100 nm fs laser excitation (spot size ≈5 mm) in transmission mode, a strong SHG response was observed only from the odd-numbered *E*-isomer lines, while the even-numbered *Z*-isomer lines remained SHG inactive (Figure 5f and Supplementary Figure 36-39a-c and 40). The re-patternability of the same device was demonstrated by shifting the photomask by 1 mm (equivalent to a line width) and re-irradiating with solar light, inducing *Z*↔*E* region conversion. The SHG active regions shifted accordingly, demonstrating the erasable and reusable operation of the microdevice (Figure 5f,h and Supplementary Movie 2). It should be noted that the variations in the intensities of the observed light patterns arise from the Gaussian spatial intensity profile of the incident fundamental beam.

SHG bistability was performed in five selected regions on a grid patterned sample (*E*-isomer; $\chi^{(2)} \neq 0$; labelled 1,2,3,4, and 5), each comprising four 1×1 $mm^2$ square pixels (Figure 5i). Excitation of region 1 at 1100 nm displayed area-selective SHG from all active pixels, and similar SHG responses were confirmed across regions 2-5 (Figure 5i, Supplementary Figure 39d-f and 41 and Supplementary Movie 3). In *Z*-isomer (unmasked) regions, the SHG signal was absent, whereas a weak TPL ($\lambda_{max}$ = 590 nm; yellow emission) was observed under 800 nm excitation. Interestingly, the device demonstrated reversible switching between TPL (at ω = 800 nm) and SHG (at ω = 1100 nm) and maintained stable performance over more than 14 recycling cycles (Figure 5d and Supplementary Movie 4,5).

The NLO switching behavior of photoisomerizable BTDPA is shown in a fully connected $K_4$ graph, linking four distinct optical states (Figure 5j). The horizontal axis (left to right) represents the fs laser–induced switching, whereas the vertical axis (top to bottom) represents the solar-driven photoisomerization (Figure 5j). This unified graphical model correlates the molecular photo-response with macroscopic NLO tunability. Consequently, a flexible, sunlight-driven NLO switching device based on BTDPA was successfully demonstrated, highlighting photoisomerization-assisted optical modulation in organic crystalline films (Supplementary Figure 42).

**Time-resolved NLO angular gating for information generation, transmission, and detection:** For real-time NLO data transfer based on binary (1/0) operations, the pixelated flexible thin-film device was angularly oscillated between 0° and 5° using a motor driven by ON–OFF keying (OOK) modulation (Figure 6a). Rapid milli-second scale angular device oscillation enabled controlled exposure of individual pixels to the laser beam 1100 nm with a 100 μm diameter, generating SHG at 5°corresponding to the ON state (1), while no SHG signal was detected at an angle of 0° when the beam fell outside the pixel, defining OFF state (0) (Figure 6b and Supplementary Figure 43). The SHG signals were recorded in reflection mode using a Si photodiode (PD) equipped with boxcar integrator (SR 280). Synchronized film oscillation, programmed at 200 ms, produced a 0-1-0 cycle with an average velocity of 0.05°/ms (Figure 6c, Supplementary Eq. (4,5)).

For binary signal transmission, the least significant bit (LSB)-first protocol was employed. Further, trigger and feedback voltages were applied to the motor for accurate angular positioning of the device (Supplementary Figure 44). The initial experimental state was set to 1 (ON state). Therefore, the film oscillation during laser irradiation produced square waves for the 1 state for a duration of 100 ms (50 ms each for back-and-forth motion) (Figure 6b). Oscillations of 0 to 1, 1 to 0, 1 to 1, and 0 to 0 were conceptualized for angle versus time, where the 0 to 1 or 1 to 0 were achieved using a maximum angular velocity of 0.0785°/ms (Figure 6c, Supplementary Figure 45a, and Supplementary Eq. (6,7)). However, the 0 to 0 and 1 to 1 transitions had a stationary state of 100 ms between the total transition of 200ms (Figure 6c and Supplementary Figure 45a). Using this approach, the text transmission "*Nonlinear* *optical communication using an organic crystalline thin film."* was successfully implemented through NLO signal string generation (encoding) and detection (decoding) (Figure 6d, Supplementary Code 1, and Supplementary Movie 6-8).

The formation of square waves across various modulation states was quantitatively validated by the binary density distribution, and the Kernel Density Estimation (KDE) plot exhibited a sharp bimodal profile, confirming that the signal remained strictly confined to the required logic-level thresholds with minimal inter-state deviation (Supplementary Figure 45b and Supplementary Code 2-3). The stability of the device was confirmed by the NLO response over a prolonged acquisition at a fixed pixel (Supplementary Figure 45c-d). A slight intensity variation caused by an error in the feedback mechanism was corrected by adjusting the cut-off amplitude to % at the full-wave half-maximum (Figure 6e). Two of the text characters (p and o) are shown as detailed square waveforms during the transition between the 0 and 1 states and vice versa, and their correlation with an LSB binary sequence matching those characters demonstrates real-time string generation (Figure 6f). The full waveforms were converted to binaries subsequently to text using the Python 3.13 (Figure 6d-f, Supplementary Code 4-5). Similar experiments of text transmission using the controlled oscillatory motion of two separate *E* and *Z* thin films did not perform NLO communication (Supplementary Movie 9,10). The mechano-optical gating experiment performed on the solar light-induced isomerized BTDPA thin film, demonstrates the real application of NLO switching in waveform generation and detection for NLO signal transmission.

## Discussion

We demonstrated a sunlight-driven *E*/*Z* photoisomerizable cyanostilbene-based pixelated organic nonlinear optical (NLO) thin-film switch fabricated on a flexible substrate.

The *E*-isomer pixels displayed distinct SHG signal, whereas the *Z*-isomer remained SHG-inactive owing to dissimilar polar and nonpolar molecular arrangements within their crystal packings. Solar radiation-induced photoisomerization enabled the realization of reversible NLO switching from the SHG-active ON state (binary 1) to the SHG-silent OFF state (binary 0), which can be dynamically accessed using 1100 nm fs laser excitation. programmable mechanical oscillation of the device at controlled angles allowed selective angular exposure of pixels and controlled readout of binary states, enabling the transmission of text messages encoded as NLO signal strings. This study establishes a new paradigm material platform for sun-light operable, flexible, and low-cost solid-state NLO switches with potential applications in all-optical switching, optical diodes, and optical computing architectures. Future progress will depend on addressing the key challenges in device integration with existing technologies for the practical deployment of organic NLO switches.

## Methods

### 1. Materials

2-(4-(4-bromophenyl)thiazol-2-yl)acetonitrile (CAS No.: 94833-31-5, 97% purity) was purchased from Sigma Aldrich, 4-(dimethylamino)benzaldehyde (CAS No.: 100-10-7, >98% purity), and sodium methoxide (CAS No.: 124-41-4, >96.0% purity) were purchased from TCI Chemicals. HPLC-grade and deuterated solvents from Supelco were used without prior purification for the synthesis, experiment and other characterization.

### 2. Synthesis of BTDPA:

BTDPA was synthesized via a Knoevenagel condensation reaction. 2-(4-(4-bromophenyl)thiazol-2-yl)acetonitrile and 4-(dimethylamino)benzaldehyde were reacted in methanol using sodium methoxide as the base. The reaction afforded *E*-BTDPA as a green solid with an isolated yield of 68%.

The slow crystallization of the crude product from a 1:1 (v/v) $CH_3OH/CH_2Cl_2$ solvent mixture under ambient conditions yielded two distinct isomeric crystalline forms: E-BTDPA (green crystals) and Z-BTDPA (orange crystals). The thermal stabilities of the isolated isomers were evaluated using thermogravimetric analysis, revealing decomposition temperatures ($T_d$) of 303 °C for E-BTDPA and 319 °C for Z-BTDPA.

### 3. Molecular and photophysical Characterization

$^1H$ (500 MHz) and $^{13}C$ (125 MHz) NMR spectra were recorded on a Bruker DPX spectrometer with a solvent proton as an internal standard ($CDCl_3$:$^1H$: 7.26 ppm, $^{13}C$: 77.16 ppm). Commercially available deuterated $CDCl_3$ was used in this study. Solid-state $^{13}C$ (400 MHz) NMR spectra of the mixture of the compound and KBr were recorded on a Bruker DPX spectrometer. The chemical shifts (δ) were expressed in parts per million (ppm). High-resolution spectra were recorded using Bruker ESI-TOF. FT-IR spectra were recorded using an iD7−ATR detector attached to a Thermo Scientific iS5 spectrometer.

UV-visible absorption was assessed with a quartz cuvette having a 1 cm path length for liquid samples, while a 1 cm diameter quartz was utilized on a JASCO V-750 spectrometer in diffuse reflectance UV-visible (DR-UV-vis) mode for solid samples. Emission spectra in the solid state were obtained using a JASCO FP-8500 spectrofluorometer. The photoluminescence quantum yield was determined with a Horiba F1-3C.

### 4. FE-SEM

The morphology of the thin film was analyzed with a Zeiss field-emission scanning electron microscope (FE-SEM) set to operate at 5 kV.

### 5. µ-Raman Analysis

The Raman spectra for both the single crystal and thin films were obtained through a Labram HR evolution microscope system. This setup utilized a 632 nm wavelength excitation laser and a 50× objective lens, which provided an estimated laser spot size of approximately 1.18 µm. All Raman measurements were conducted with a standard input power of 5 mW.

### 6. AFM/Profilometer

Thickness profile measurements were performed on thin films of different thicknesses using an Ambios XP200 profilometer and an Oxford Asylum MFP-3D Origin AFM.

### 7. Confocal laser scanning microscope

Experiments involving optics on an individual microcrystal were conducted utilizing the backscattering mode configuration of the Wi-Tec alpha 300 AR laser confocal optical microscope (LCOM), which is fitted with a CCD detector cooled by a peltier module. For the optical excitation of the crystalline thin film, a 405 nm diode laser source was employed along with a 60× objective. Unless stated differently, a 20× objective lens was utilized for collecting spectra and images.

### 8. Powder XRD

The PXRD patterns for the two photoisomers were obtained with a Bruker D8 advanced diffractometer from Bruker-AXS, utilizing Cu Kα radiation with a wavelength of 1.5406 Å.

### 9. Single Crystal X-ray diffraction

Single-crystal X-ray diffraction were obtained at 279 K using a Rigaku diffractometer, which features dual Cu/Mo at zero Eos. The experiment employed monochromatic Cu-Kα radiation, with the beam size set to 100 µm.

## 10. Solar simulator studies

Photoisomerization was performed using an Enitech solar simulator equipped with a xenon (Xe) lamp operating at a maximum irradiation power of 450 W. Photomasks were used to create the line and pixel patterns.

## 11. Nonlinear optical measurements

Nonlinear optical studies of both thin films and single crystals were carried out using a Ti:Sapphire femtosecond (fs) laser (3-box system: MaiTai, Empower, Spitfire Ace; repetition rate: 1 KHz; pulse width: 80 fs; average power: 4.5 mW). An optical parametric amplifier (OPA) (TOPAS Prime, Light Conversion) is integrated into the system to tune the laser wavelength across a broad range (400 nm–2400 nm). The experimental setup incorporated multiple optical components, including a long-pass filter (Thorlabs FELH900) to block wavelengths below 900 nm, a short-pass filter to suppress wavelengths above 700 nm, optical density filters to precisely control the laser power, and a set of lenses for beam focusing and defocusing.

## 12. NLO Communication Setup

A femtosecond laser (central wavelengths 1100, 1200 and 1300 nm) was used as an optical transmitter. The Gaussian beam was focused on the sample using a $CaF_2$ lens of 20 cm focal length to achieve a spot diameter of approximately 100 μm (Supplementary Figure 24). Optical gating was implemented using a 32-bit LX7 dual-core microcontroller coupled to a power amplification circuit based on an IRLZ44N MOSFET. The MOSFET was controlled by the digital output pin of the microcontroller and was powered by an external 3.3 V supply to actuate the angular optical gate.

Bit-level modulation was performed using on–off keying (OOK), with the bit duration defined in the software (typically, 100 ms). Each transmitted byte comprised a start bit, eight data bits transmitted in least-significant-bit (LSB)-first order, followed by a stop bit. The second-harmonic signals generated by the sample were detected using a Si PIN photodiode. The voltage signal from the Silicon detector is collected using a boxcar integrator synchronized with the femtosecond laser TTL trigger. The output from the boxcar was collected by a DAQ system configured with a sampling rate of 2000 Hz and 200 samples per acquisition, triggered directly by the laser TTL pulse to ensure precise synchronization with the signal. The signal was digitized and recorded using LabVIEW software on a data-acquisition system, employing a threshold-based comparator for signal discrimination.

For data reconstruction, the transitions in the digitized signals were interpreted as logical bit changes. Character synchronization and decoding were performed offline using a custom Python script to convert the recovered binary into text.

**13. Components and Software**

All electronic components were obtained from online sources, as detailed in Supplementary Table 2. A 32-bit LX7 dual-core processor-based microcontroller was used for programming the power and controlling the optical angular gating. Python 3.13 was employed for serial communication and host-side information generation on a Windows 11 operating system. Supplementary Codes 1–6 contain the codes for all modes.

**14. Theoretical Calculation**

Theoretical calculations were performed using the Gaussian 16 program based on the single-crystal structure of *E*/*Z*-BTDPA.[39] Single-point energies and first-order hyperpolarizabilities were computed at the ωB97XD level of theory with a 6-311++G(2d,2p) basis set.[40] Raman spectra were computed using density functional theory (DFT) with the harmonic approximation, employing an excitation wavelength of 633 nm and a temperature of 298.15 K. The electrostatic potential (ESP) surface and molecular orbitals were visualized using GaussView 5.0.8, while Multiwfn was used for subsequent Raman data analysis.

**Data availability**

The crystallographic data for E-BDTPA and Z-BDTPA has been deposited at the Cambridge Crystallographic Data Centre (CCDC) with deposition numbers 2502331 (296.6 K) and 2504145 (296.6 K), respectively. These data can be accessed free of charge from The CCDC at www.ccdc.cam.ac.uk/structures/. Source data accompany this paper. The corresponding codes for this study are included in the supplementary information files and are also available from the corresponding author upon request.

**Acknowledgements**

R.C. acknowledges SERB-New Delhi [CRG/2023/003911] for financial support. A.K. & B.K.D thank PMRF for the PhD fellowships.

**Author information**

These authors contributed equally: Ankur Khapre, Avulu Vinod Kumar.

Authors and Affiliations

1. School of Chemistry and Centre for Nanotechnology, University of Hyderabad, Gachibowli, Hyderabad, India

Ankur Khapre, Avulu Vinod Kumar, Haseeba Nazreen, Biswajit Kumar Barman & Rajadurai Chandrasekar

2. School of Physics, University of Hyderabad, Gachibowli, Hyderabad, India

Bedanta Kumar Deka, Shilpa Mangalassery & Sri Ram Gopal Naraharishetty

**Author contribution**

AK and RC conceived the idea of thin-film fabrication, angular oscillation experiment set-up and NLO communication. AVK and HN synthesized the phase changing material, performed single crystal X-ray and photo-physical studies. AK fabricated the thin-film, investigated the solid-state photoisomerization reaction and built the angular oscillation set-up. BKB carried out the theoretical calculations and computational analyses. AK, AVK, HN and BKB worked under the supervision of RC. AK, BKD and SM conducted the NLO experiments under the supervision of SRGN and RC. AK, BKB, AVK and RC analyzed data and wrote the manuscript. AK and AVK contributed equally to this work. All authors discussed the results, contributed to data interpretation and approved the final manuscript.

Corresponding author

Correspondence to Rajadurai Chandrasekar. Mail: r.chandrasekar@uohyd.ac.in

**Ethics declarations**

Competing interests

The authors declare no competing interests.

**Additional information**

Supplementary information is available.

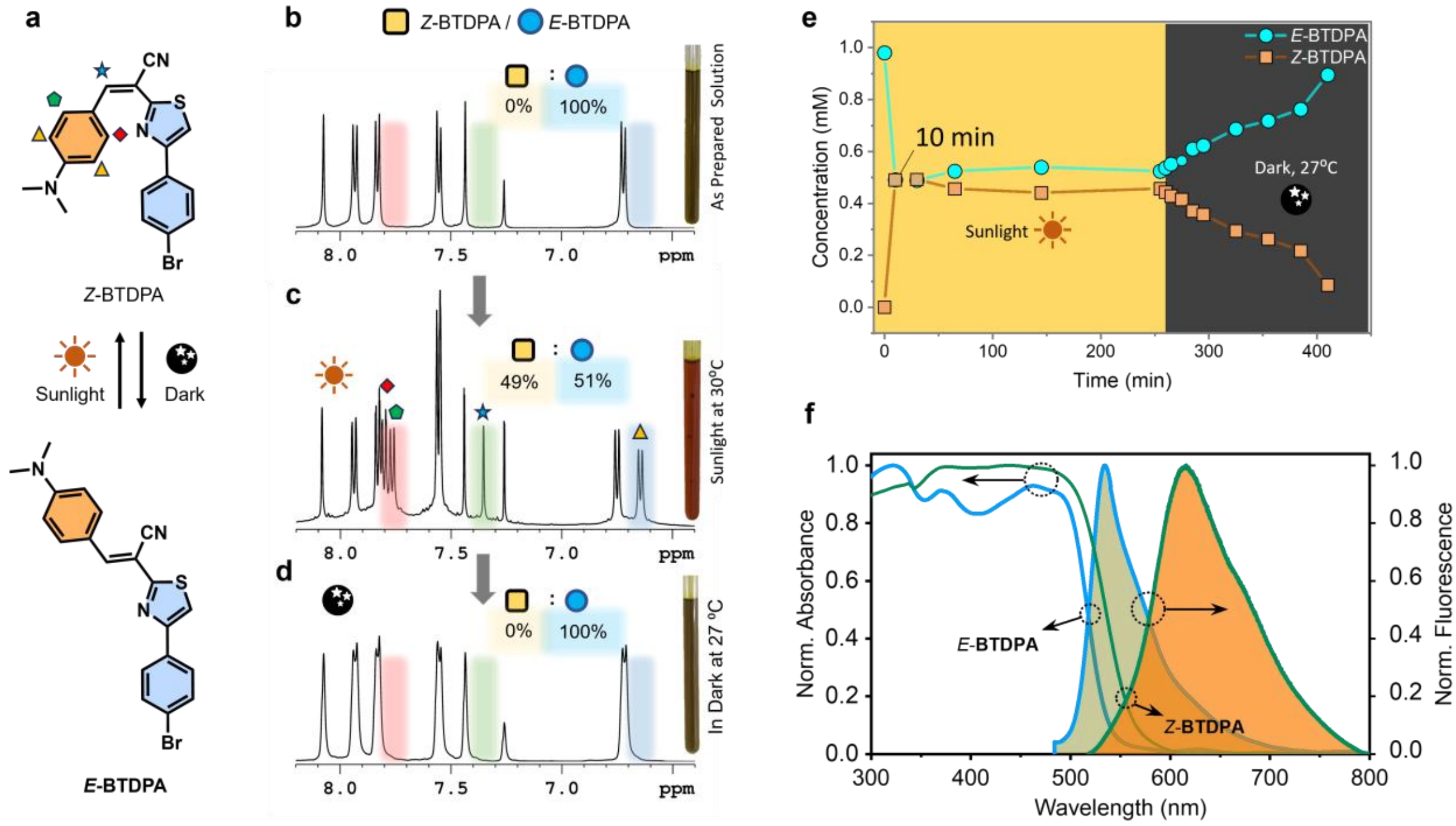

**Figure 1: Photophysical studies of the BTDPA molecule.** (a) Illustration of reversible isomerization of the BTDPA molecules between *E*- and *Z*- states. (b-d) Reversible isomerization (*Z*-BTDPA↔*E*-BTDPA) studies of BTDPA using $^{1}$H NMR spectroscopy ($CDCl_3$, 500 MHz) in sunlight and dark conditions (inset images show the color conversion: green to orange and back to green). (e) The concentration of *E*-*Z* isomers versus isomerization time plot for sunlight irradiation and dark conditions (cyan and orange lines show *Z*-BTDPA and *E*-BTDPA concentrations, respectively). (f) Solid-state UV-Vis and optical emission spectra of *E*-BTDPA and *Z*-BTDPA.

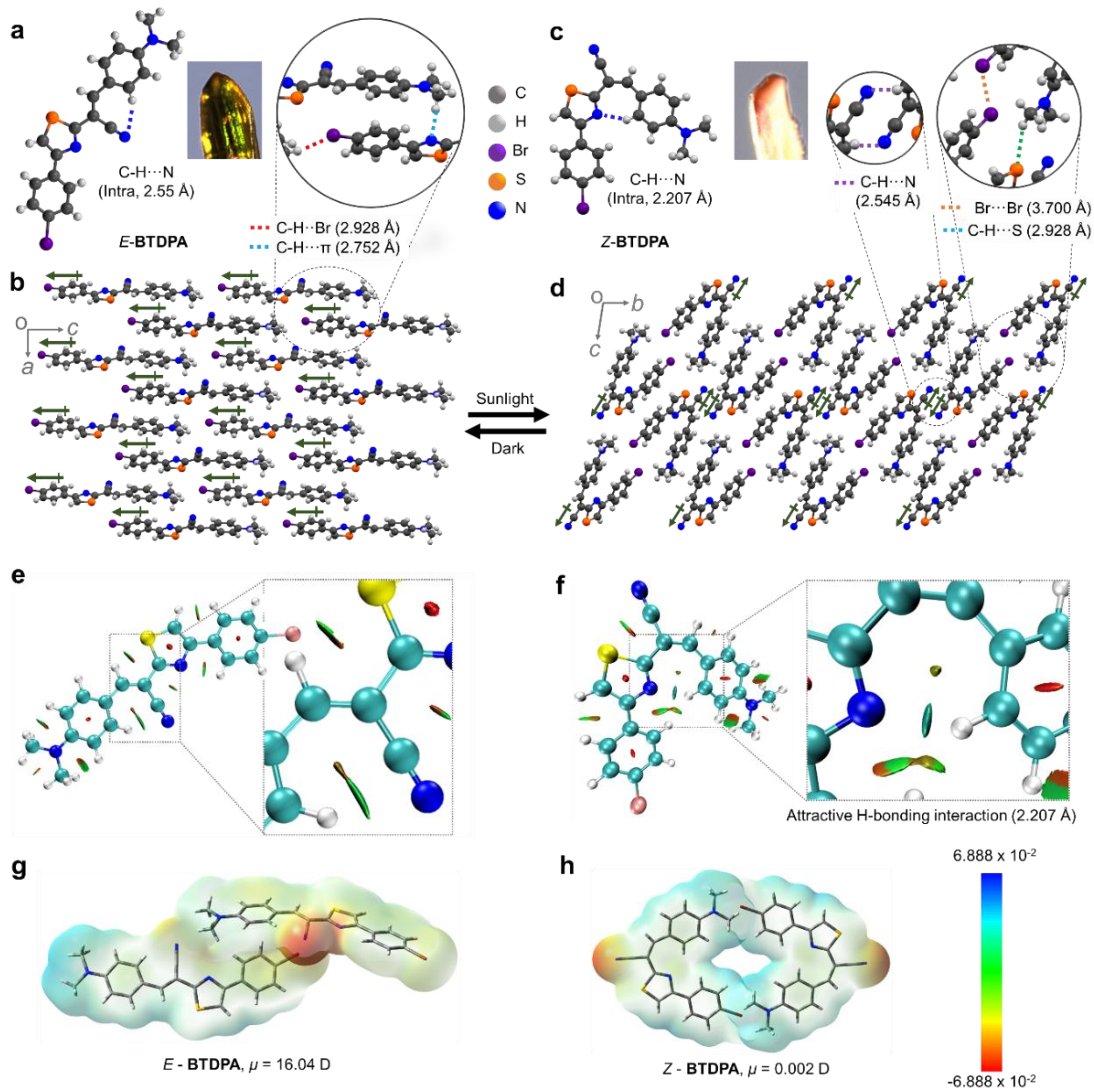

**Figure 2: Crystal structures, molecular packing, noncovalent interaction (NCI) plots, and electrostatic potential (ESP) surface analyses of *E*- and *Z*-BTDPA isomers. a,** Single-crystal X-ray structure of *E*-BTDPA showing intramolecular C–H···N hydrogen-bonding interactions. **b,** Head-to-tail packing of *E*-BTDPA molecules featuring C–H···π and C–H···Br interactions along the crystallographic *c-axis*. **c,** Single-crystal X-ray structure of *Z*-BTDPA displaying intramolecular C–H···N hydrogen bonding. **d,** Molecular packing of *Z*-BTDPA along the crystallographic *a*-axis, stabilized by intermolecular C–H···N, Br···Br, and C–H···S interactions. **e, f,** Three-dimensional color-filled NCI plots illustrating the distribution of noncovalent interactions in *E*- and *Z*-BTDPA, respectively. The cyan and green regions indicate strong and medium-to-weak attractive interactions, respectively, while the red regions denote repulsive contacts. **g, h,** Electrostatic potential surface maps of *E*- and *Z*-BTDPA obtained from molecular simulations showing the ground-state unit-cell dipole moments. The

isovalue for the ESP surfaces was set to 0.0006 atomic unit. The color scale in an ESP plot represents the electrostatic potential energy in the atomic unit.

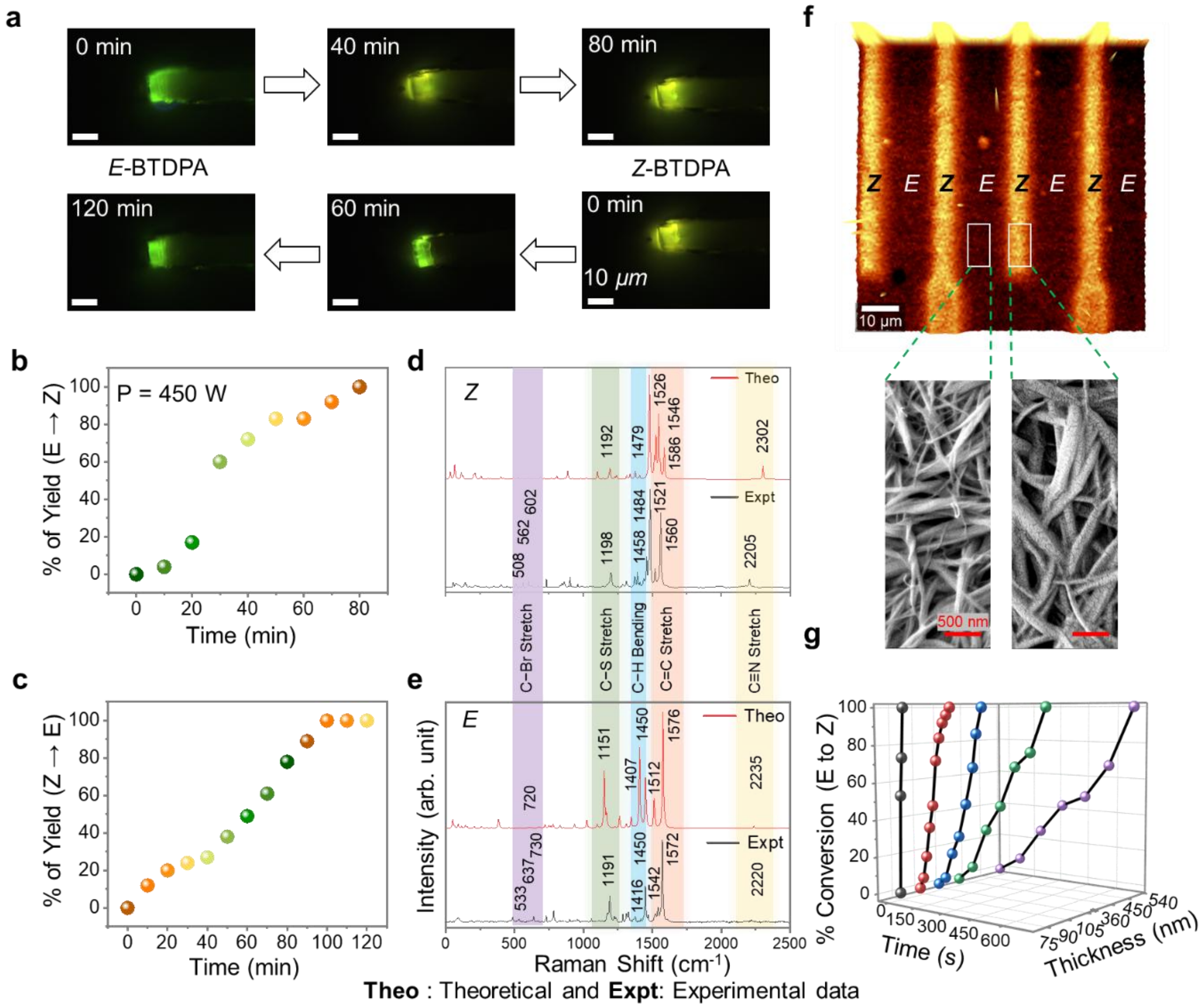

**Figure 3: Photoisomerization studies and characterization of solid-state BTDPA molecules. a**, FL images of single-crystal *E/Z*-BTDPA reversible isomerization under light from a solar simulator and in the dark. **b,c**, Normalized optical and FL spectra recorded at 10 min intervals during (b) the E → Z isomer conversion under sunlight and (c) the Z → E isomer conversion in the dark. **d,e**, Experimental (black line) and calculated Raman spectra (red line) of (d) *Z*-BTDPA and (e) *E*-BTDPA. f, FL imaging of the solid-state thin film of the photo-masked Z/E isomer pattern of BTDPA. The insects show magnified FE-SEM images of both isomers. **g,** The solid-state thin film thickness vs yield of conversion efficiency (from *E* to *Z* isomer with time at a constant power of 450 mW).

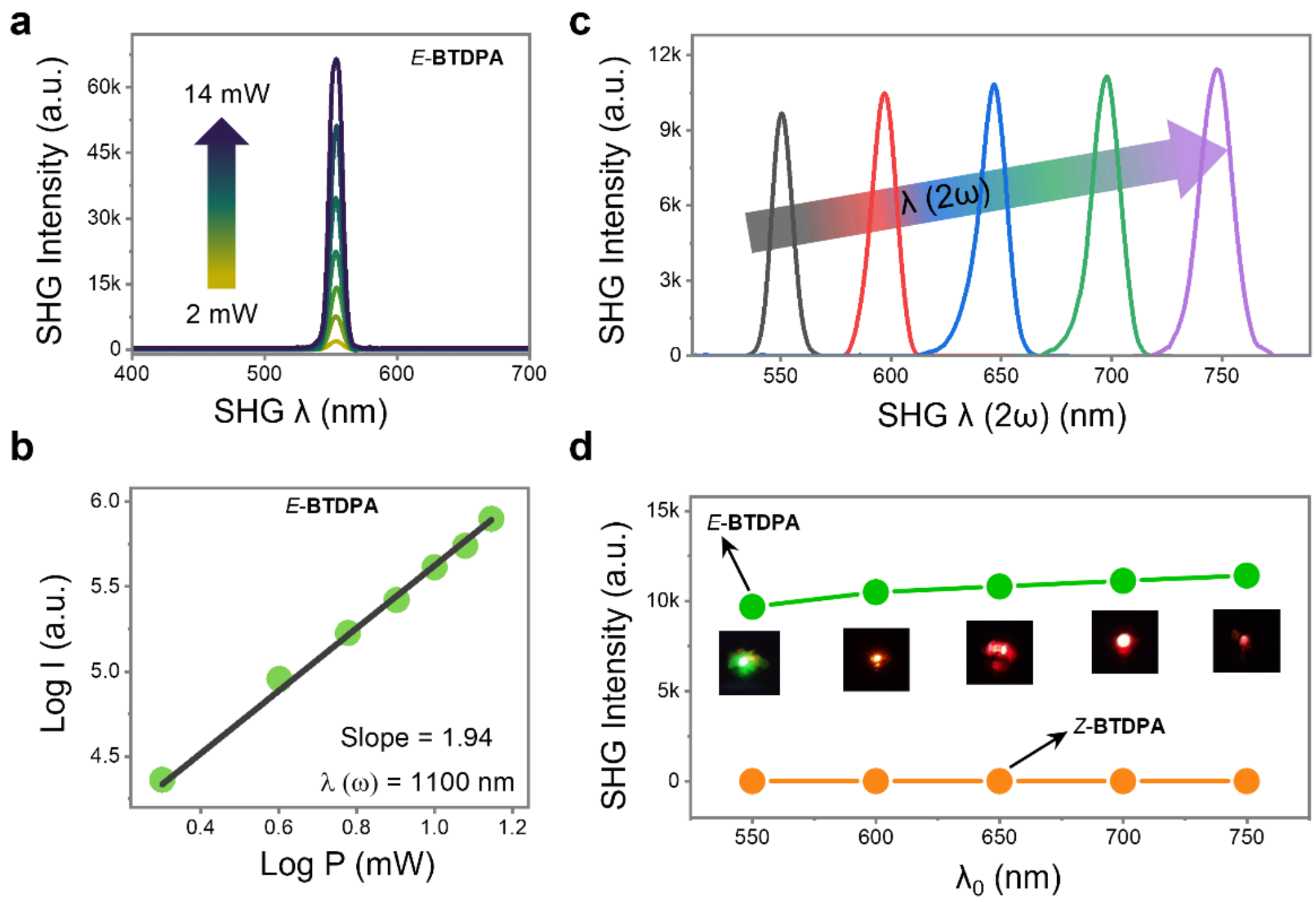

**Figure 4: Nonlinear optical studies of *the E*/*Z*-BTDPA single crystals. a**, Power-dependent SHG optical spectra of the E-BTDPA crystal at a fundamental wavelength of 1100 nm. **b**, Log-log plot of the integrated SHG intensity versus the pump power of E-BTDPA (the black line shows the SHG fit). **c**, SHG intensity vs. wavelength (noncentrosymmetric crystal) using near IR fundamental wavelength (**λ**) increases from 1100 to 1500 nm. **d**, Comparison plot between intensity (SHG ON/OFF) of versus wavelengths for *E*- and *Z*-**BTDPA**.

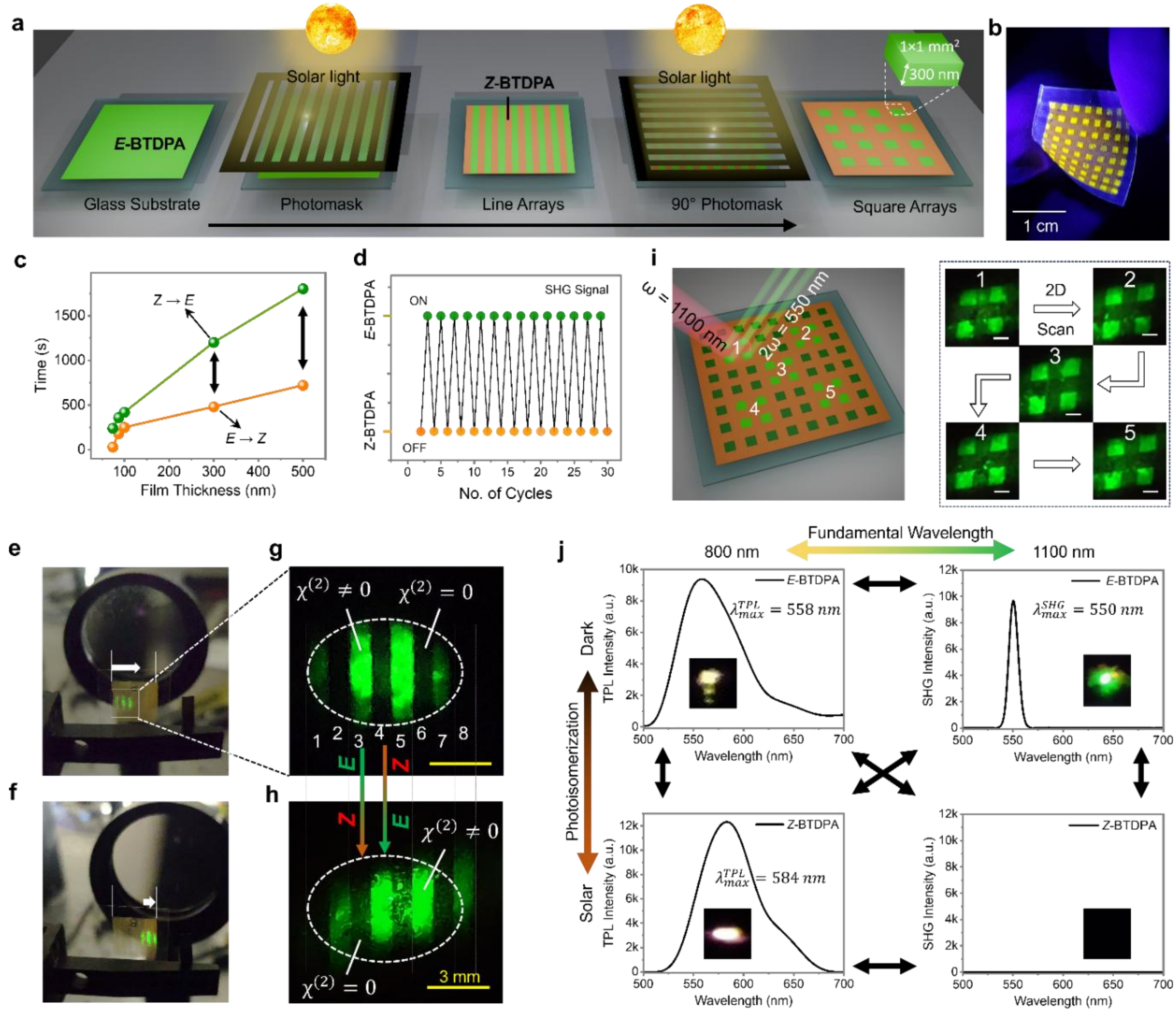

**Figure 5: Photoisomerization-induced nonlinear optical tunability of BTDPA thin-film.** **a**, Graphical illustration of the patterning of Z-**BTDPA** on a PET/glass substrate. **b**, Photograph of a thin, flexible, patterned **BTDPA** film on a PET substrate under UV illumination. **c**, E to Z isomerization of **BTDPA** in a thin film with respect to time. **d**, NLO optical switching performance of the *E/Z*-**BTDPA** thin films. **(e, f)** Translation motion of the thin-film array producing the SHG from the selected pixels. **g,h**, Photographs depict the line-patterned thin film of E/Z isomers with reversible SHG signal appearance with the time of studied devices with the application of 1100 nm fundamental wavelength. **i**, Graphical illustration of the patterned thin-film substrate of *E*-**BTDPA** pixels surrounded by Z-**BTDPA** (each pixel has a 1 x1 $mm^2$ area). The SHG emission images were captured at various positions (1–5) during translational motion with exposure to a 1100 nm fundamental wavelength of a femtosecond laser. Scale bar, 1 mm. **j**, Fully connected K4 representation of all organic nonlinear optical switching processes assisted by photoisomerization (top to bottom) and fundamental-laser switching (left to right).

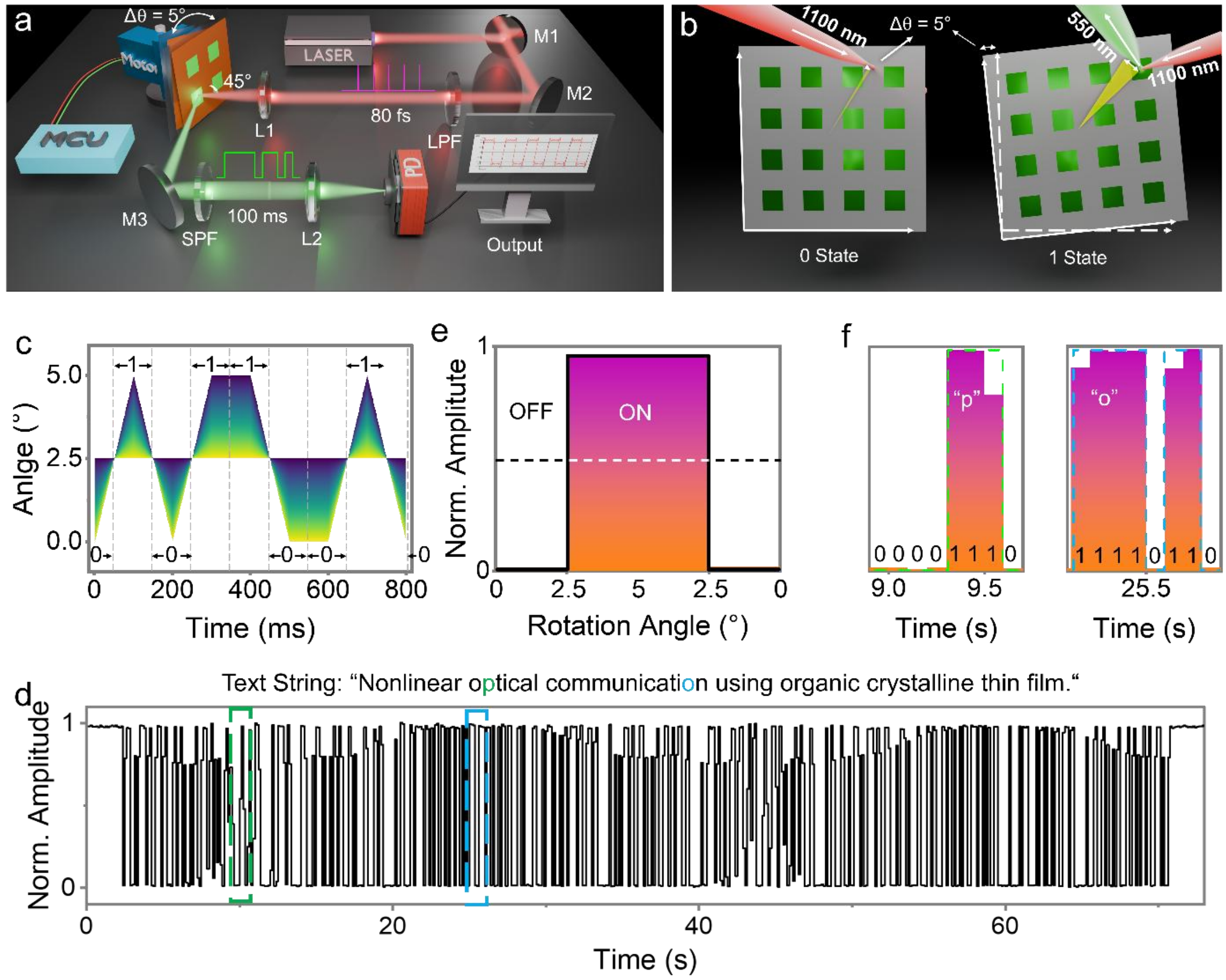

**Figure 6. Time-resolved SHG intensity showing binary modulation synchronized with angular gating. a,** Schematic illustration of the experimental setup. A fundamental beam ($\omega$ ) is directed onto the nonlinear crystalline pixel, while an angularly oscillated optical gate periodically on and off the SHG. The gate is driven through a controlled angular sweep between 0° and 5°, producing deterministic temporal modulation of the optical excitation and the resulting second-harmonic signal ($2\omega$). **b,** The bidirectional rotary chopper (1 mm pixel scintillates 0°-5°-0°, blocking laser 0-2.5° (OFF, bit 0; 50 ms), twitting 2.5°-5° (ON, bit 1; 50 ms) for 50% duty cycle. **c,** Angle-time mapping of the angular gate. **d,** The SHG intensity (colored) was synchronized with the ideal transmission (black), confirming half-duty-cycle modulation for NLO signal detection from the pixelated film. Scale bars: angular range of 0°–5°. **e,** Real-time trace of the received NLO signal (light green) of *the E*-BTDPA pixel surrounded by *Z-BTDPA-based* mechanical-to-photonic transduction and its comparison with the time-domain trace of the full *E*-form (green) and *Z*-form (black) binary sequences (red) for a text string. **f,** A detailed view of the waveform transitions, revealing a mechanical rise of 5 ms, and the least significant bit first concept square wave for "p" and "o" characters for the string.

Supplementary information

# Large-Area Patternable Solar-Powered Bistable Organic Crystalline Film for Nonlinear Optical Communication

Ankur Khapre[1,3], Avulu Vinod Kumar[1,3], Haseeba Nasreen Punathil[1], Biswajit Kumar Barman[1], Bedanta Kumar Deka[2], Shilpa Mangalassery[2], Sri Ram G Naraharishetty[2], Rajadurai Chandrasekar*[1]

1. Advanced Photonic Materials and Technology Laboratory, School of Chemistry and Centre for Nanotechnology, University of Hyderabad, Prof. C. R. Rao Road, Gachibowli, Hyderabad 500046, Telangana, India
2. School of Physics, University of Hyderabad, Prof. C. R. Rao Road, Gachibowli, Hyderabad 500046, Telangana, India
3. Equal contribution of authors

*E-mail: r.chandrasekar@uohyd.ac.in

## Table of Contents:

## 1. Instrumental Methods

### a) High Resolution Mass Spectromatry:

High-resolution mass spectra (HRMS) were recorded using a Bruker Maxis quadrupole time-of-flight (Q-TOF) mass spectrometer (Bruker Daltonics, Bremen, Germany) operated in electrospray ionization (ESI) mode. The instrument was utilized under the following conditions: a capillary voltage of 4500 V, a nebulizer pressure of 0.6 bar, and a dry gas flow rate of 4.0 L $min^{-1}$ at a temperature of 180 °C. Mass calibration was performed using sodium format clusters. Bruker Compass Data Analysis software was used for data processing.

### b) Thermogravimetric analysis:

The thermogravimetric analysis experiment was done in PerkinElmer STA 6000 instrument with 10°C/min, and the data was plotted using origin 2024a software.

### c) µ-Raman Analysis:

For each Raman spectrum, ten accumulations were collected with an acquisition time of 2 s per accumulation. Raman peaks were fitted using Lorentzian functions.

### d) Powder X-ray Diffraction:

The instrument functioned at an accelerating voltage of 40 kV and a current of 30 mA. Data acquisition occurred over a 2θ range of 5–50° with a scan rate of 5° $min^{-1}$.

### e) Single crystal X-ray diffraction:

The crystal structure was determined utilizing Olex2 (v1.2.9) in conjunction with the SHELXT structure solution program, employing the intrinsic phasing method. Subsequent refinement was conducted using SHELXL through least-squares minimization. Anisotropic displacement parameters were applied to all non-hydrogen atoms during refinement. Crystal packing diagrams were generated using Mercury (v4.2.0).

### f) Confocal micro spectroscopy studies:

Spectra were acquired using a CCD detector equipped with a 300 and 600 grooves $mm^{-1}$ grating. The accumulation time was set to 30 s with an integration time of 0.5 s. Each spectrum represents the average of ten accumulations. All measurements were performed under ambient conditions.

### g) Photoluminescence (PL) mappings:

Micro-photoluminescence (µ-PL) imaging experiments were performed in a backscattering configuration using a confocal Raman microscope system (Alpha300 RAS, WITec GmbH) equipped with a CCD detector (Peltier-cooled). Spectra were acquired using a BLZ = 500 nm and 600 grooves $mm^{-1}$ grating, with an accumulation period of 10 s with an integration period of 0.5 s. Each spectrum represents the average of ten accumulations. Optical excitation was provided by a 532 nm Nb:YAG diode laser through a 100× objective. All measurements were performed under ambient conditions.

### h) Fluorescence lifetime imaging microscopy studies:

Fluorescence lifetime (FL) decay profiles were acquired using a confocal fluorescence lifetime imaging microscope (PicoQuant MicroTime 200) in time-resolved manner and integrated with an invertoscope (Olympus IX71). Crystals were dispersed onto a coverslip and excited with a 405 nm solid-state picosecond pulsed laser (5 µW output power, 20 MHz repetition rate, FWHM: 176 pm). Emission was collected using a water-immersion Olympus UPlanApo 60× objective (NA 1.2), and the excitation beam was blocked with a 430 nm high pass filter. Signal acquisition was carried out using a time-correlated single-photon counting module (PicoHarp 300). The data were analyzed using SymPhoTime software. All measurements were performed at room temperature.

### i) Field-Emission scanning electron microscopy studies

All samples dispersed solution were dropcasted on carbon coated copper grid. Before imaging, the gold of an approximately 10 nm thick layer was coated to reduce the charging.

## 2. Synthesis of BTDPA:

**Supplementary Figure 1.** Synthetic process of BTDPA compound.

100 mg, (0.358 mmol) of 4-(4-Bromophenyl)-2-thiazoleacetonitrile, NaOMe (29 mg, 1.5 eq.) was added in round bottom flask containing 15mL of dry methanol and stirred for 20mins. To this, 4-(dimethylamino)benzaldehyde 59 mg, (1.1 eq.) was slowly added. The above mixture was stirred at 60 ºC (oil bath) for 4 h. Upon completion of the reaction, the product mixture was cooled, resulting in the formation of a yellow crystalline precipitate. This precipitate was subsequently filtered, washed with cold methanol, and air-dried for 12 hours to yield a greenish-yellow solid of BTDPA, with a yield of 68%. $^{1}$H NMR (500MHz, $CDCl_3$): δ 8.07(s, 1H), 7.93(d, J=8.5 Hz, 2H), 7.83(d, J=8.5 Hz, 2H), 7.55(d, J=8.5 Hz, 2H), 7.43(s, 1H), 6.72(d, J=8.5 Hz, 2H), 3.09(s, 6H). $^{13}$C{$^{1}$H} NMR (125 MHz, $CDCl_3$) δ (ppm) 164.61, 161.228, 155.09, 154.97, 152.60, 152.48, 148.16, 144.91, 133.68, 133.12, 133.00, 132.66, 131.98, 131.89, 128.07, 127.97, 122.44, 122.38, 120.65, 120.56, 120.31, 118.42, 114.72, 112.75, 111.67, 110.99, 99.89, 97.49 FTIR (Neat; $\underline{\nu}$ in $cm^{-1}$): 2216 (-C≡N), 1607 (-C=C-) HRMS (ESI-TOF) (m/z): Calculated for $[M]^+$, $[M+H]^+$ (410.0326); Found 410.0327.

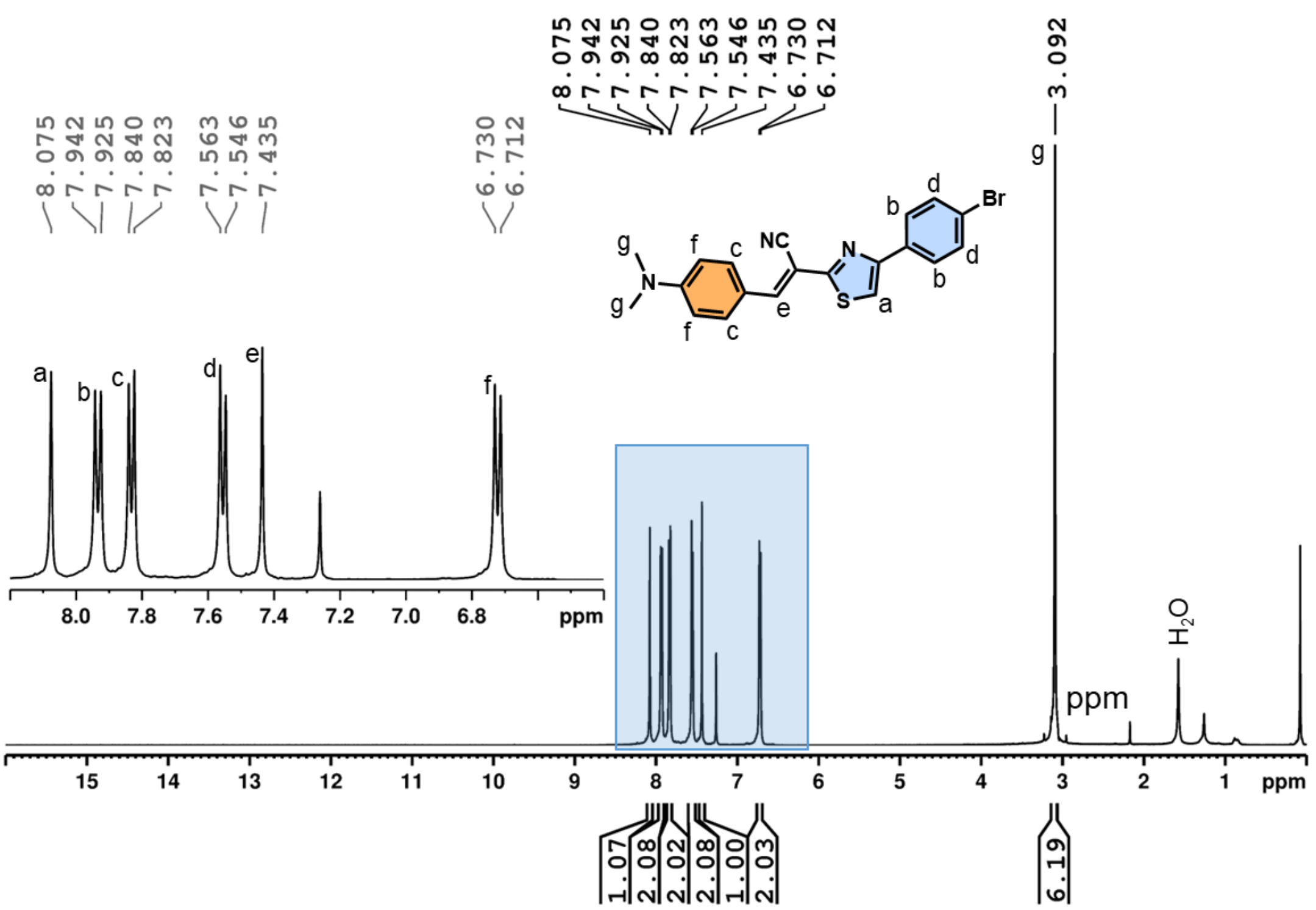

**Supplementary Figure 2.** $^{1}$H-NMR spectrum of *Z*-BTDPA in $CDCl_3$ Solvent.

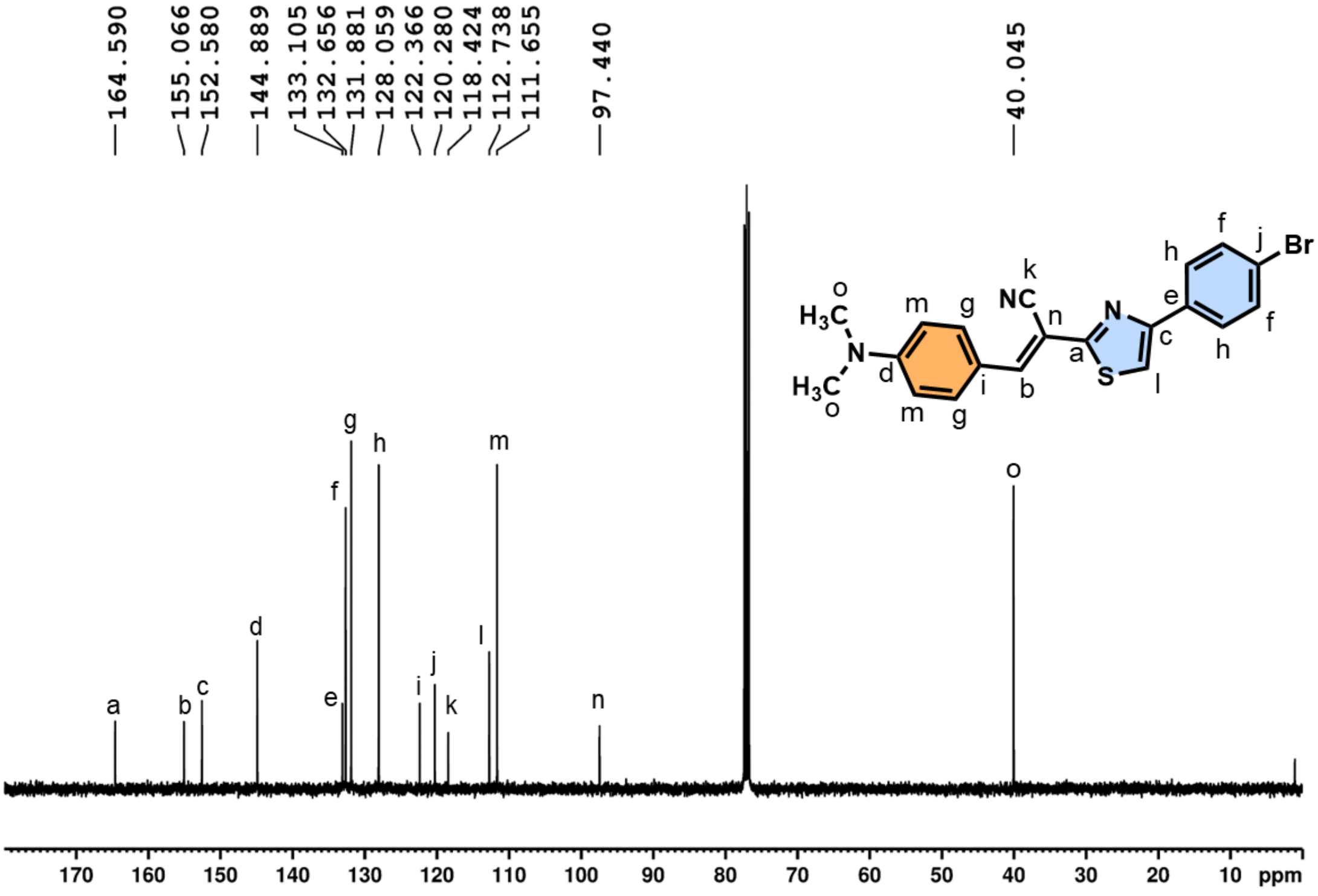

**Supplementary Figure 3.** $^{13}$C-NMR spectrum of *Z*-BTDPA in $CDCl_3$ Solvent.

## 3. Isolation of Isomers

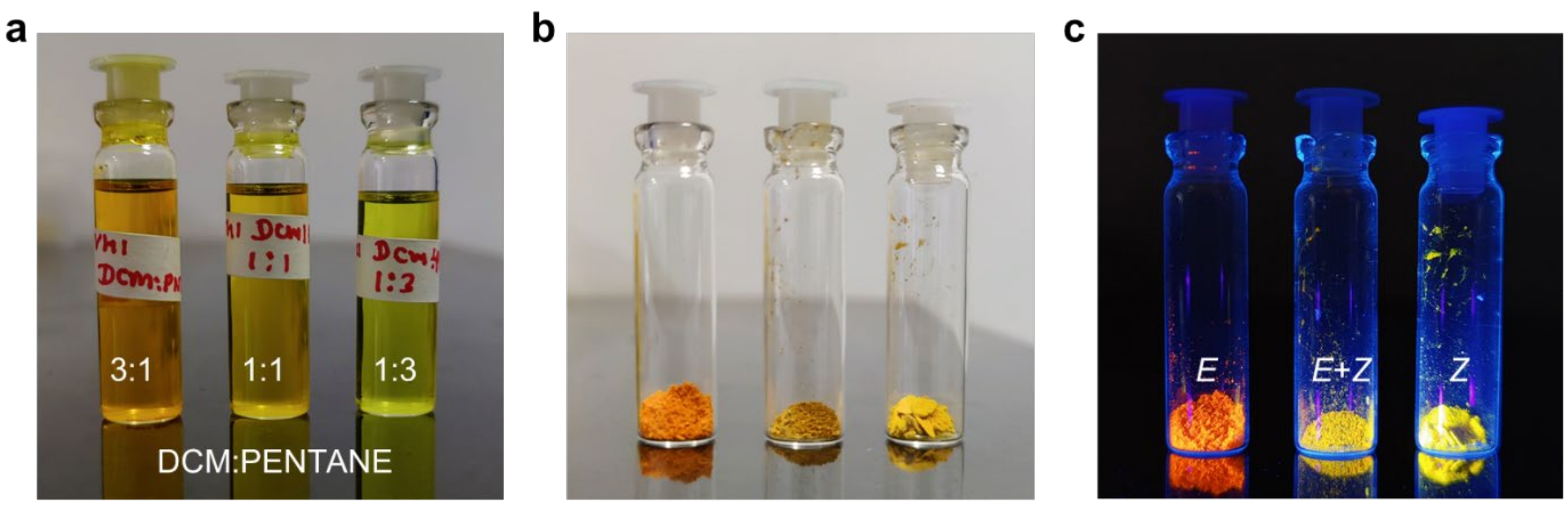

**Supplementary Figure 4. a**, Photograph of BTDPA compound solution in dichloromethane: pentane 3:1(left), 1:1(mid) and 1:3(right) solutions to obtain an orangish yellow(left), a mixture of both (mid) and yellowish green(right) isomers, respectively. **b**,**c**, Camera photographs in visible light and UV-light of vials containing orangish yellow(left), a mixture of both (mid), and yellowish green(right) isomers, respectively.

## 4. Thermal stability studies

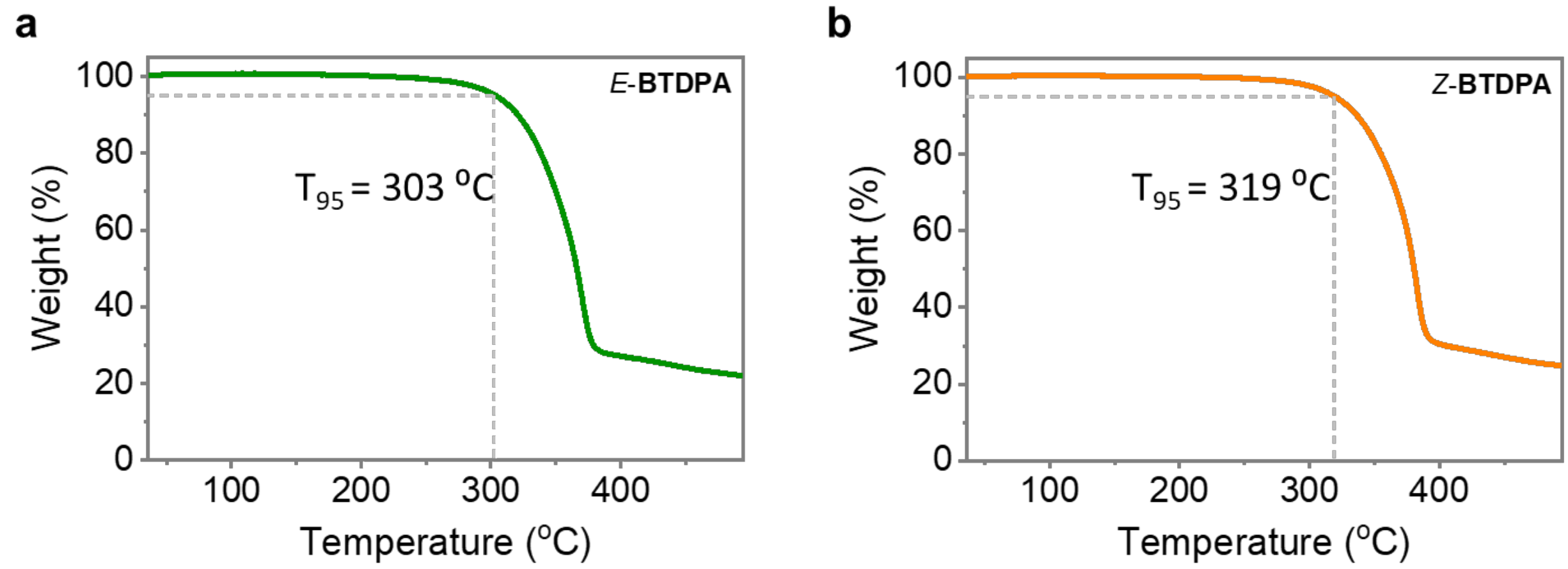

**Supplementary Figure 5: a-b**, TGA plots for *E*-BTDPA and *Z*-BTDPA isomers ($T_{95}$ represents the temperature where 5% of compound degrades).

## 5. Photoisomerization studies using NMR spectroscopy

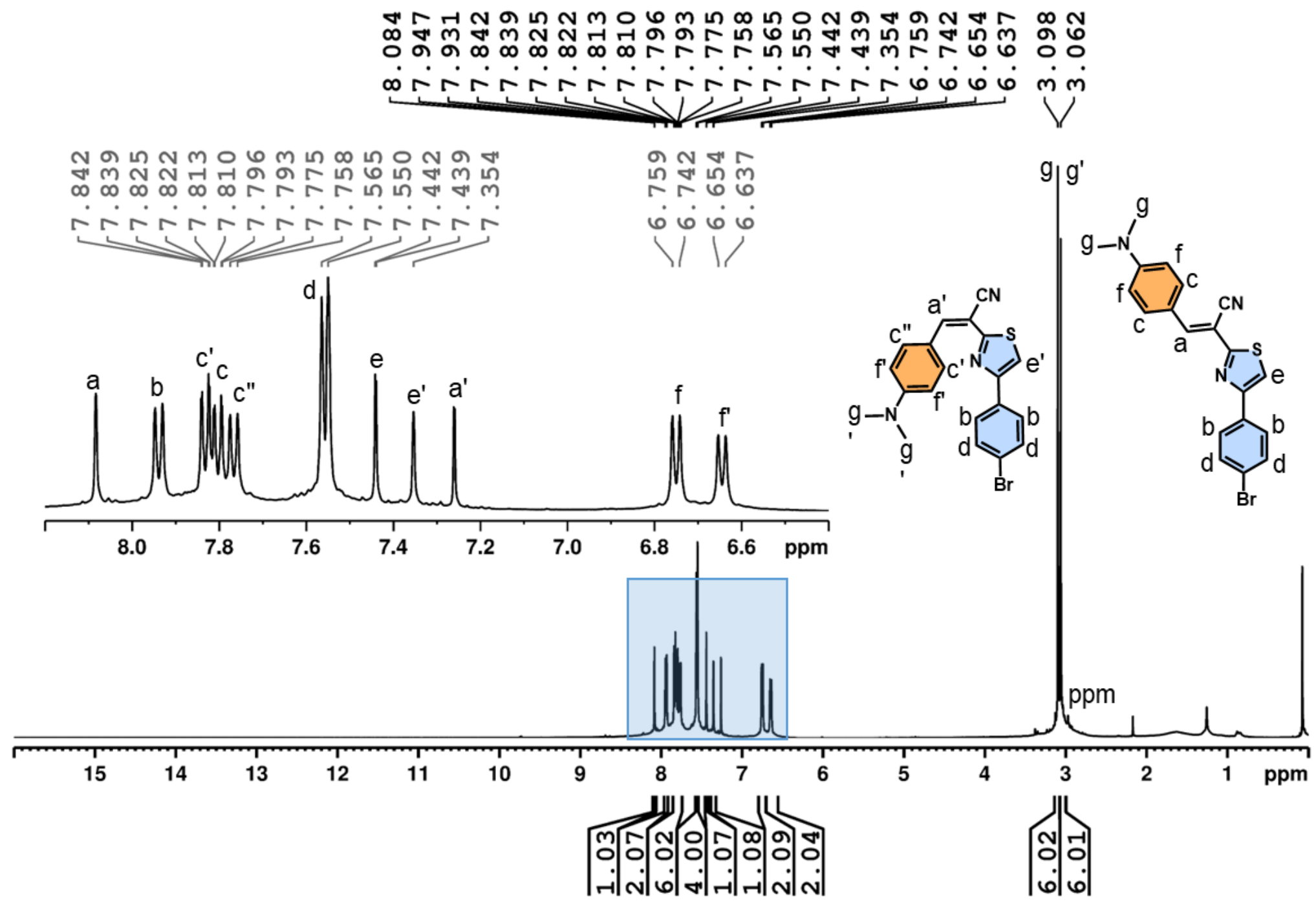

**Supplementary Figure 6**. $^{1}$H-NMR spectrum of $E \rightarrow Z$ conversion BTDPA in $CDCl_3$ Solvent.

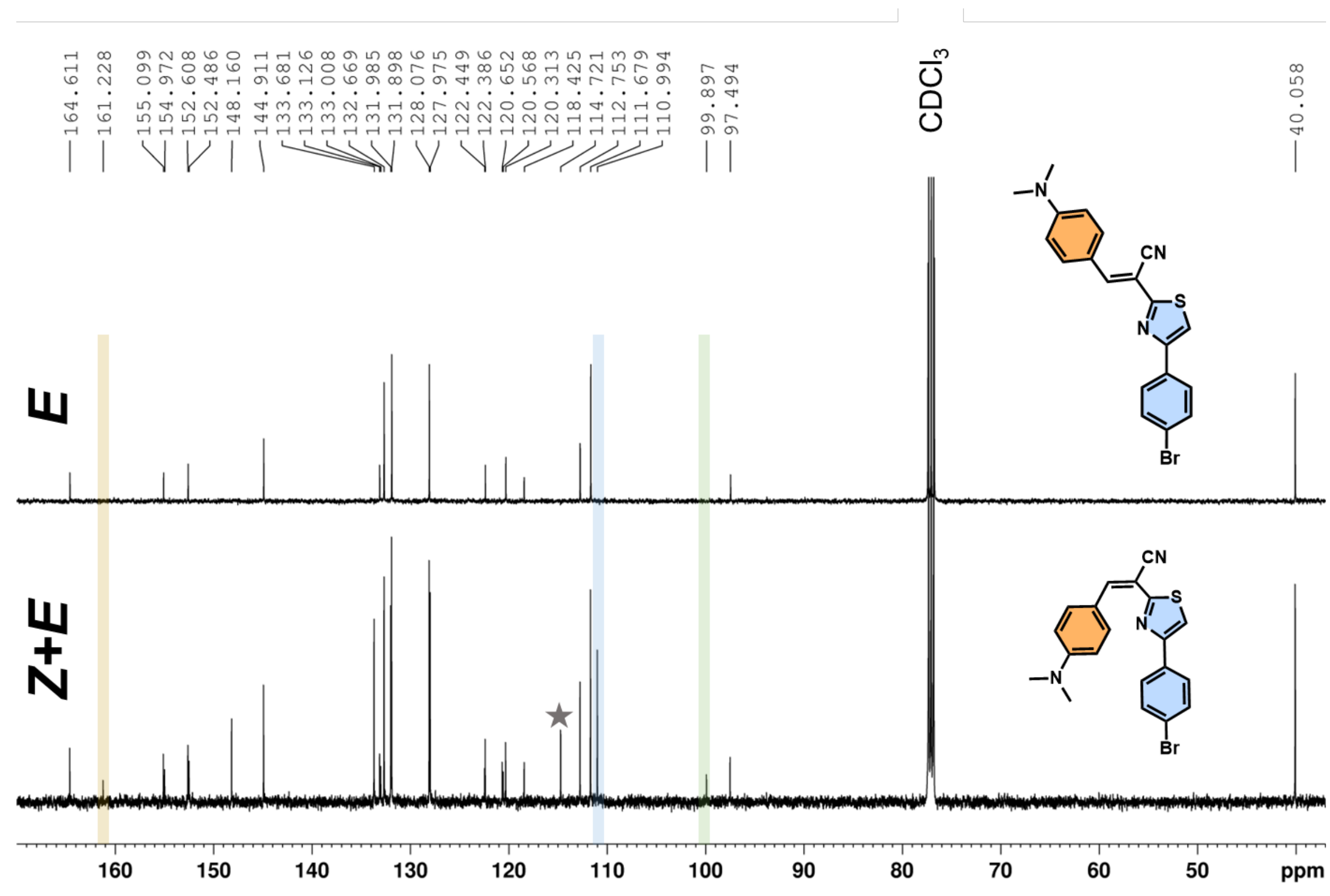

**Supplementary Figure 7**. $^{13}$C-NMR of *E*-**BTDPA** and *Z*-**BTDPA** compounds in $CDCl_3$ solvent. Gray star represents $CH_3CN$. impurity.

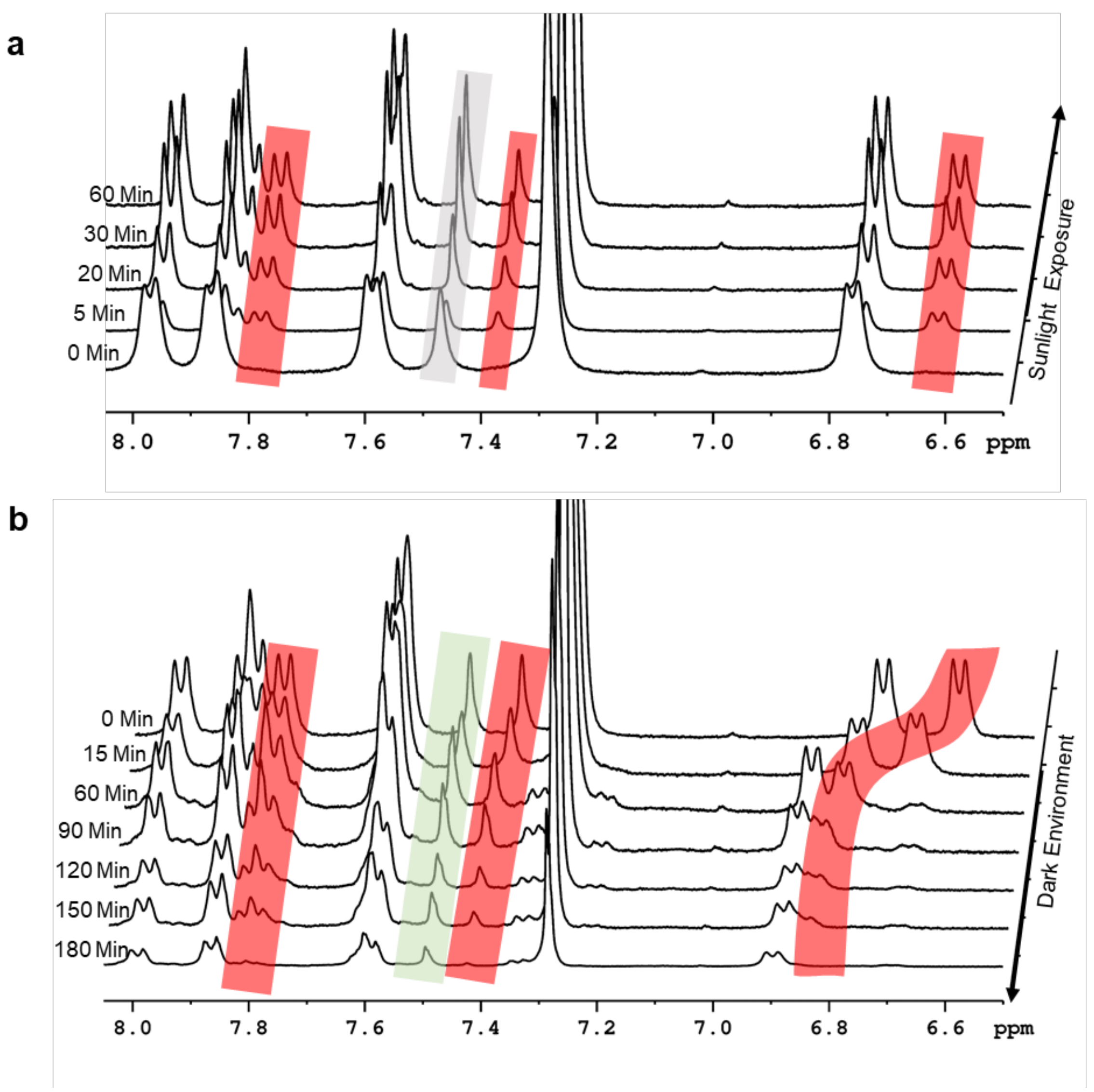

**Supplementary Figure 8**. **a**, Sunlight induced photoisomerization studies from *E*-**BTDPA** → (*Z*+*E*)-**BTDPA** using $^{1}$H-NMR studies (0 to 60 min (maximum conversion 1:1)). **b**, $^{1}$H-NMR studies for the reversibility under dark condition at RT from (*Z*+*E*)-**BTDPA** → *E*-**BTDPA** (0 to 180 min).

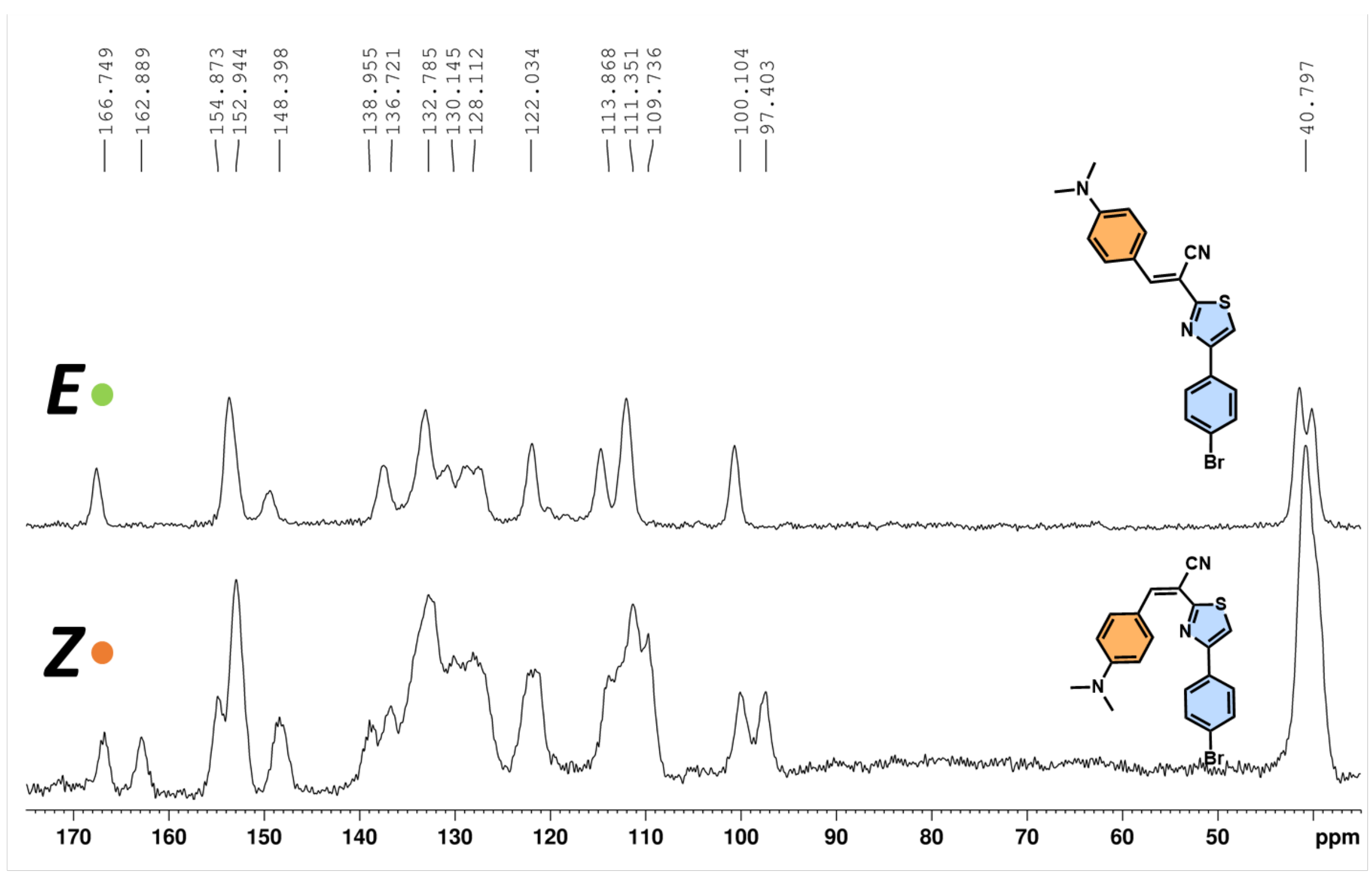

**Supplementary Figure 9.** $^{13}$C-NMR of *E*-**BTDPA** and *Z*-**BTDPA** compounds in the solid state.

## 6. Solution-state optical studies of *E*/*Z*-BTDPA

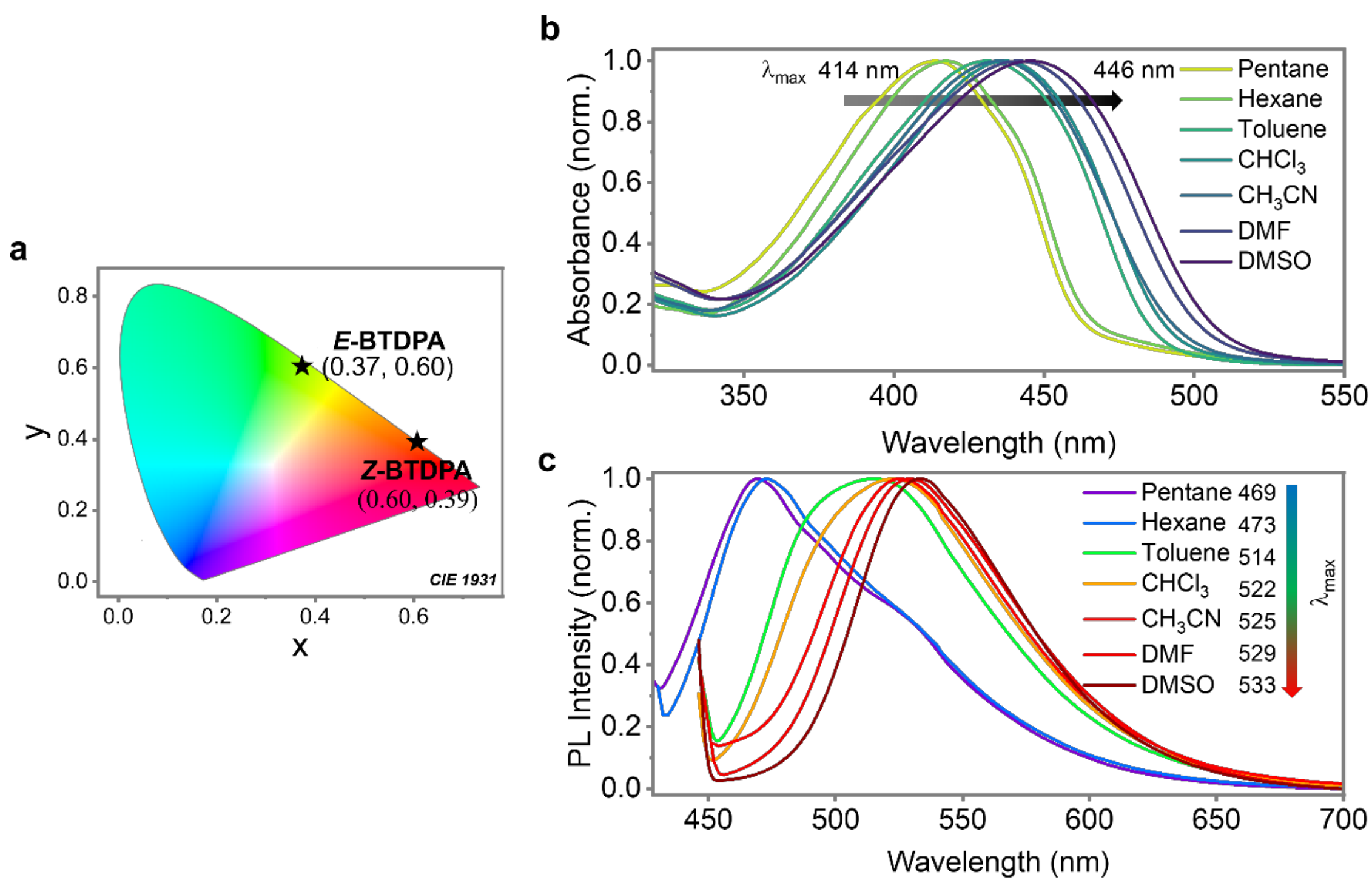

**Supplementary Figure 10: a,** Commission Internationale de L'Eclairage (CIE) coordinates of *E* & *Z*-BTDPA in solid state. **b, c**, Solvatochromic effect of *E*-BTDPA molecules in different solvents, **b**, normalized absorption spectra and **c**, Normalized emission spectra.

## 7. Theoretical calculations of *E*/*Z*-BTDPA:

To further understand the molecular level stability of *E/Z*-**BTDPA**, theoretical calculations have been performed using the WB97XD level of theory and 6-311++g (2d,2p) basis set.[1-2,6] The HOMO-LUMO energy band gap aids in determining the molecule's stability. The energy gap was determined to be 6.58 eV for the Z isomer and 6.32 eV for the E isomer, indicating that the Z isomer is more stable under its unit-cell packing conditions. This interpretation is further supported by the more negative Gibbs free energy (G) of the *Z*-isomer unit cell (−20,519,932.8535 kJ/mol) compared with that of the *E*-isomer unit cell (−20,519,464.91847 kJ/mol). The distributions of the frontier molecular orbitals demonstrated that HOMO of *E*-**BTDPA** molecules is mainly localized on the electron-rich dimethylamino group which extended to the benzene ring of bromophenyl group. In contrast, the Z-**BTDPA** molecule exhibits a more delocalized HOMO distribution between both molecules in the unit cell, with the donor group contributing significantly to the frontier molecular orbital distribution, while the acceptor group contributes minimally. This stabilization of HOMO energy increases the band gap of the Z-**BTDPA** molecules. The LUMO of *E*-**BTDPA** molecules are mainly localized on the thiazolacrylonitrile group of other molecules of the unit cell. In the case of the *Z*-**BTDPA** molecule, LUMO distribution is mostly on the thiazolacrylonitrile group with little extension to the central phenyl ring of the bromophenyl group, which hinders charge delocalization between two molecules of the unit cell (Supplementary Fig. 11). Due to that, both molecules have nearly the same LUMO energy in the range of –0.67 eV to –0.69 eV, suggesting the LUMO levels' small influence on the energy band gaps.

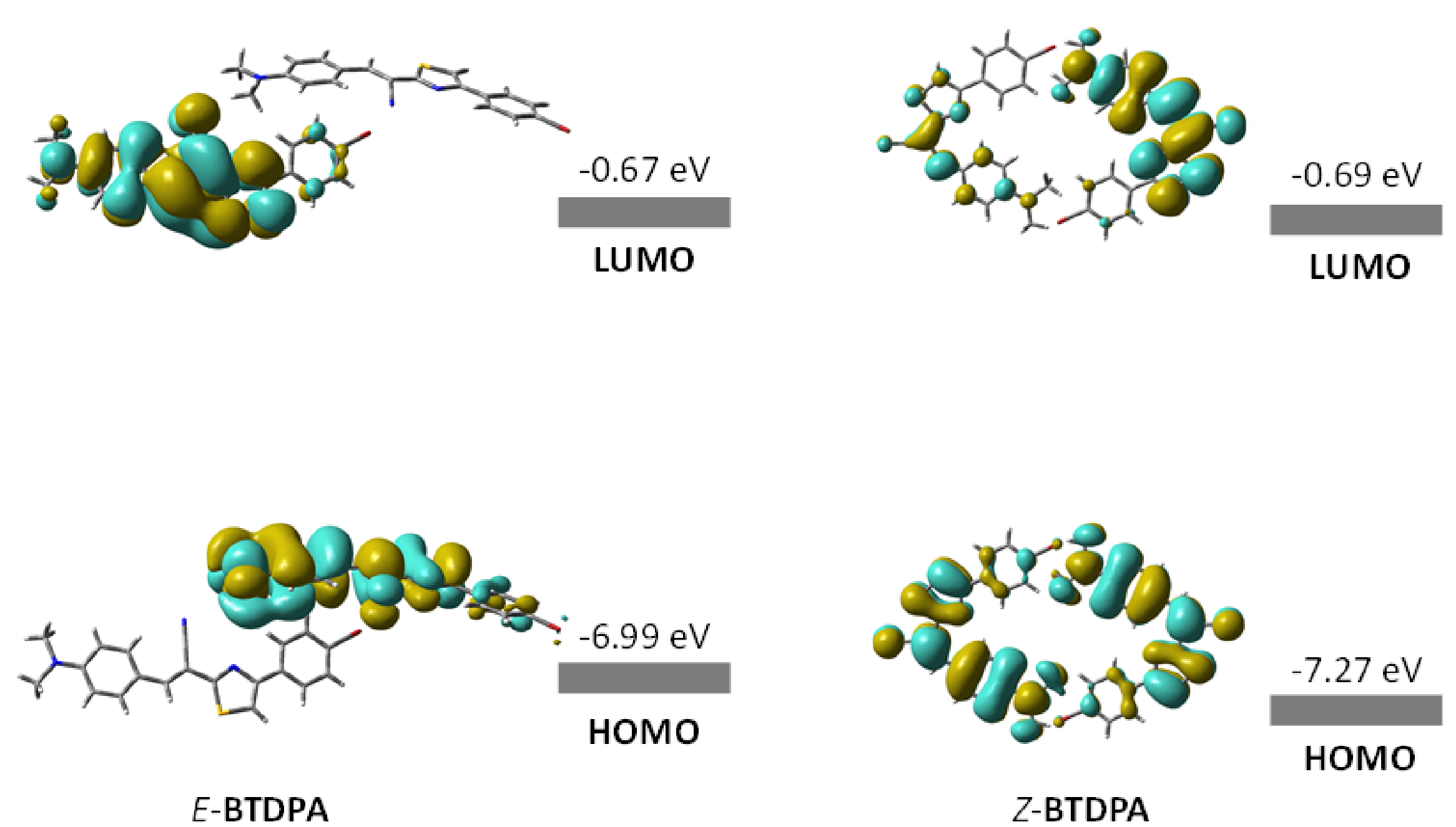

**Supplementary Figure 11.** Spatial plots of frontier molecular orbital energies of HOMO and LUMO of the *E*/*Z*-**BTDPA** compounds at the WB97XD/6-311++g (2d,2p) level. The isovalue used for frontier molecular orbital visualization is 0.01.

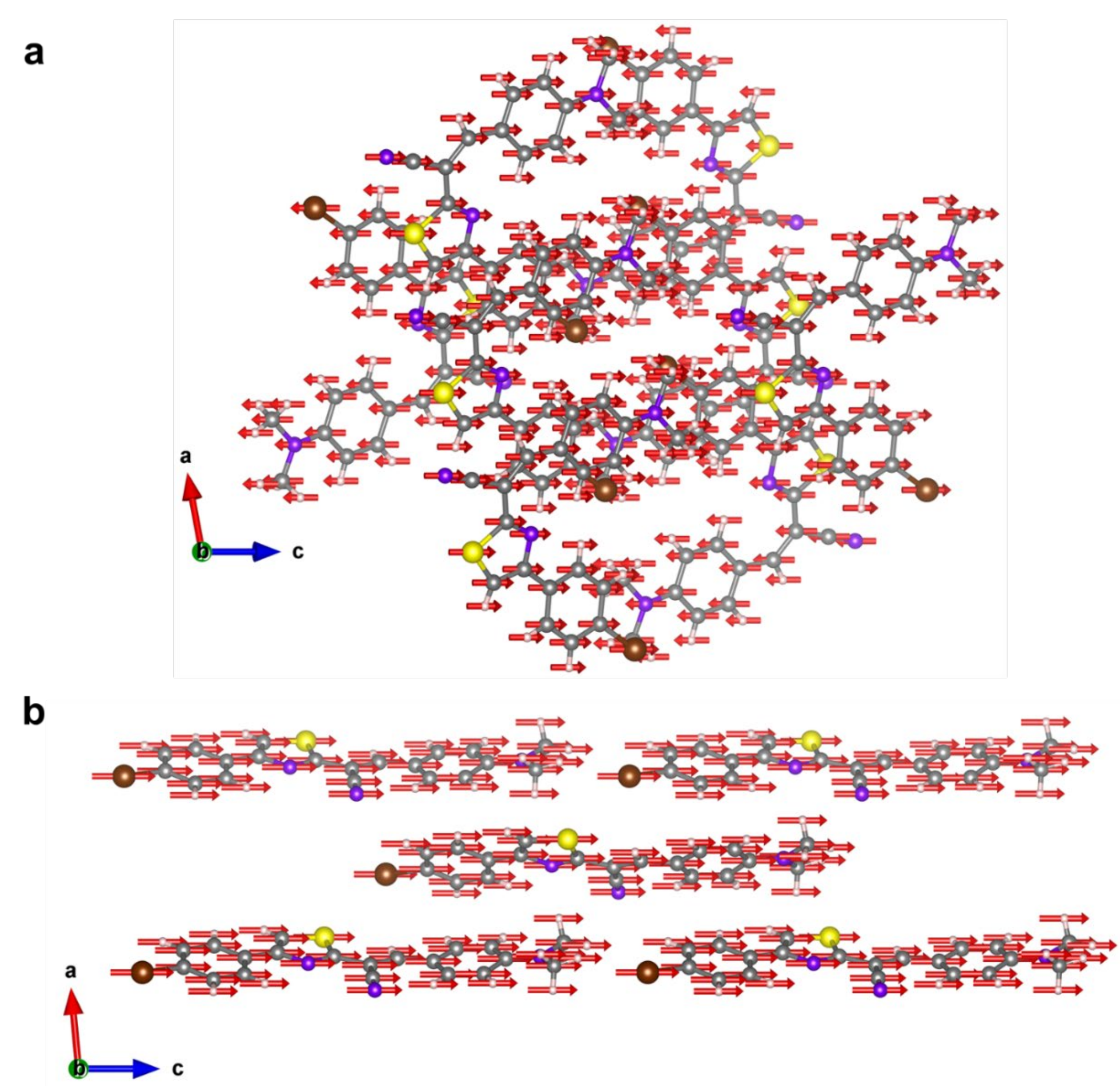

**Supplementary Figure 12: a**, Molecular arrangement of *Z*-**BDTPA** crystal and **b**, *E*-**BDTPA** crystal viewed along the crystallographic b direction. The red arrows show the orientation of the polar vectors.

## 8. X-ray diffraction studies of *E*/*Z*-BTDPA

| Name | *E*-BTDPA | *Z*-BTDPA |
|---|---|---|
| Molecular name | (E)-2-(4-(4-bromophenyl)thiazol-2-yl)-3-(4-(dimethylamino)phenyl)acrylonitrile | (Z)-2-(4-(4-bromophenyl)thiazol-2-yl)-3-(4-(dimethylamino)phenyl)acrylonitrile |
| CCDC | 2502331 | 2504145 |
| Molecular formula | $C_{20}H_{16}BrN_3S$ | $C_{20}H_{16}BrN_3S$ |
| Crystal system | Monoclinic | Triclinic |
| Space group | Pn (7) | $P\bar{1}$ |
| Cell length(Å) | a 7.6609 | a 6.1329 |

|  | b 6.1148<br>c 19.6526 | b 12.1990<br>c 12.4396 |
|---|---|---|
| Cell angles(⁰) | α 90<br>β 95.335<br>γ 90 | α 104.613<br>*β* 97.811<br>γ 98.589 |
| Cell volume($Å^3$) | 916.635 | 875.69 |
| Density ($Å^{-3}$) | 1.487 g/cm³ | 1.556 g/cm³ |
| z | 2 | 2 |

**Supplementary Table 2.** Single crystal X-ray unit cell parameter data for *E*/*Z*-BTDPA.

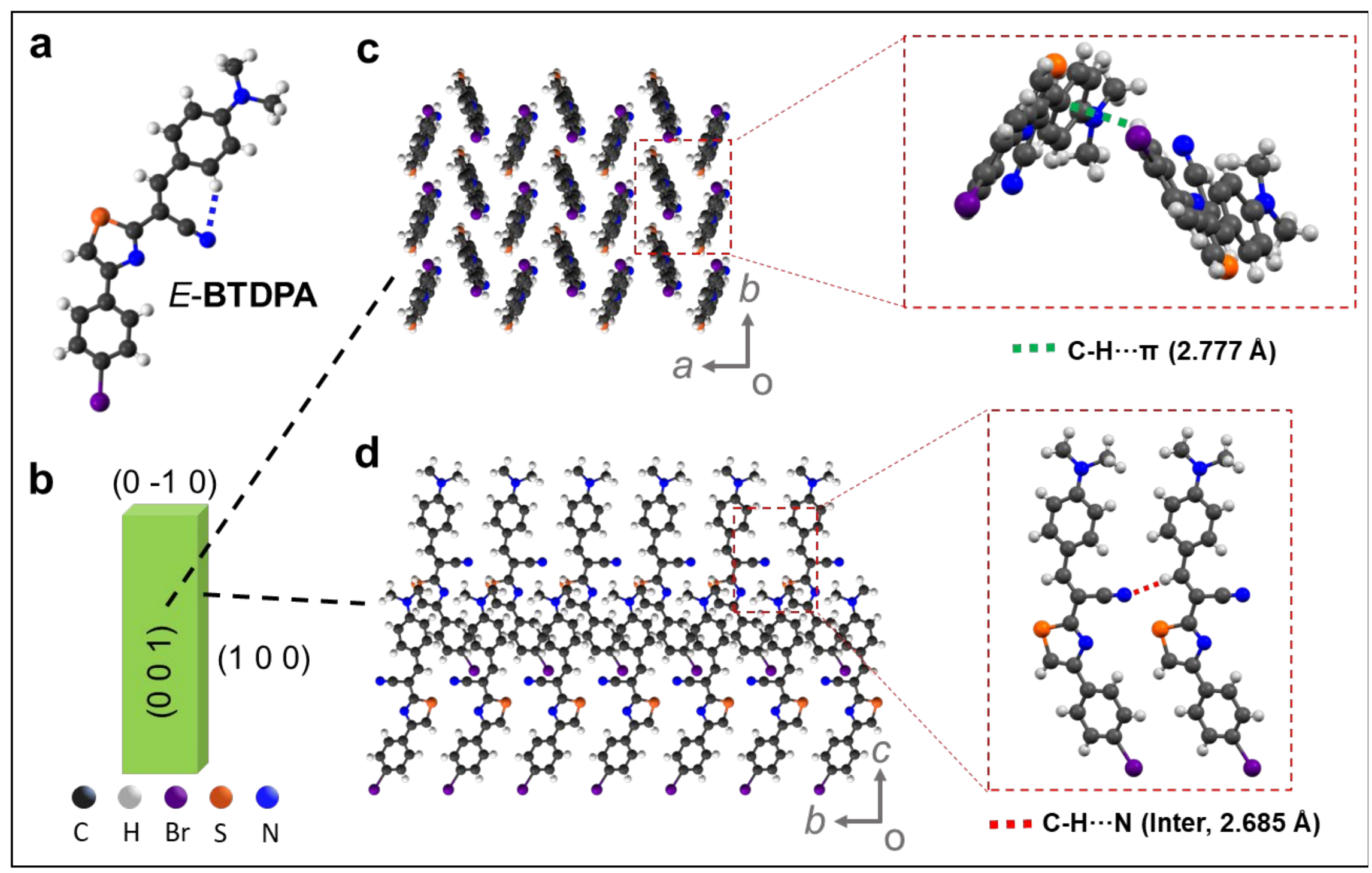

**Supplementary Figure 13.** **a**, *E*-BTDPA SCXRD structure having intramolecular H-bonding. **b**, Graphical representation of *E*-BTDPA and hkl values with its corresponding plans. **c**, Intermolecular C-H···π interactions between adjacent molecules along with *c*-axis. **d**, Intermolecular C-H···N interactions along with *a*-axis for E-BTDPA.

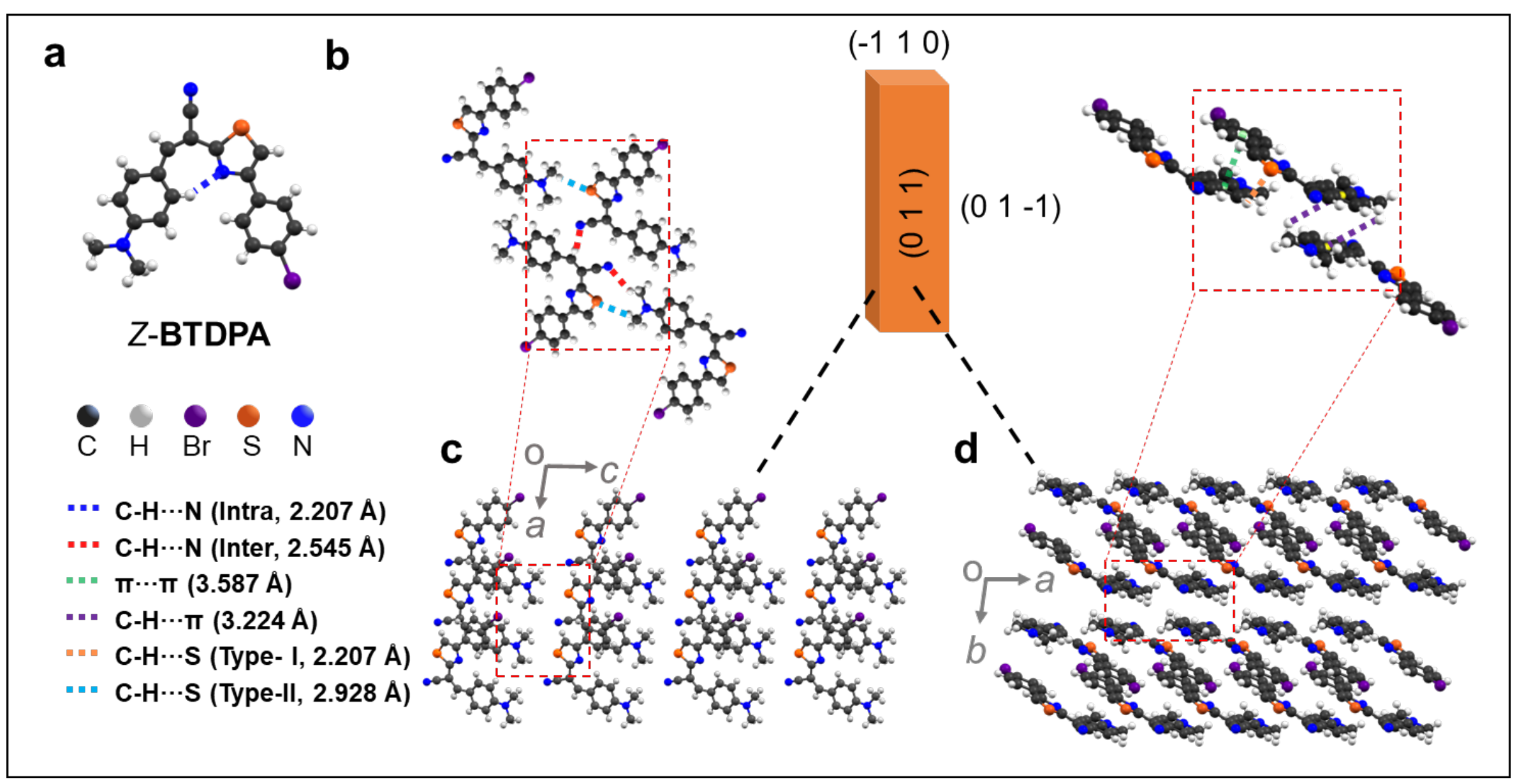

**Supplementary Figure 14. a**, *Z*-BTDPA SCXRD structure having intramolecular H-bonding. **b**, Graphical representation of *Z*-BTDPA and hkl values with its corresponding plans. **c**, Intermolecular interactions between adjacent molecules along with *b*-axis. **d**, Intermolecular interactions along with *c*-axis for *Z*-BTDPA.

## 9. Photoisomerization in single crystal

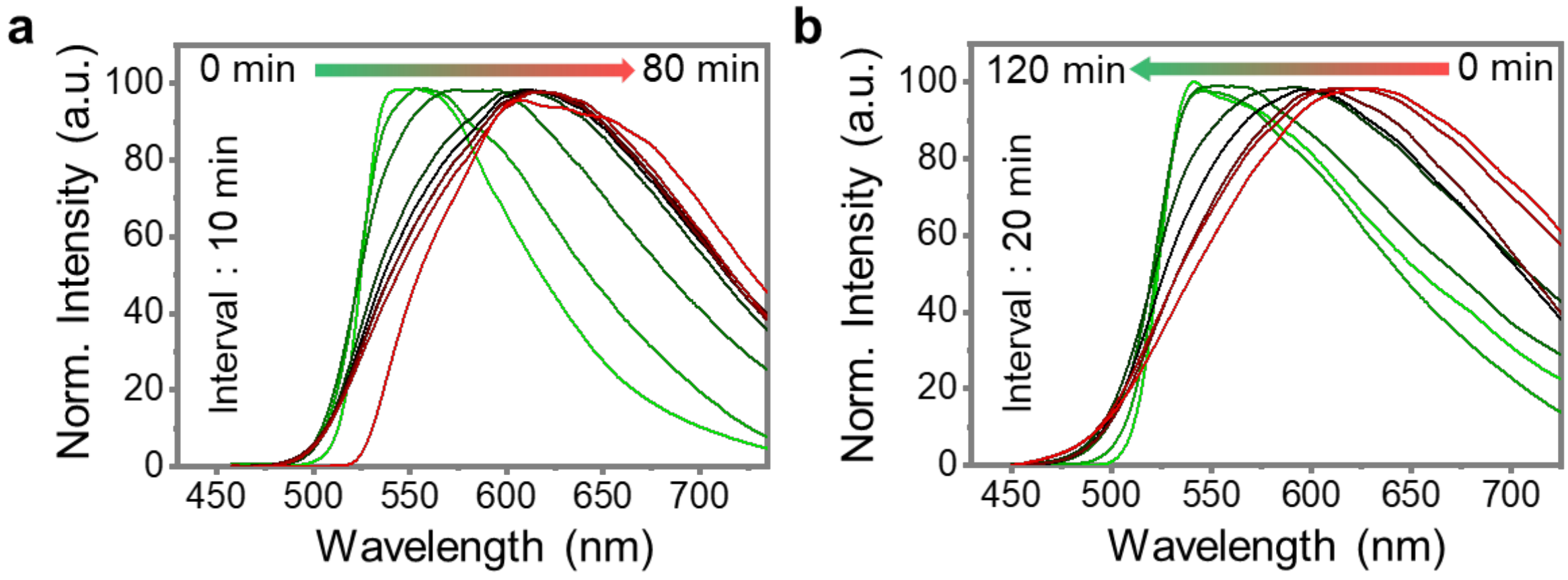

**Supplementary Figure 15. a**, Optical waveguiding during the interval of 10 min of sunlight exposure. **b**, Optical waveguiding under dark condition with interval of 20 min.

## 10. Thin-film device fabrications and its linear and nonlinear optical waveguiding studies

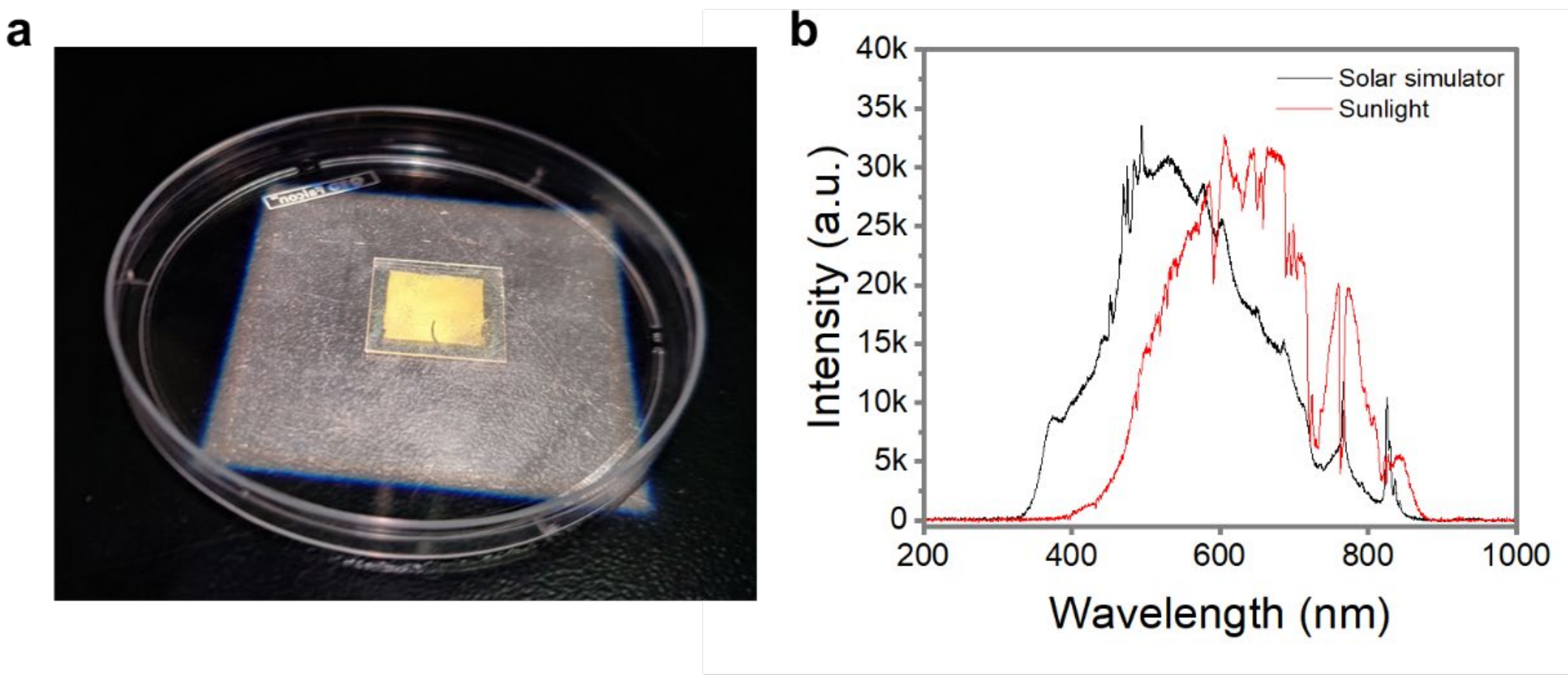

**Supplementary Figure 16. a**, Thin film device pictures during light exposure using a solar simulator. **b**, Solar light spectrum generated by the solar simulator and sunlight spectrum.

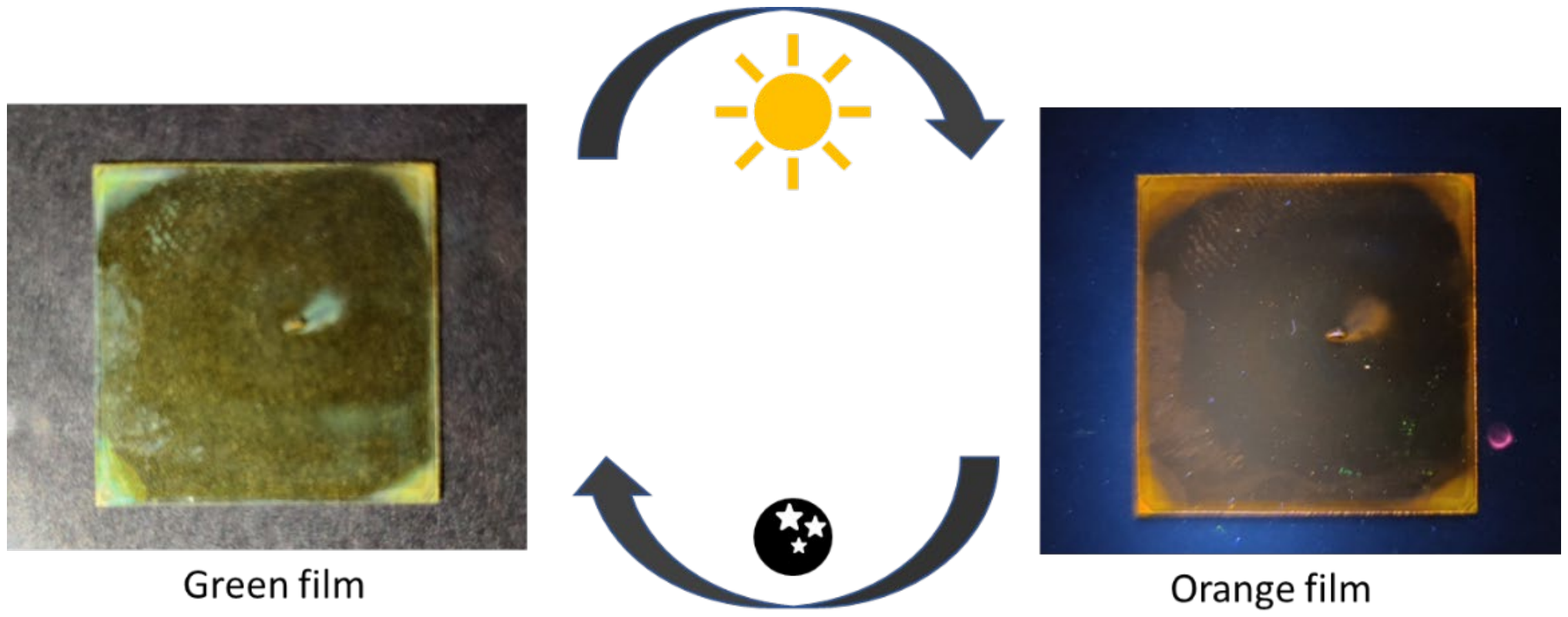

**Supplementary Figure 17.** Spin-coated thin film of BDTPA in IPA (1 mg / 200 μL) conversion in sunlight.

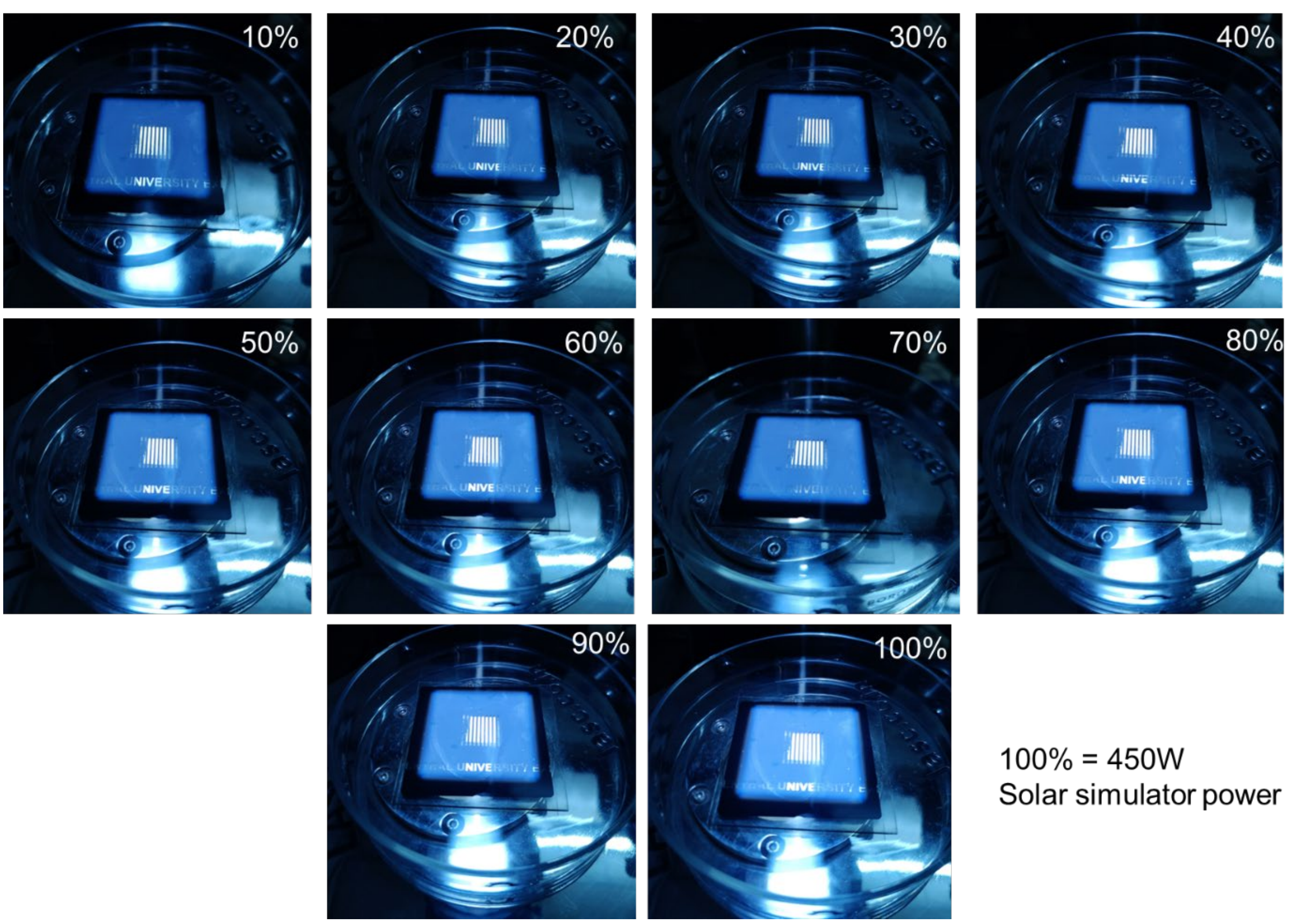

**Supplementary Figure 18.** Photographs of the solar simulator light irradiation of photo-masked BDTPA thin film with 10-100% solar pawer with interval of 10%.

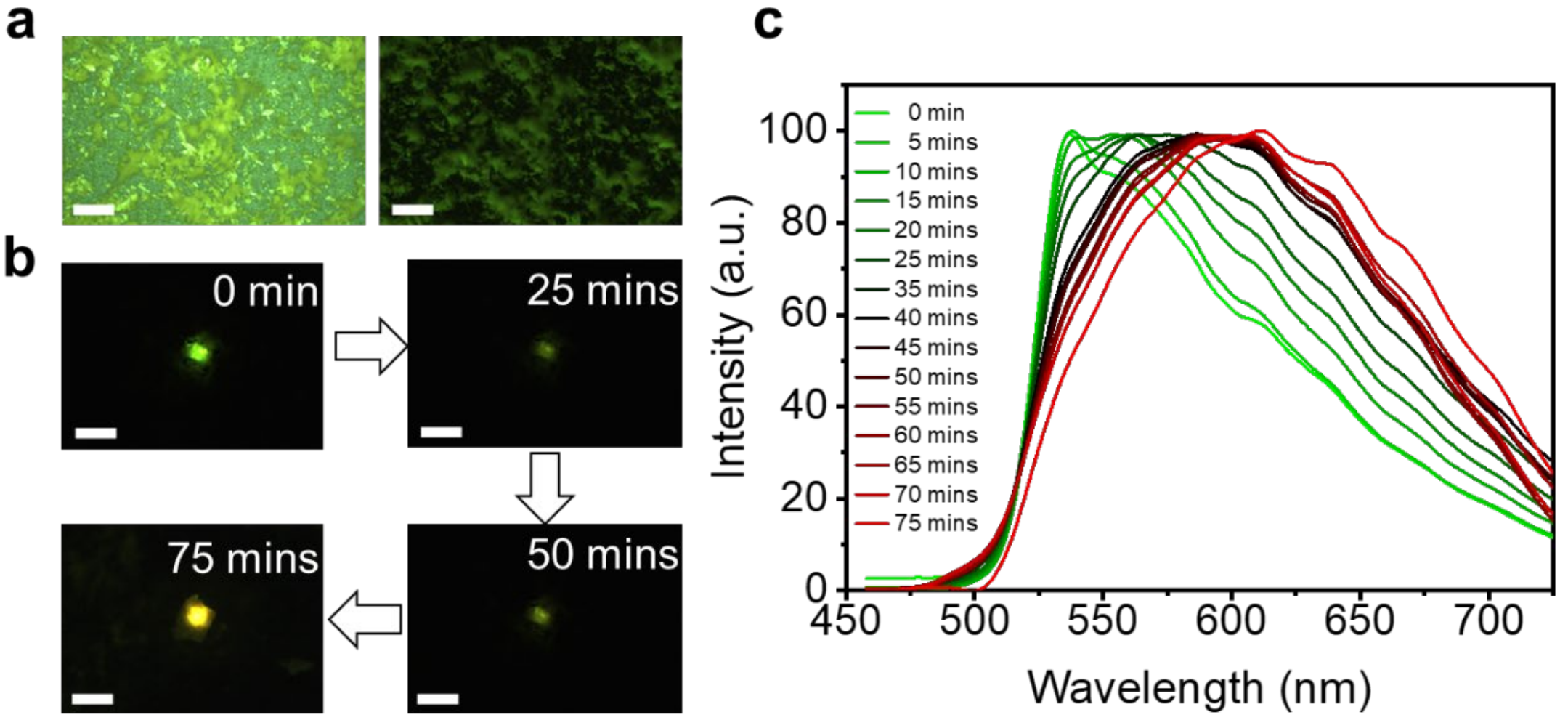

**Supplementary Figure 19.** **a**, Confocal optical and FL images of E-BTDPA thin film. **b**, Optical FL Images of the time interval during solar irradiation exposure of maximum 450 W. **c**, Optical waveguiding spectra during E- to Z-BTDPA transformation. (Scale bar: 10 μm)

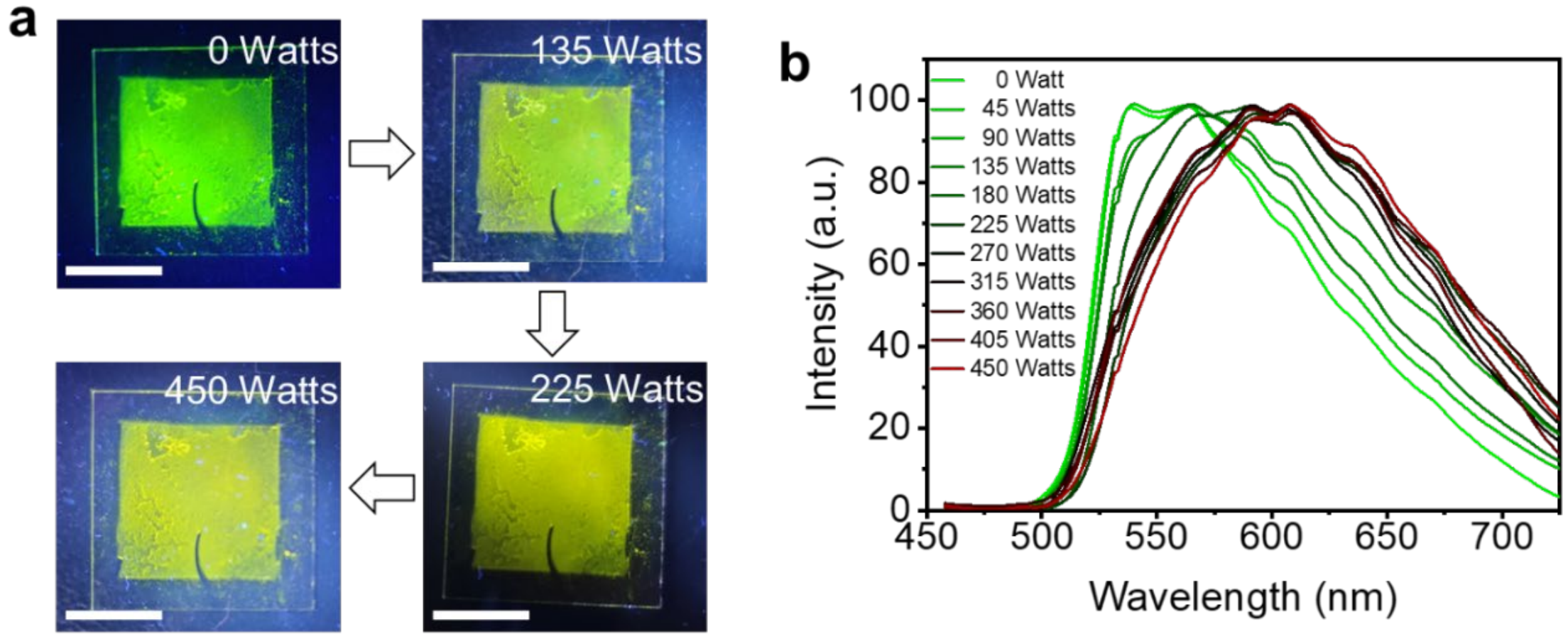

**Supplementary Figure 20. a**, Photographs under UV light exposure of *E to Z*-BTDPA thin film conversion in the solar simulator with power dependent conversion. **b**, Optical spectrum recorded for each 10W for 10mins irradiation of the solar simulator.

## 11. Thin film device pattering using photomask and its characterization.

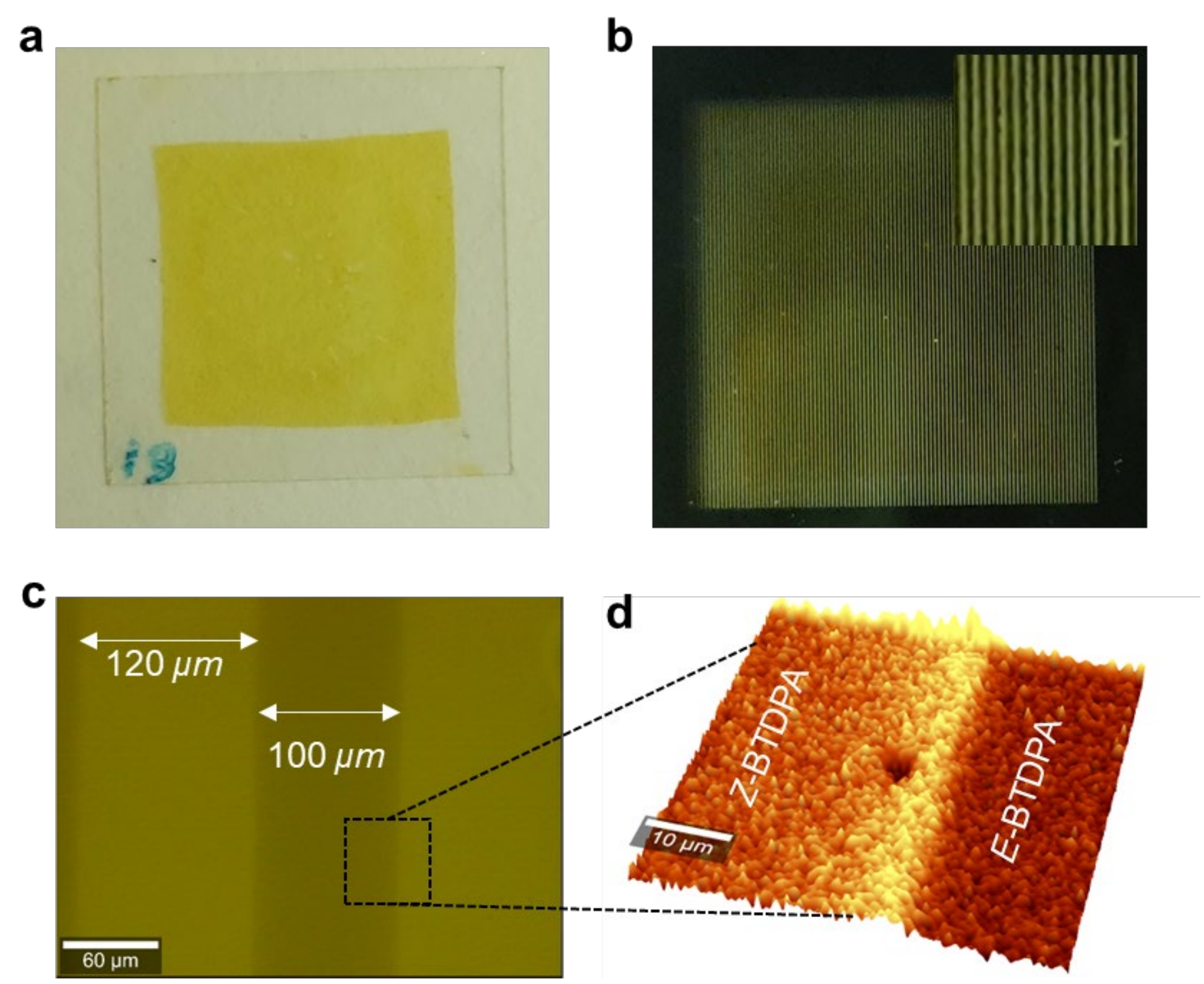

**Supplementary Figure 21. a**, Photograph of spin-coated ***E*-BTDPA** thin film. **b**, Photograph of a photomask (line width 100 μm) stacked on thin film. **c**, Optical image of the *E* and *Z*-BTDPA sunlight irradiate induced isomerized pattern. **d**, PL mapping of the patterned device 532 nm laser.

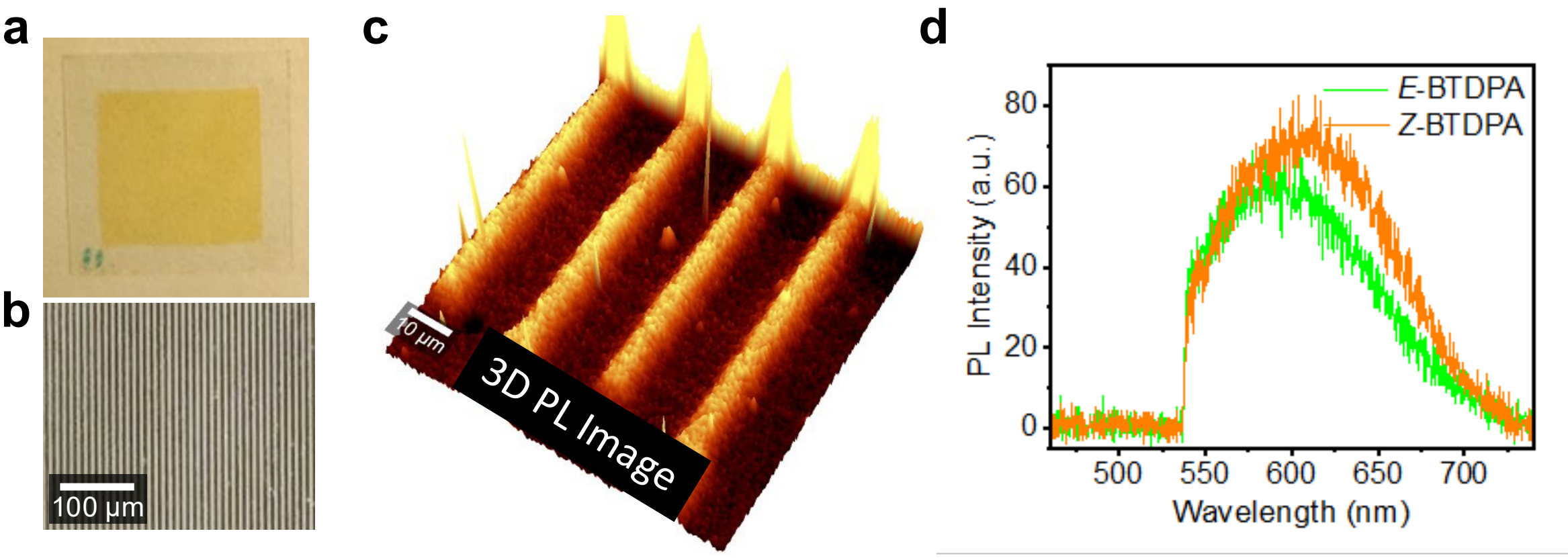

**Supplementary Figure 22.** **a**, Photograph of spin-coated ***E*-BTDPA** thin film. **b**, Photograph of a photomask stacked on thin film (line width 10 μm). **c**, 3D images of PL mapping of the patterned device with 532 nm laser excitation. **d**, Corresponding PL spectra of irradiated and non-irradiated lines.

## 12. FESEM images of both forms.

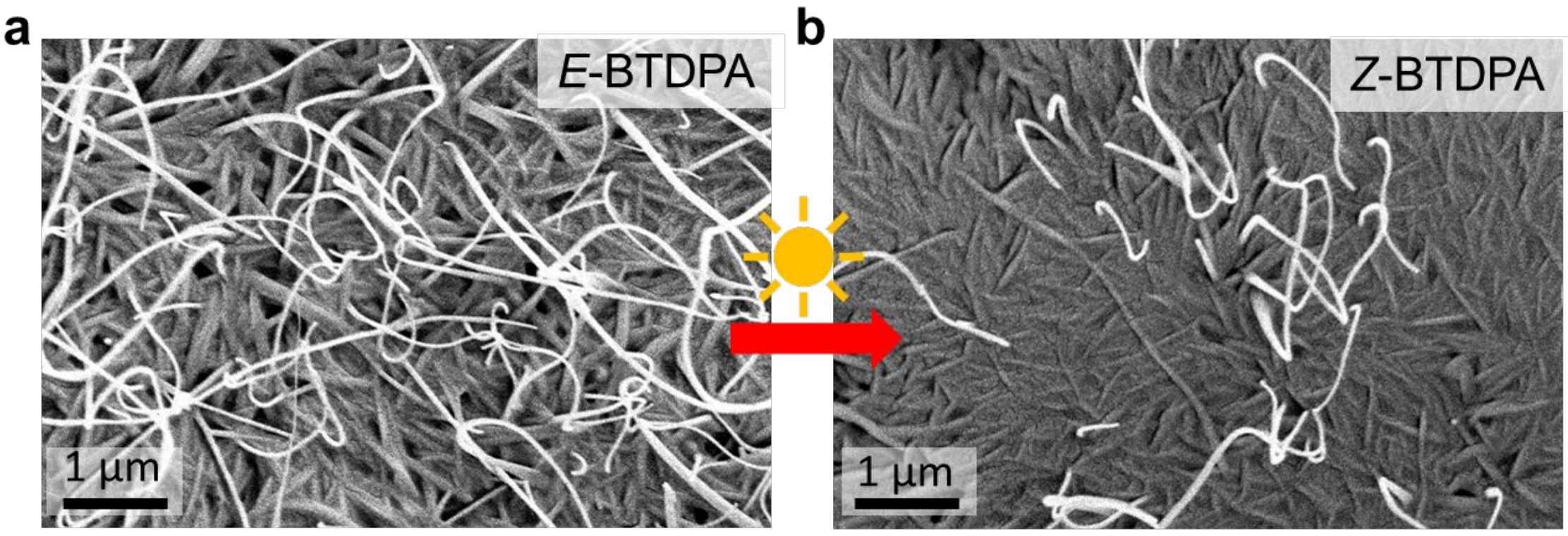

**Supplementary Figure 23.** **a**, FESEM image of spin coated *E*-BTDPA thin film surface morphology. **b**, Conversion for *E* to *Z* form after irradiation with solar light.

## 13. Thickness dependent on photoisomerization and corresponding optical studies

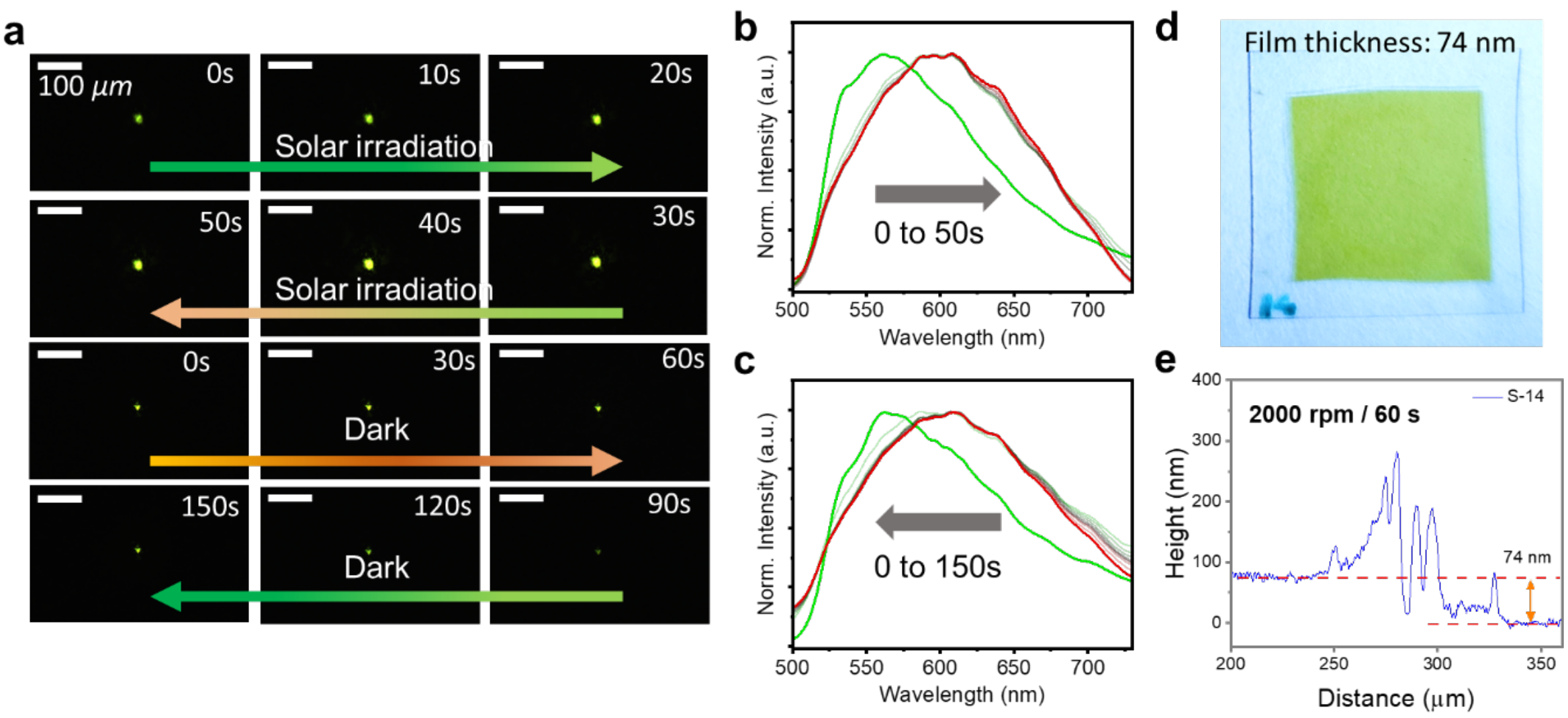

**Supplementary Figure 24. a**, FL images of PL collection with solar light exposure and dark with time interval. **b**, *E*→*Z* BTDPA spectrum collection after irradiation with solar light using confocal microscope. (Excitation wavelength 405nm). **c**, *Z*→*E* BTDPA spectrum collection after irradiation with solar light. **d**, film photograph before(left) and after(right) solar irradiation. **e**, Thickness measurement plot using profilometer shows film thickness of 74 nm.

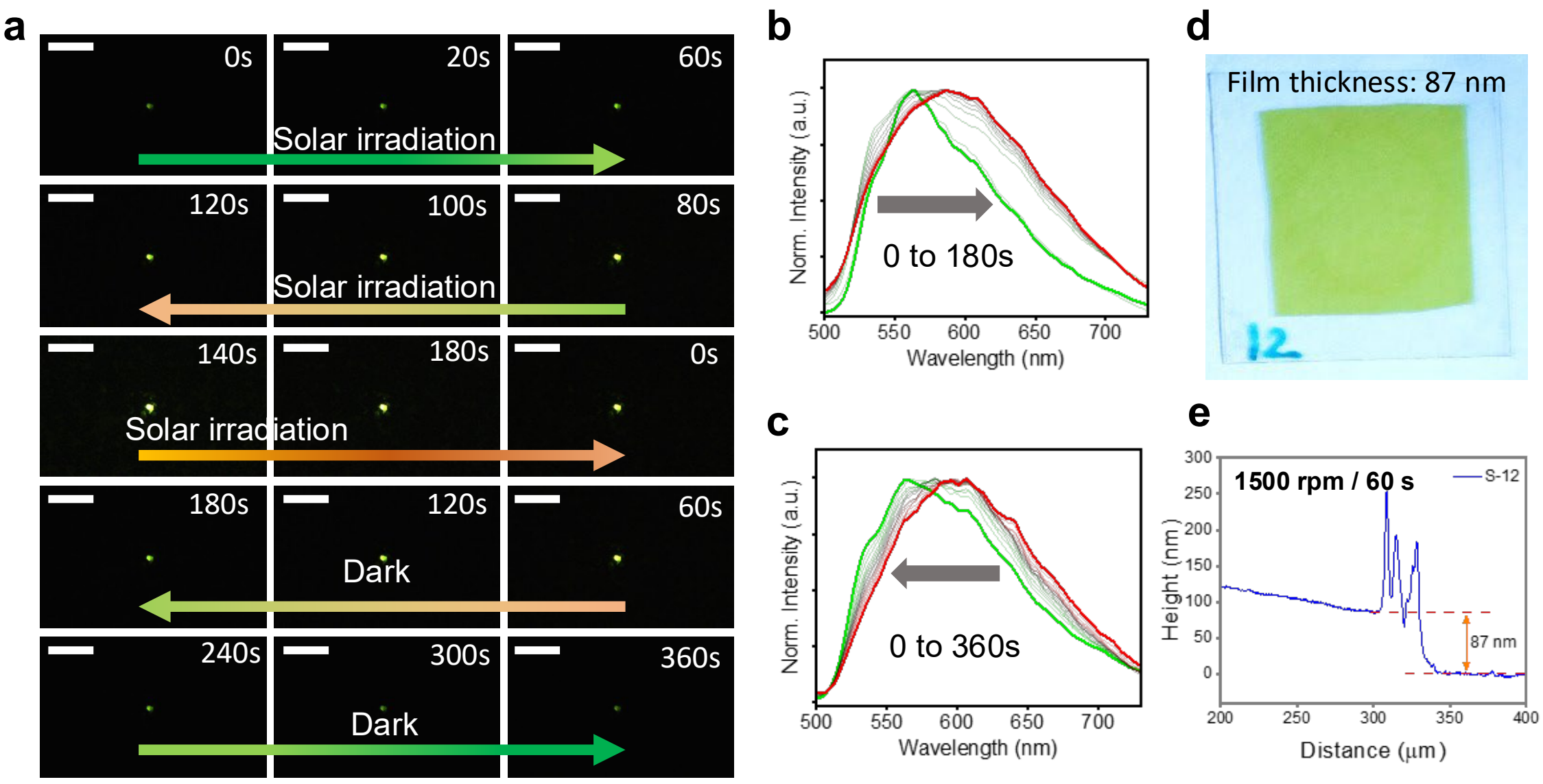

**Supplementary Figure 25. a**, FL images of PL collection with solar light exposure and dark with time interval. **b**, *E*→*Z* BTDPA spectrum collection after irradiation with solar light using confocal microscope. (excitation wavelength 405nm). **c**, *Z*→*E* BTDPA spectrum collection after irradiation with solar light. **d**, film photograph before(left) and after(right) solar irradiation. **e**, Thickness measurement plot using profilometer shows film thickness of 87 nm.

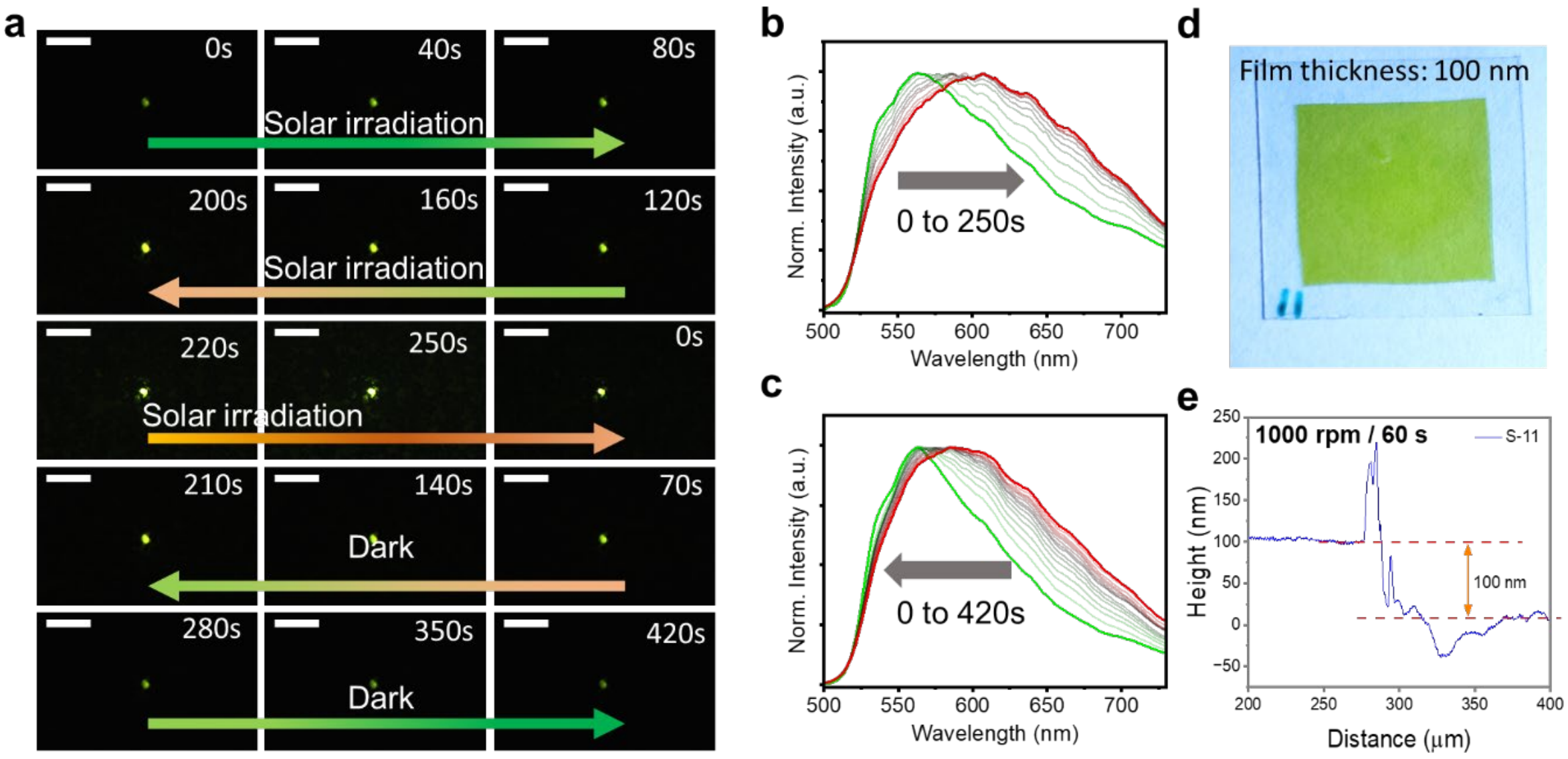

**Supplementary Figure 26. a**, FL images of PL collection with solar light exposure and dark with time interval. **b**, E→Z BTDPA spectrum collection after irradiation with solar light using confocal microscope. (Excitation wavelength 405nm). **c**, Z→E BTDPA spectrum collection after irradiation with solar light. **d**, film photograph before(left) and after(right) solar irradiation. **e**, Thickness measurement plot using profilometer shows film thickness of 100 nm.

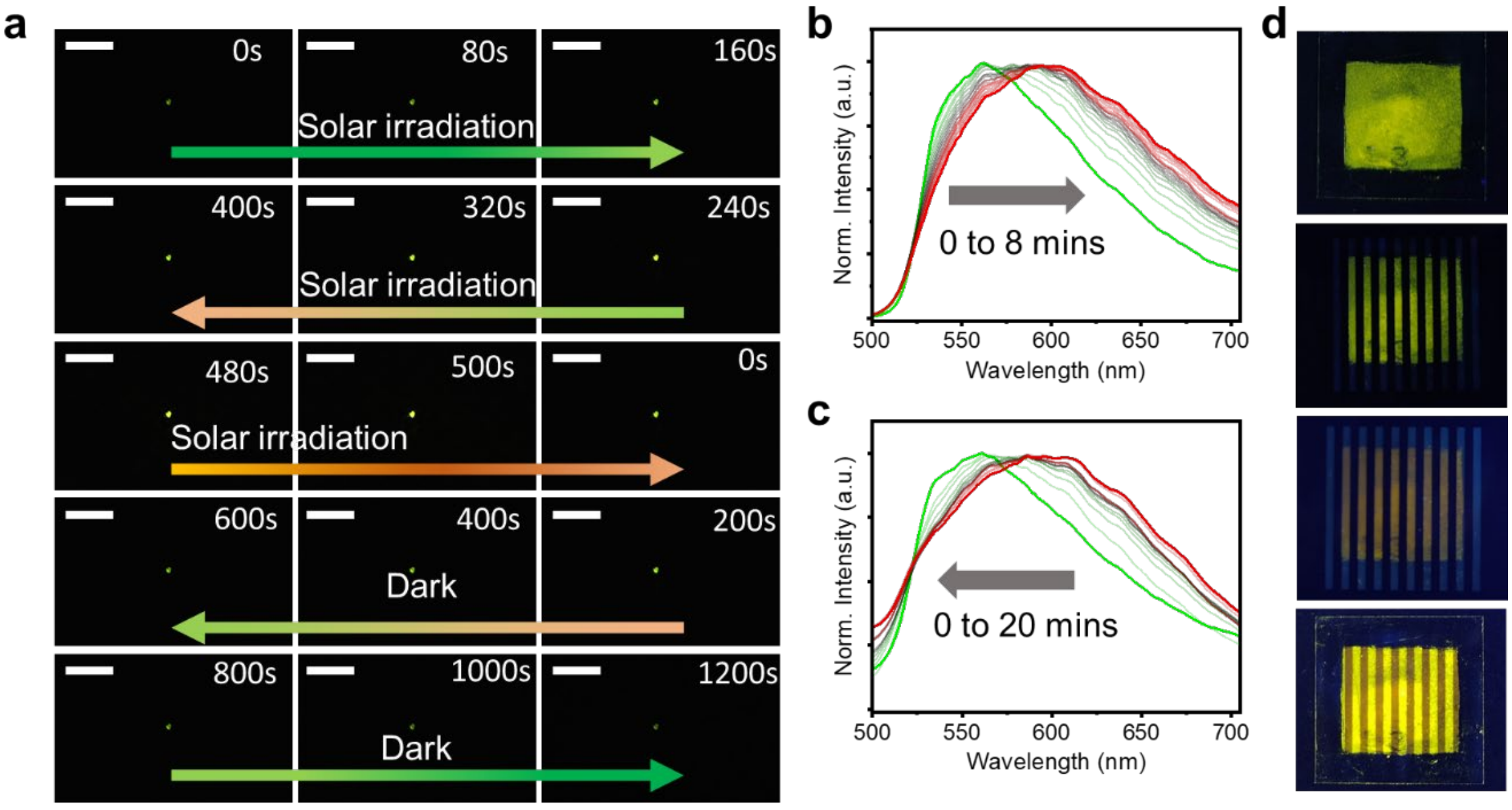

**Supplementary Figure 27. a**, FL images of PL collection with solar light exposure and dark with time interval. **b**, *E*→*Z* BTDPA spectrum collection after irradiation with solar light using confocal microscope. (excitation wavelength 405nm). **c**, *Z*→*E* BTDPA spectrum collection after irradiation with solar light. **d**, BTDPA film photograph before(left) and after(right) solar irradiation selectively by photomask. (Film thickness ≈300nm).

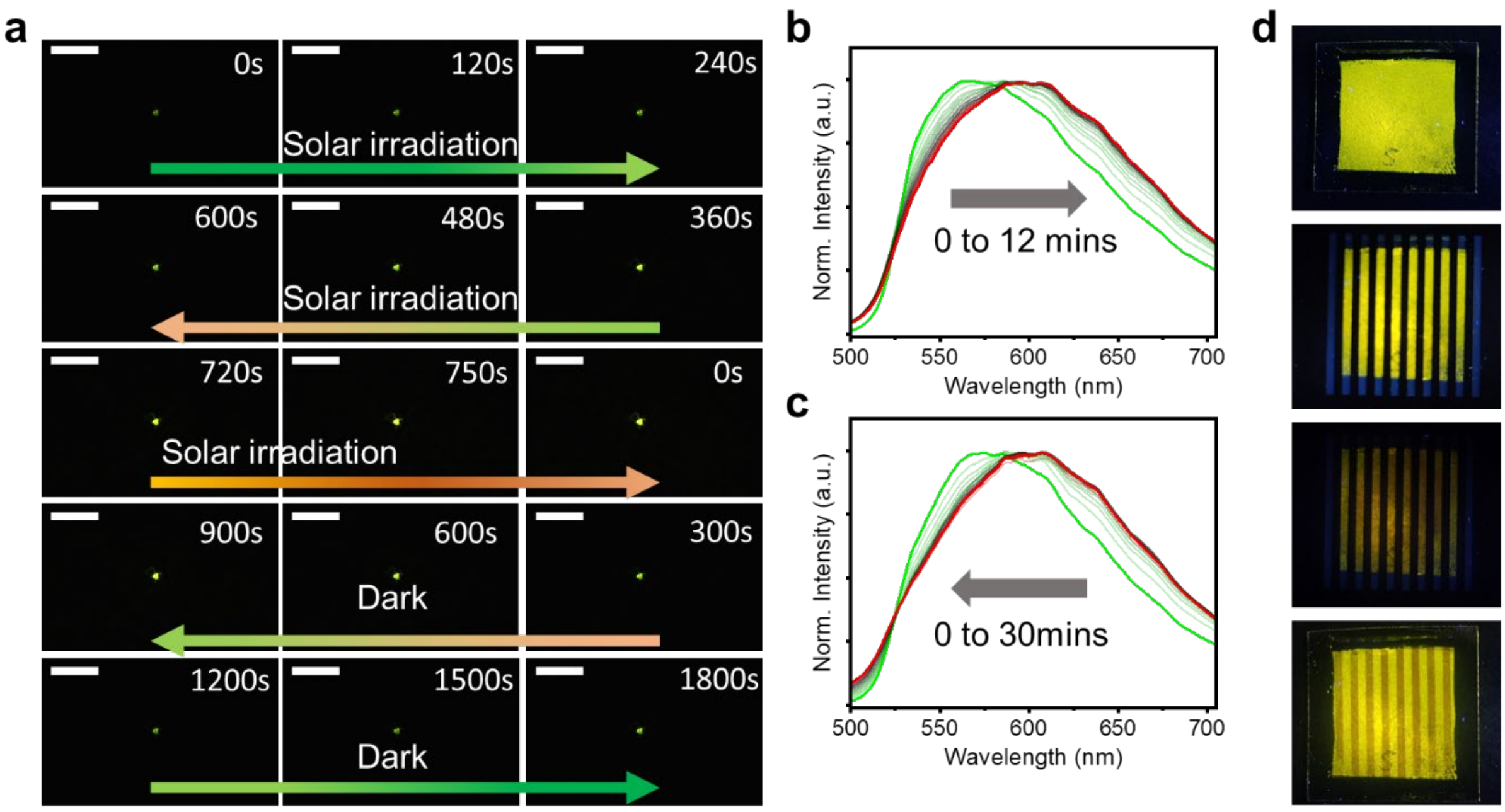

**Supplementary Figure 28. a**, FL images of PL collection with solar light exposure and dark with time interval. **b**, *E*→*Z* BTDPA spectrum collection after irradiation with solar light using confocal microscope. (excitation wavelength 405nm). **c**, *Z*→*E* BTDPA spectrum collection after irradiation with solar light. **d**, BTDPA film photograph before(left) and after(right) solar irradiation selectively by photomask. (Film thickness ≈500nm).

## 14. Powder X-ray Diffraction Studies of *E*/*Z*-BTDPA

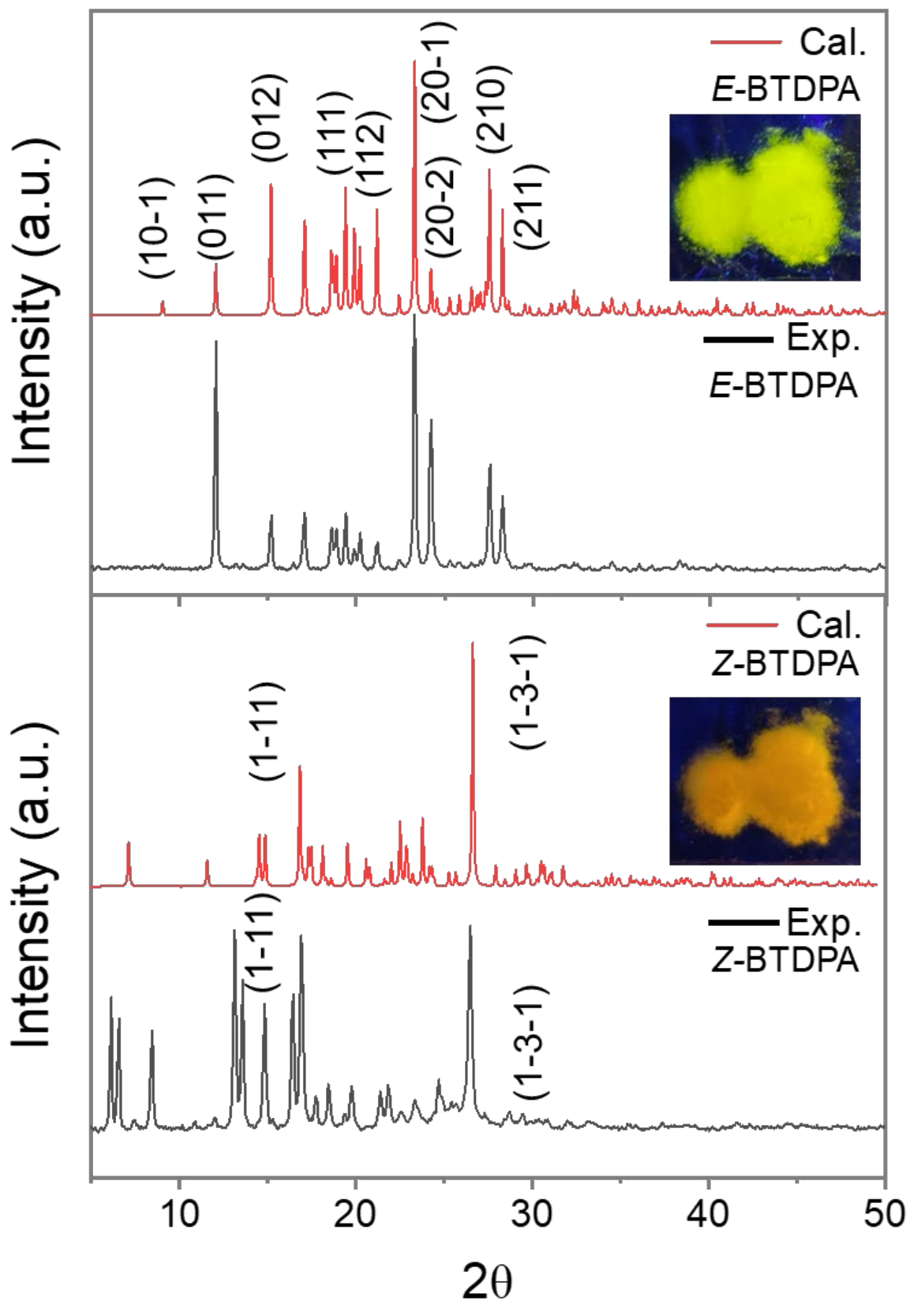

**Supplementary Figure 29.** Calculated and experimental PXRD plot for *E*-BTDPA (top) → *Z*-BTDPA (bottom) conversion with sunlight irradiation (inset represents the photograph of their respective forms).

## 15. Solid-state optical studies of *E*-BTDPA and *Z*-BTDPA

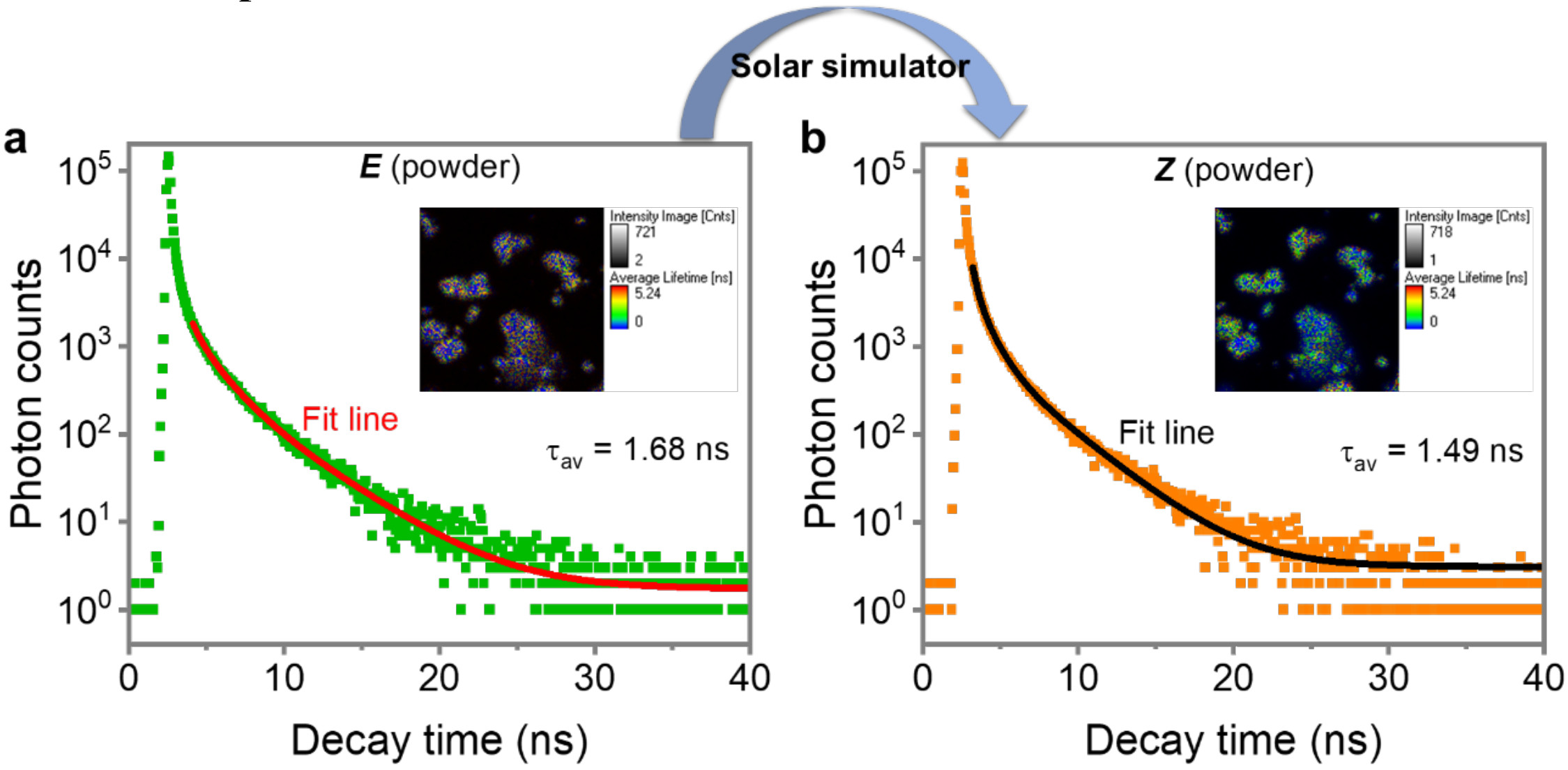

**Supplementary Figure 30.** The fluorescence decay profiles of **a**, *E*-BTDPA powder and **b**, *Z*-BTDPA powder were obtained by TCSPC. The measured photon counts were shown in green (*E*-BTDPA) and orange (*Z*-BTDPA) squares on a logarithmic scale. The decays were fitted with a tri-exponential fit (solid line). Inset photo represents the fluorescence lifetime images of respective thin films with intensity image scale bar and average lifetime scale bar.

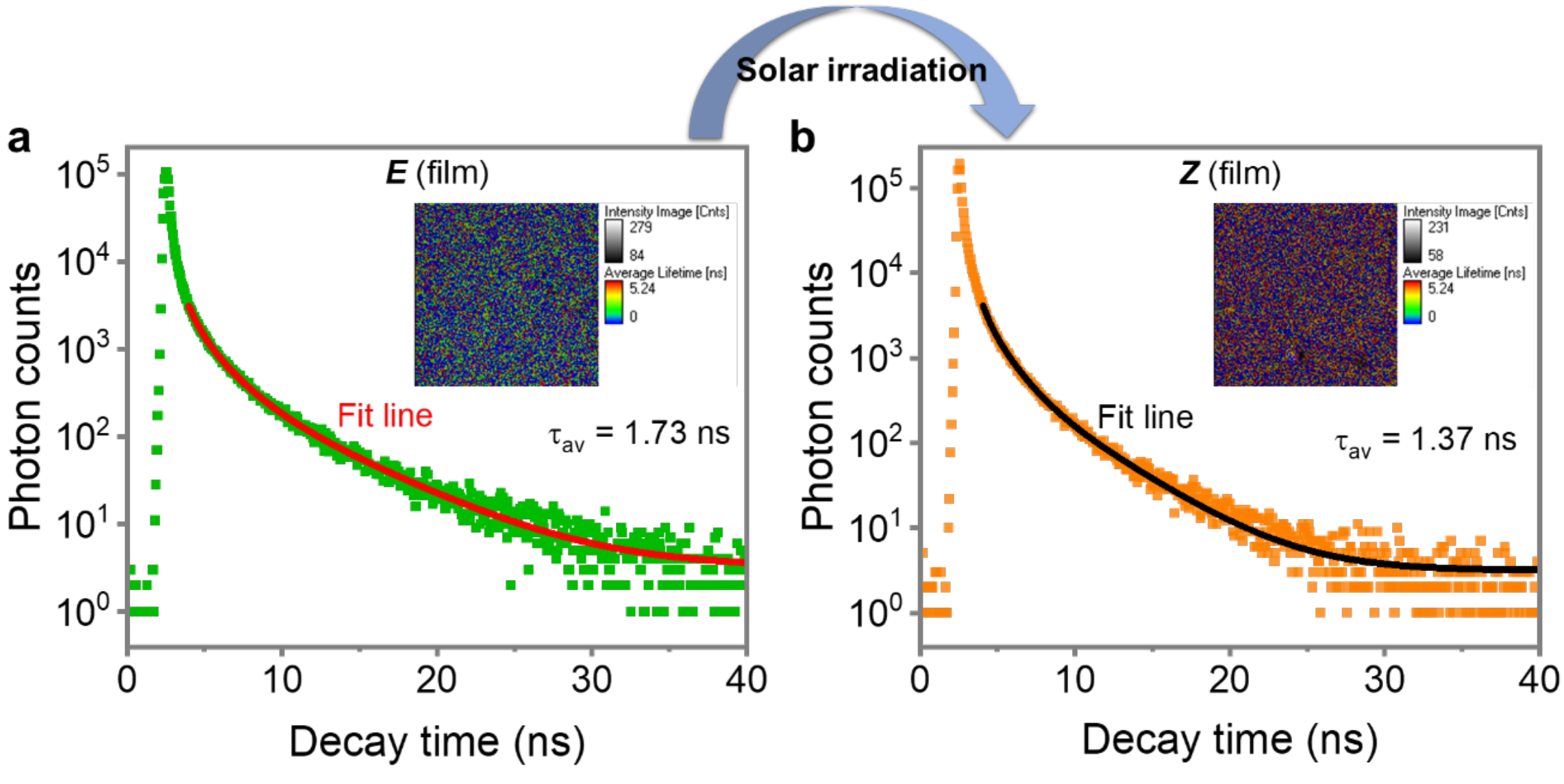

**Supplementary Figure 31:** The fluorescence decay profiles of **a**, E-BTDPA film and **b**, Z-BTDPA film were obtained by TCSPC. The measured photon counts were shown in green (E-BTDPA) and orange (Z-BTDPA) squares on a logarithmic scale. The decays were fitted with a tri-exponential fit (solid line). Inset photo represents the fluorescence lifetime images of respective thin films with intensity image scale bar and average lifetime scale bar.

**16.** **NLO studies of BTDPA crystals.**

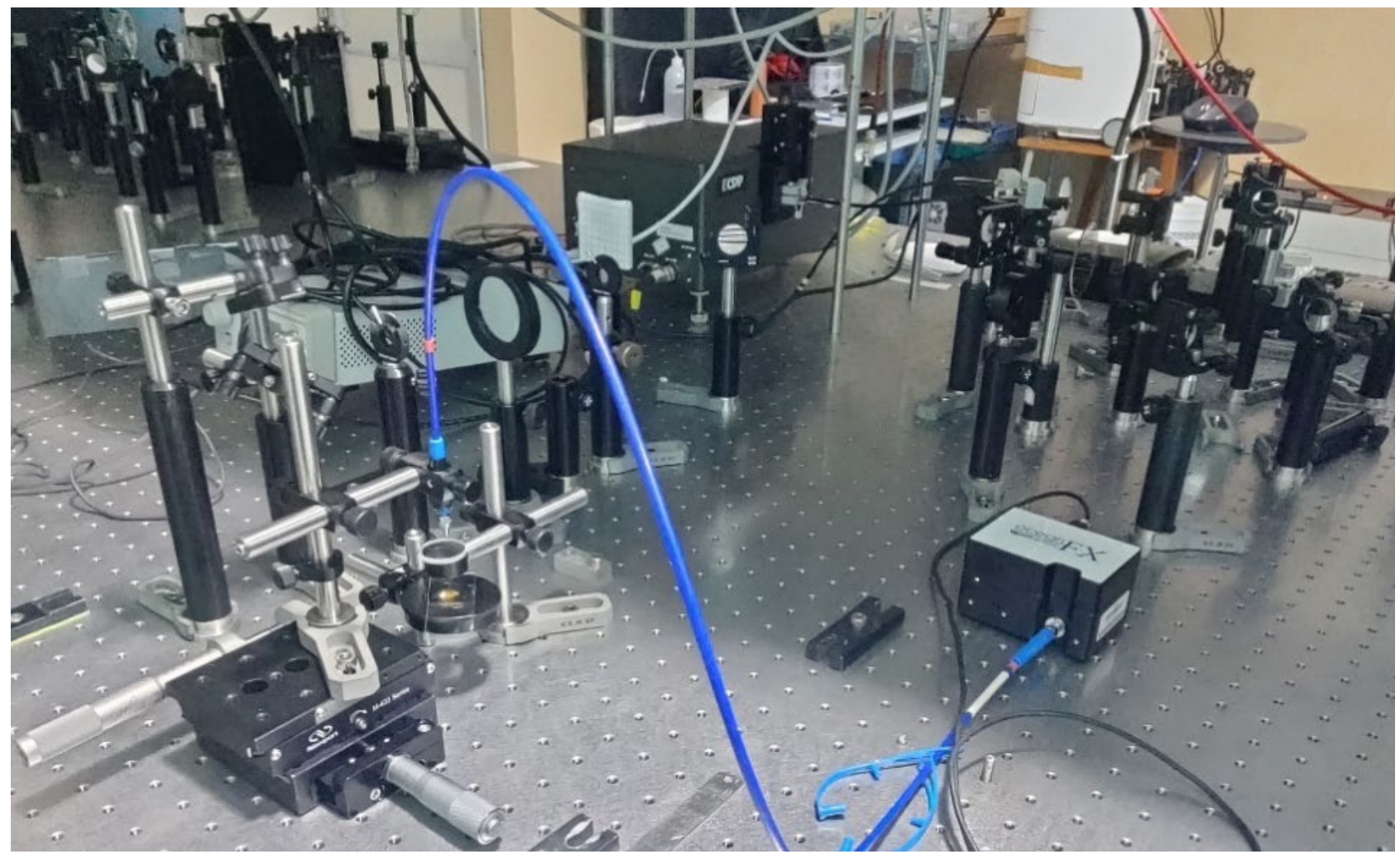

**Supplementary Figure 32.** Photograph for the power dependent and wavelength tunned the NLO study setup.

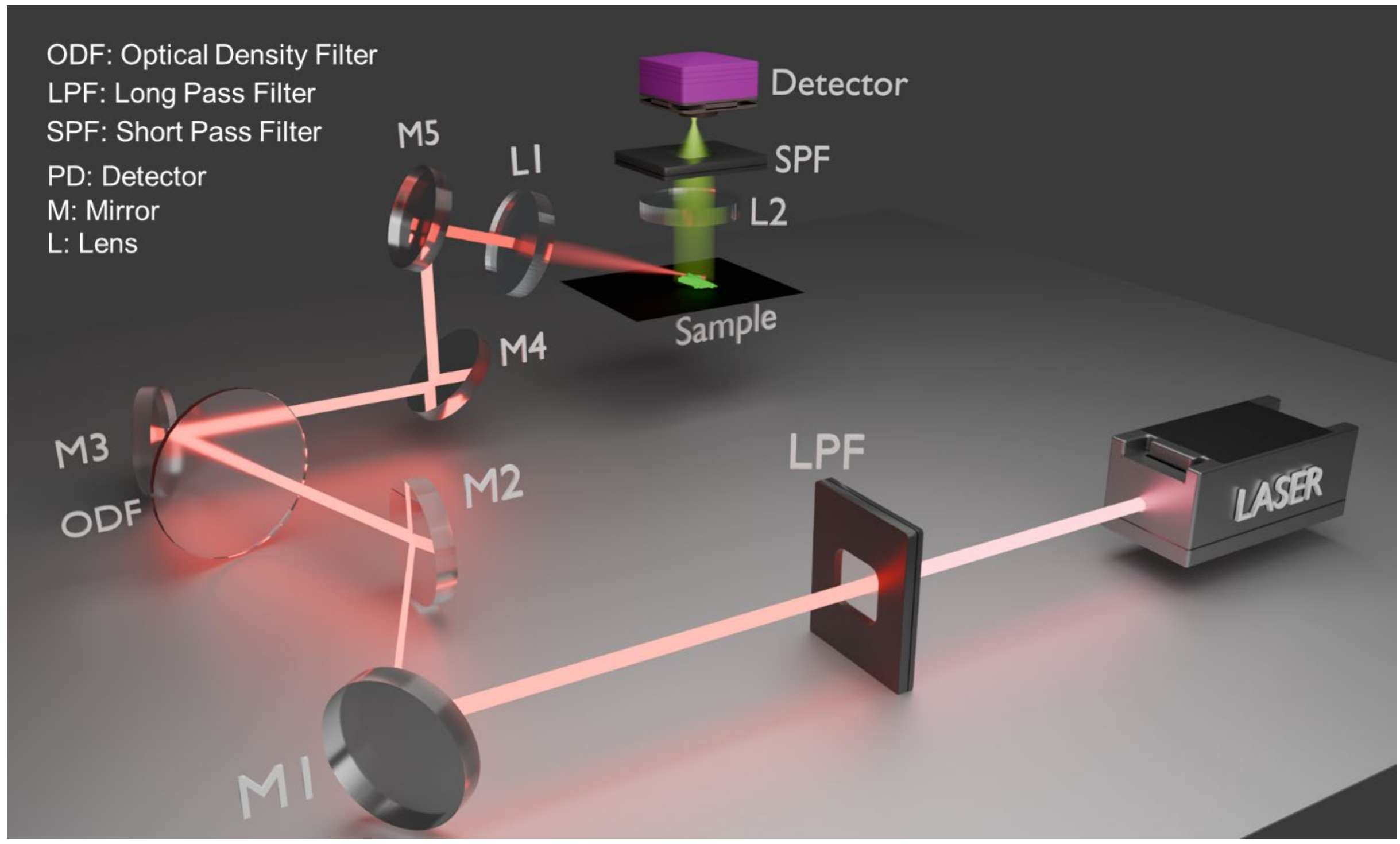

**Supplementary Figure 33.** Schematic representation for the power dependent and wavelength tunned the NLO study setup.

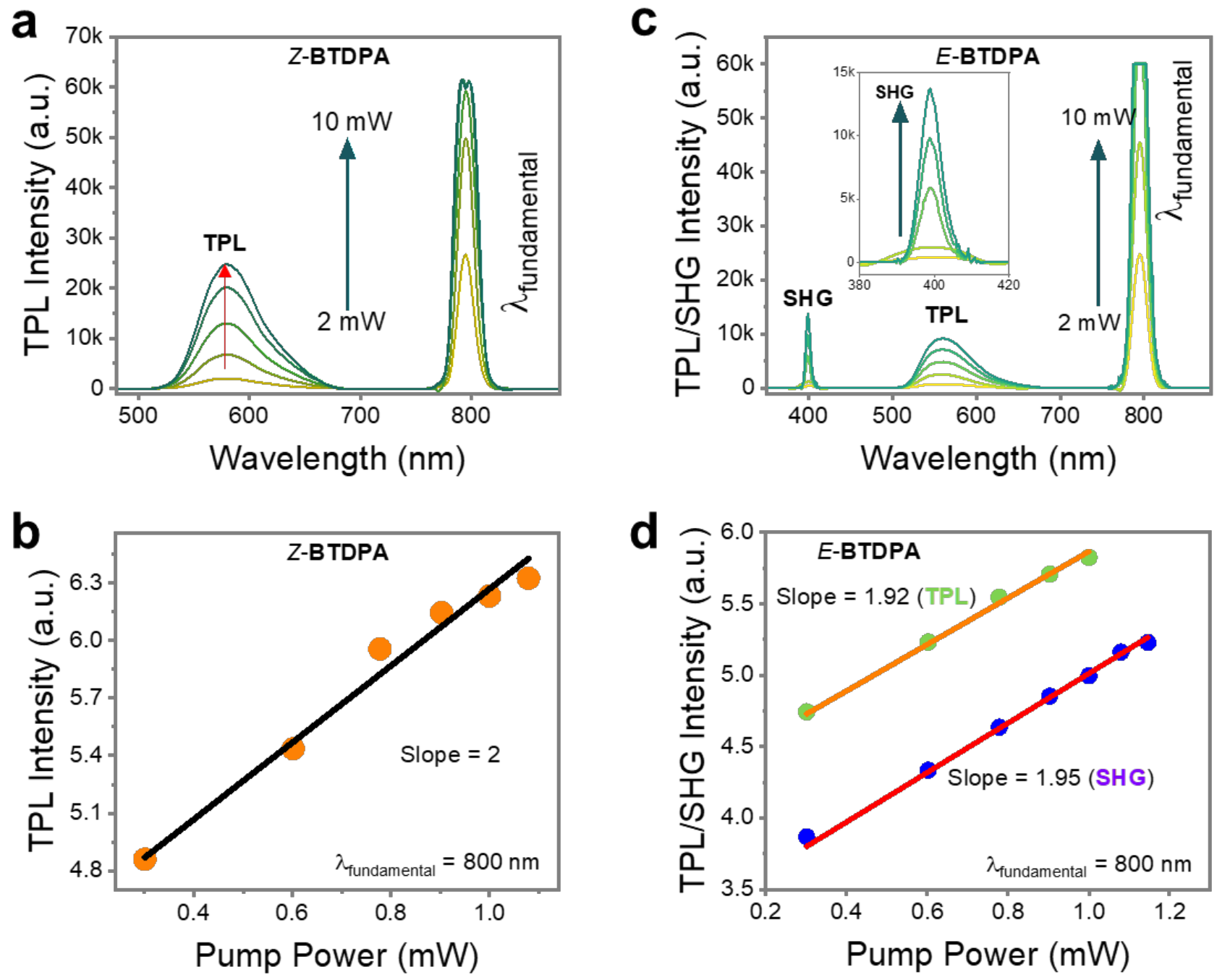

**Supplementary Figure 34. a**, TPL optical spectra of ***Z*-BDTPA** crystal power-dependent nonlinear study at 800 nm fundamental wavelength. **b**, Log-log plot between TPL intensity and pump power. TPL signal increases with pump power. The red line shows the fit curve. **c**, SHG and TPL optical spectra of ***E*-BDTPA** crystal power-dependent nonlinear study at 800 nm fundamental wavelength. **d**, Log-log plot between TPL/SHG intensity and pump power. TPL signal increases with pump power (the green and orange colours represent TPL and SHG intensity, respectively). The red colour represents the fit line.

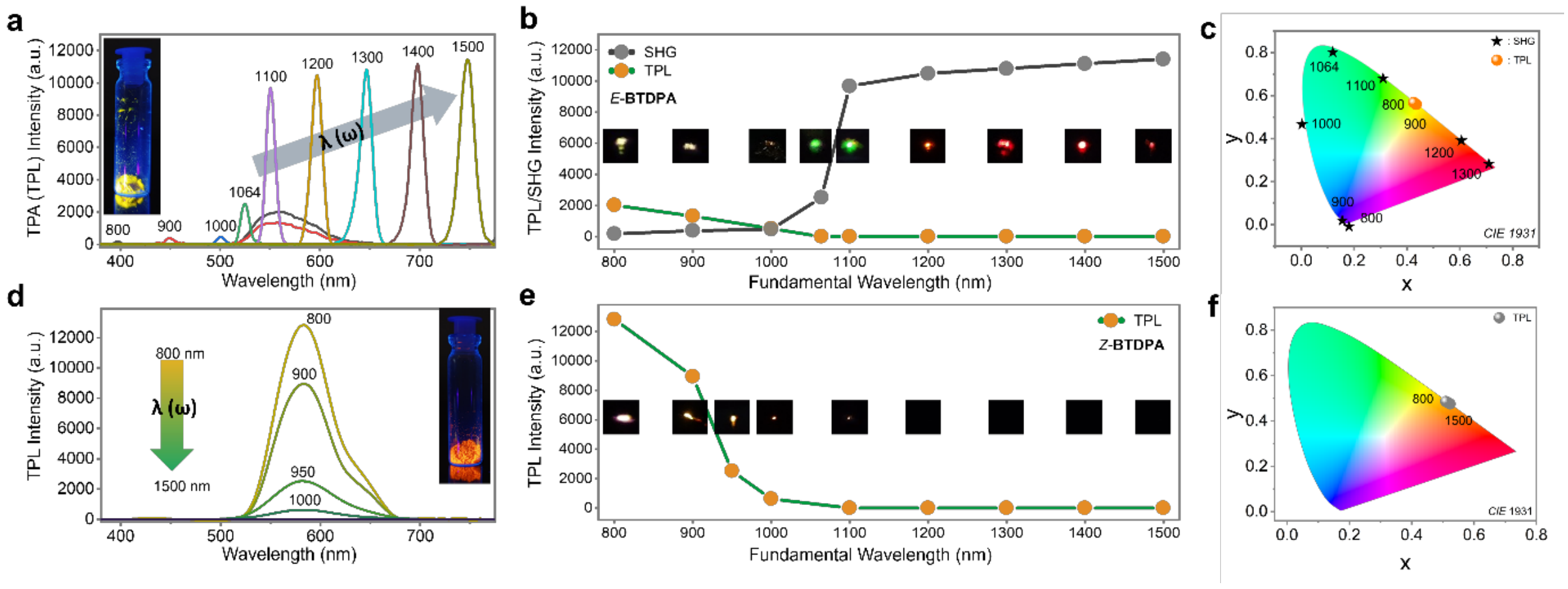

**Supplementary Figure 35: a**, TPL/SHG intensity vs wavelength spectra of *E*-BTDPA (non-centrosymmetric crystal) shows TPL (till 1064 nm). Inset shows image of green colour *E*-BTDPA compound in vial. **b**, Comparison plot between intensity (SHG and TPL) vs wavelengths of *E*-BTDPA. Inset shows NLO emission from crystals with fs laser excitations and corresponding **c**, CIE-colour plot for various NLO intensities. **d**, TPL intensity vs wavelength spectra of *Z*-BTDPA (centrosymmetric crystal). Inset orange color *Z*-BTDPA compound in vial. **e**, Comparison plot between TPL intensity vs wavelength of *Z*-BTDPA crystals. Inset shows TPL emission from crystals with fs laser excitations and corresponding **f**, CIE-colour plot form TPL intensities.

## 17. NLO study of patterned thin films

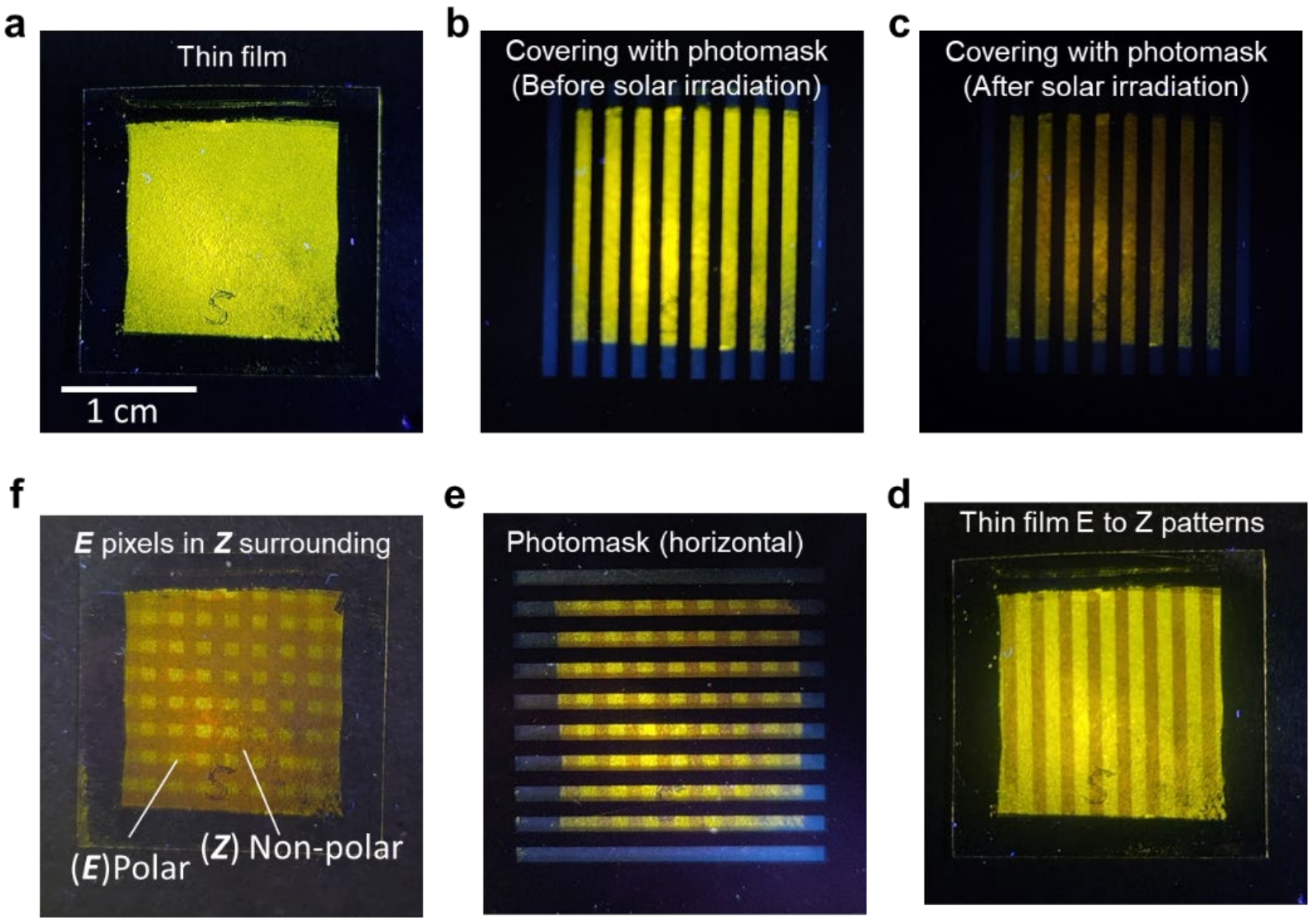

**Supplementary Figure 36. a**, E-BTDPA thin film on 22 mm glass slide. **b**, covering the thin film with the photomask in vertical direction (linewidth ≈1 mm). **c**, the film slide covered with photomask after irradiation of solar light. **d**, Photographs of lined patterned on thin film. **e**, Photograph of the film having the photomask in the horizontal direction. **f**, final patterned thin film containing E pixels with Z surrounding.

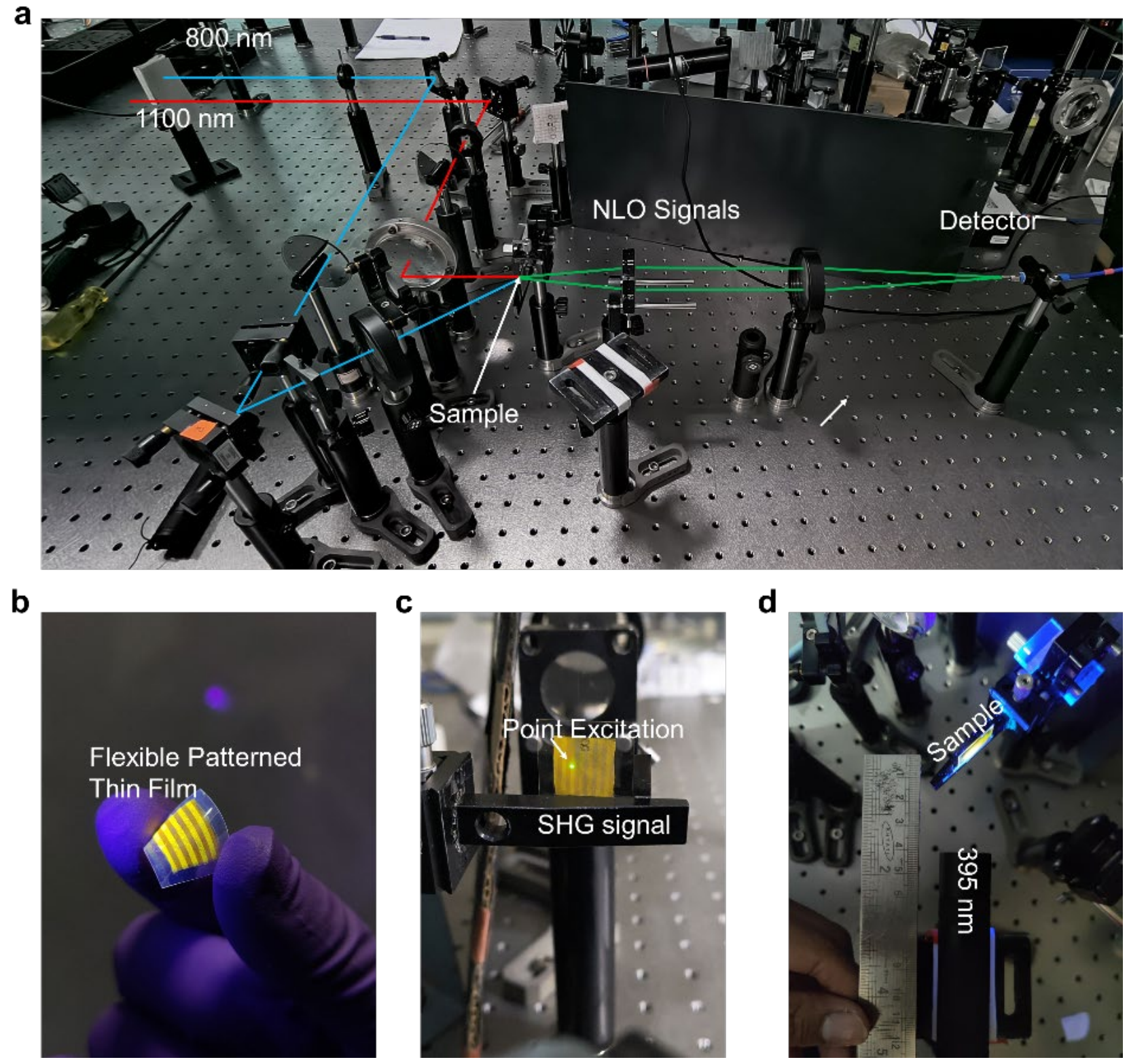

**Supplementary Figure 37. a**, Photograph of the transmission NLO studies setup. **b**, Photograph of flexible imperishable patterning of BTDPA on PET thin film. **c**, SHG generation with point excitation at the patterned thin film. **d**, Photoisomerization of BTDPA using UV light for instant switching studies.

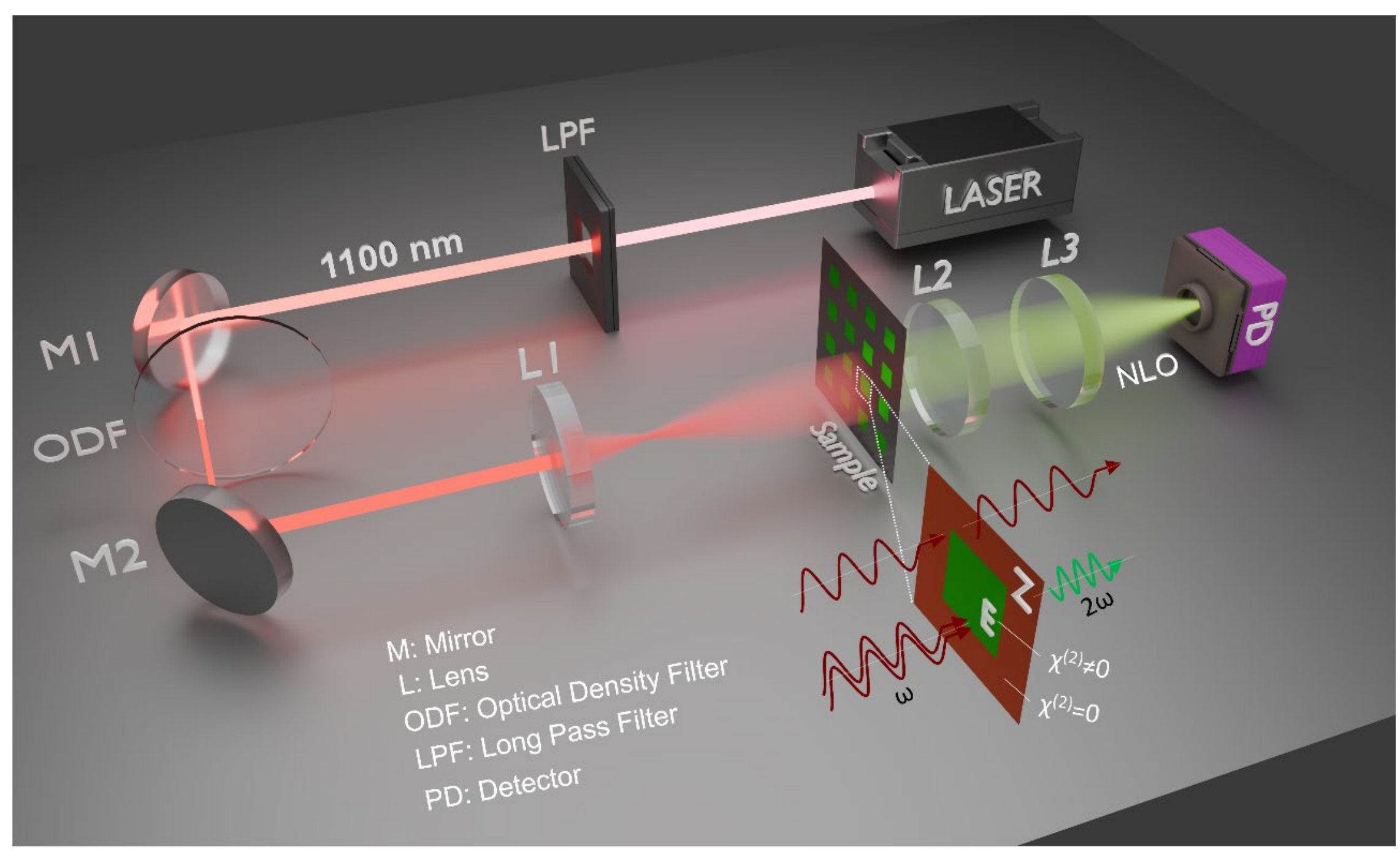

**Supplementary Figure 38.** Schematic illustration of the transmission NLO setup for the translational scanning and erasable patterning studies.

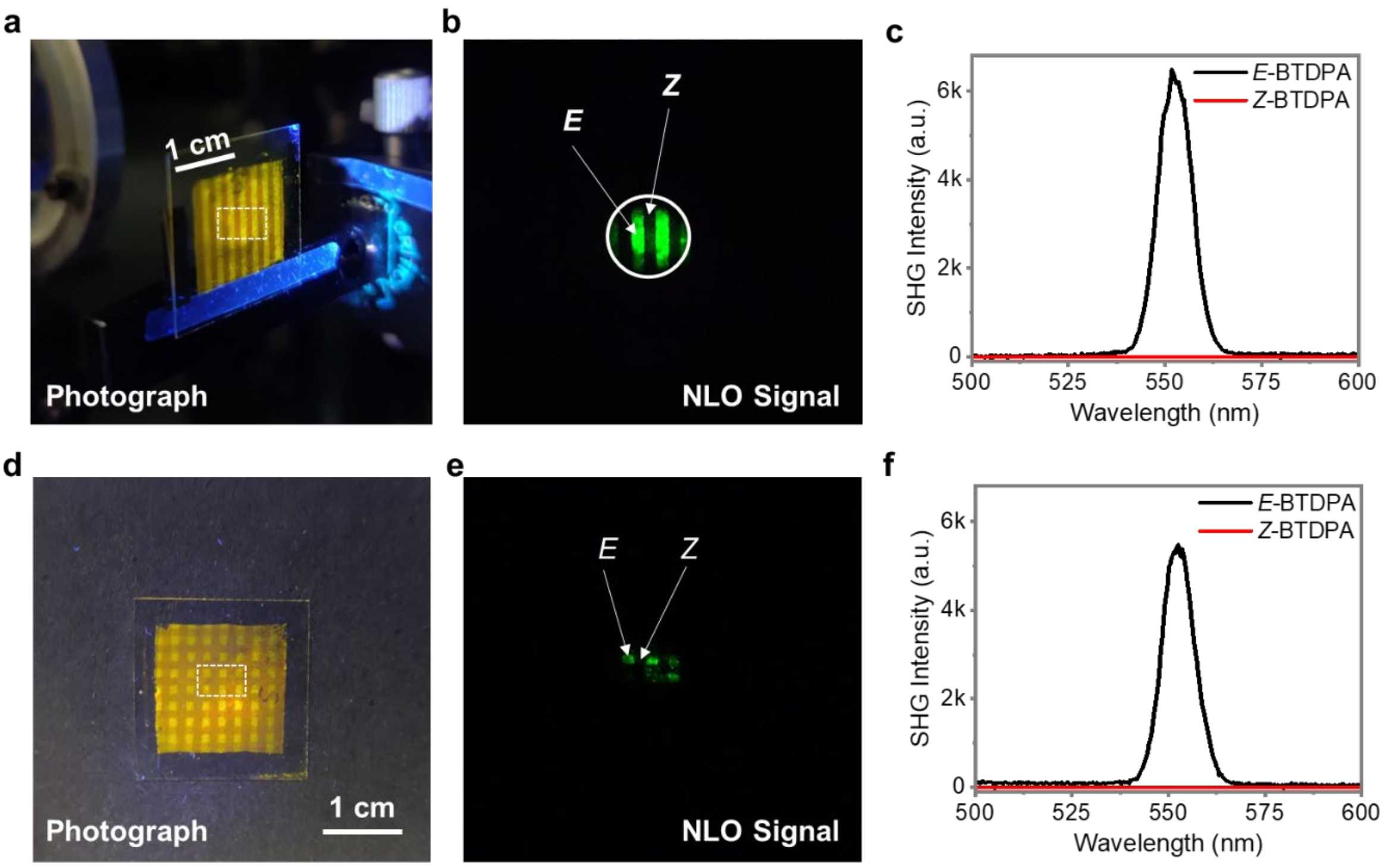

**Supplementary Figure 39. a**, Photograph of line patterned thin film mounted on the holder for NLO study. **b**, photograph of SHG optical signal generated through patterned film using fundamental of 1100 nm femtosecond pulse laser. **c**, the corresponding SHG optical signals at *E* and *Z* patterned BTDPA lines using the setup mentioned in figure S38. **d**, Photograph of pixelated patterned thin film NLO study. **e**, photograph of SHG optical signal generated through patterned film using fundamental of 1100 nm femtosecond pulse laser. **f**, the corresponding SHG optical signals at *E* pixels and *Z* surrounding to the pixels in patterned BTDPA film.

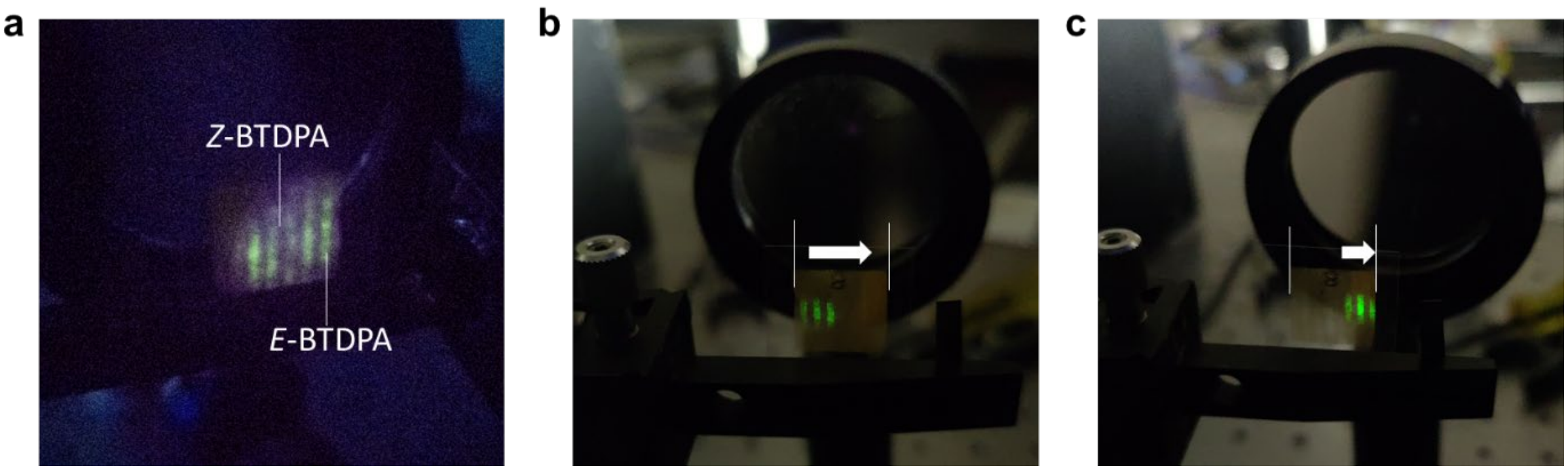

**Supplementary Figure 40. a**, photographs of the E and Z line patterned thin film NLO optical studies (optical signal shows green for SHG signal generation is *E*-BTDPA, black/no signal from *Z*-BTDPA). **b-c**, Demonstration of using photograph of the SHG from the patterned film using translational setup.

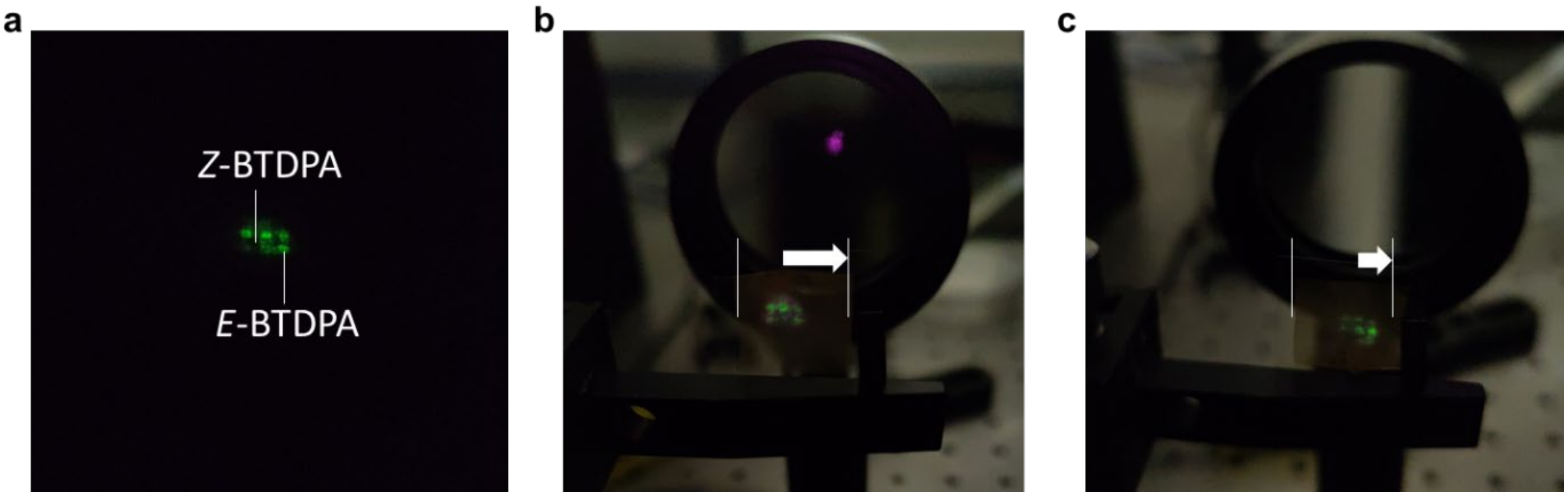

**Supplementary Figure 41. a**, photograph of the E and Z pixel patterned thin film NLO optical studies (optical signal shows green for SHG signal generation is E-BTDPA, black/no signal from Z-BTDPA). **b**, Demonstration of using photograph of the SHG from the patterned film using translational setup.

## 18. NLO signal for real time signal generation and detection

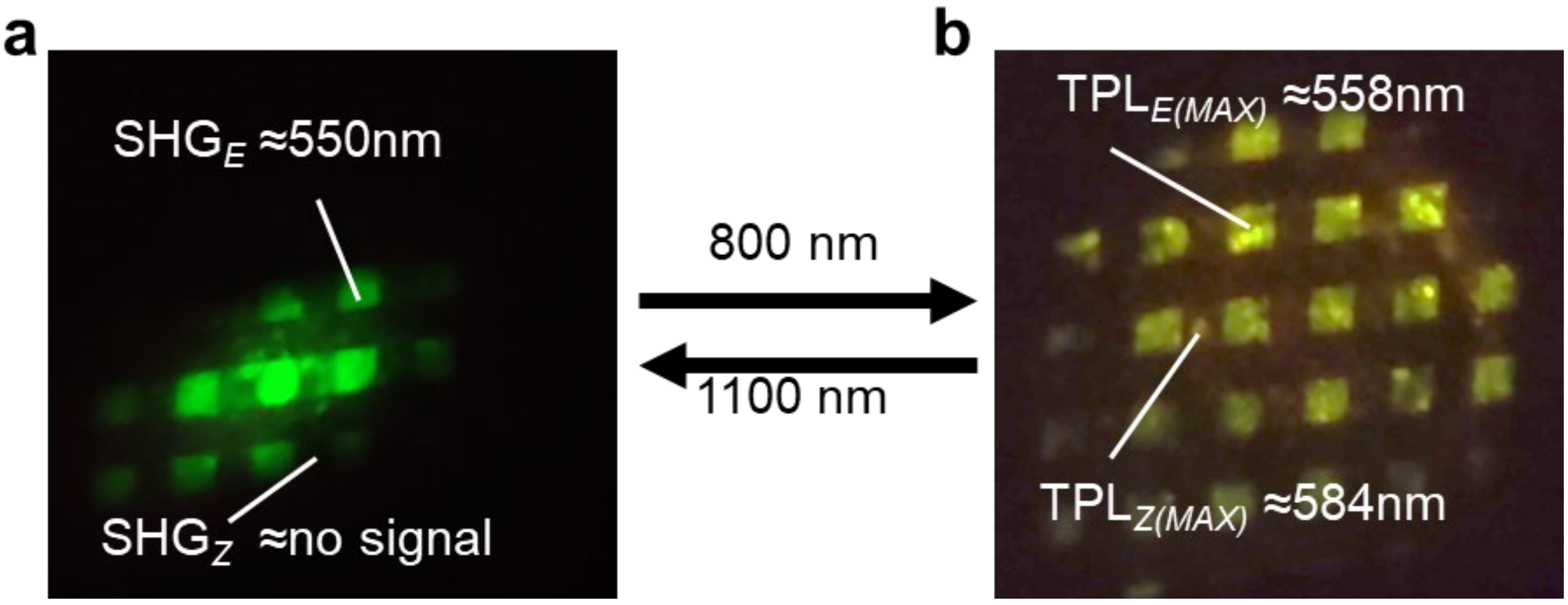

**Supplementary Figure 42. a,** Photograph of a squared *E*-form array surrounded by a thin film of the *Z*-form of BTDPA. Under 1100 nm fs excitation, the *E*-form domains generate a second-harmonic signal at 550 nm, whereas neither SHG nor TPL is detected from the surrounding Z-form. **b,** Two-photon luminescence (TPL) is observed from both *E*- and *Z*-forms under 800 nm fs excitation, with distinct emission maxima: TPL(E) ≈ 558 nm and a red-shifted TPL(Z) ≈ 584 nm.

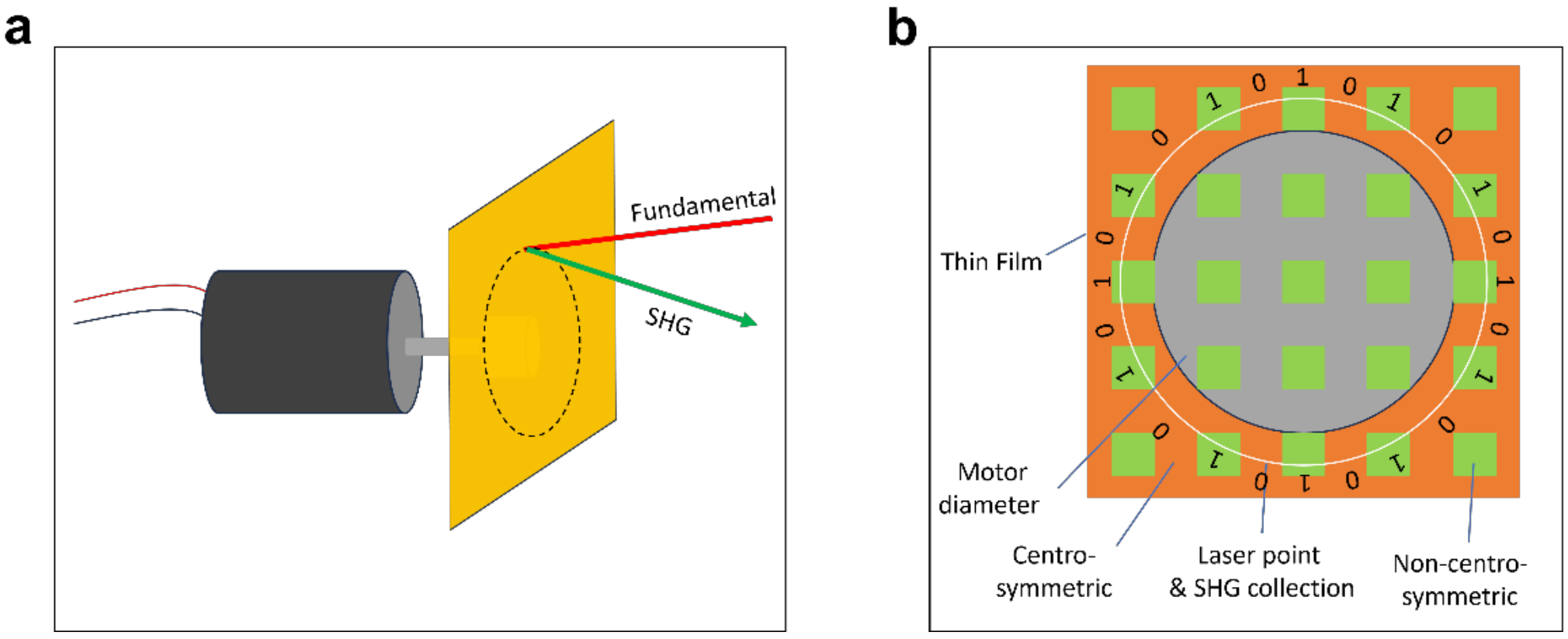

**Supplementary Figure 43.** **a**, Schematic diagram of thin film attached on motorized rotor shaft. **b**, circular rotation diameter of film and aligned FS laser diameter.

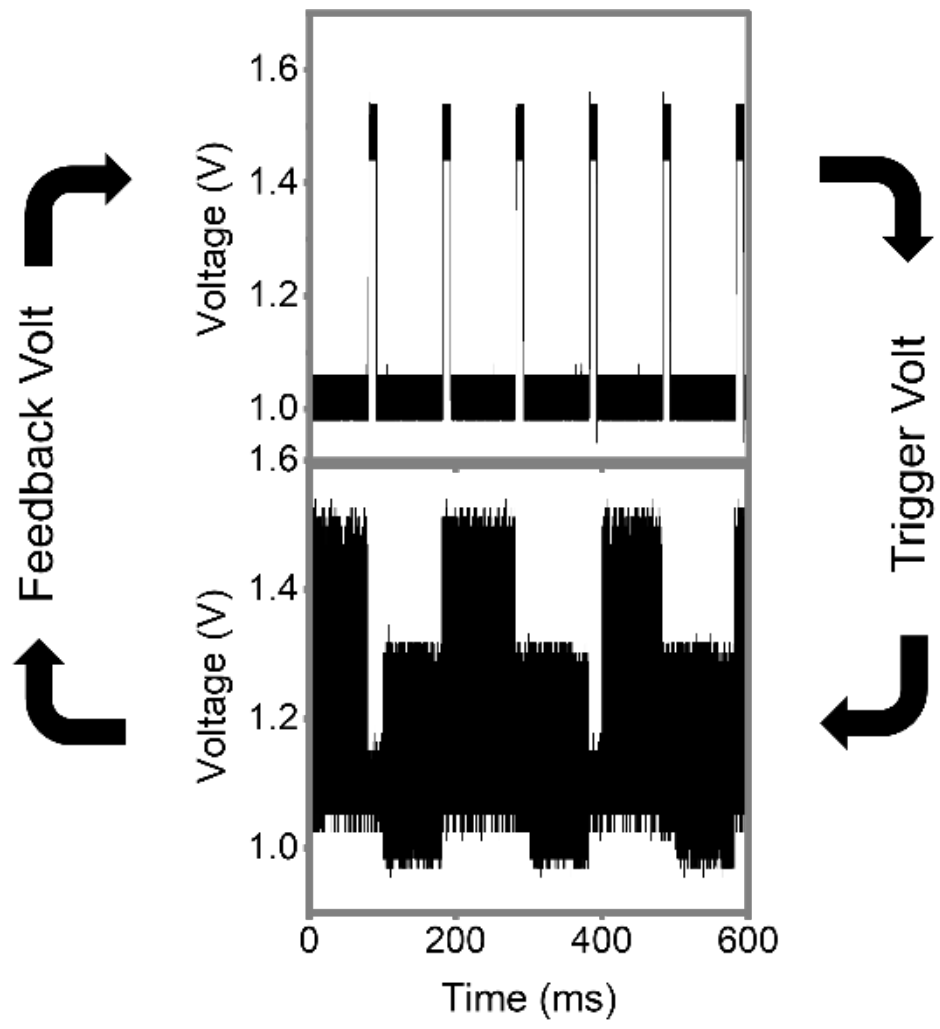

**Supplementary Figure 44.** Triggered voltage applied for short term to get an angular moment and the feedback voltage for tracking and maintaining the position accurately in loop.

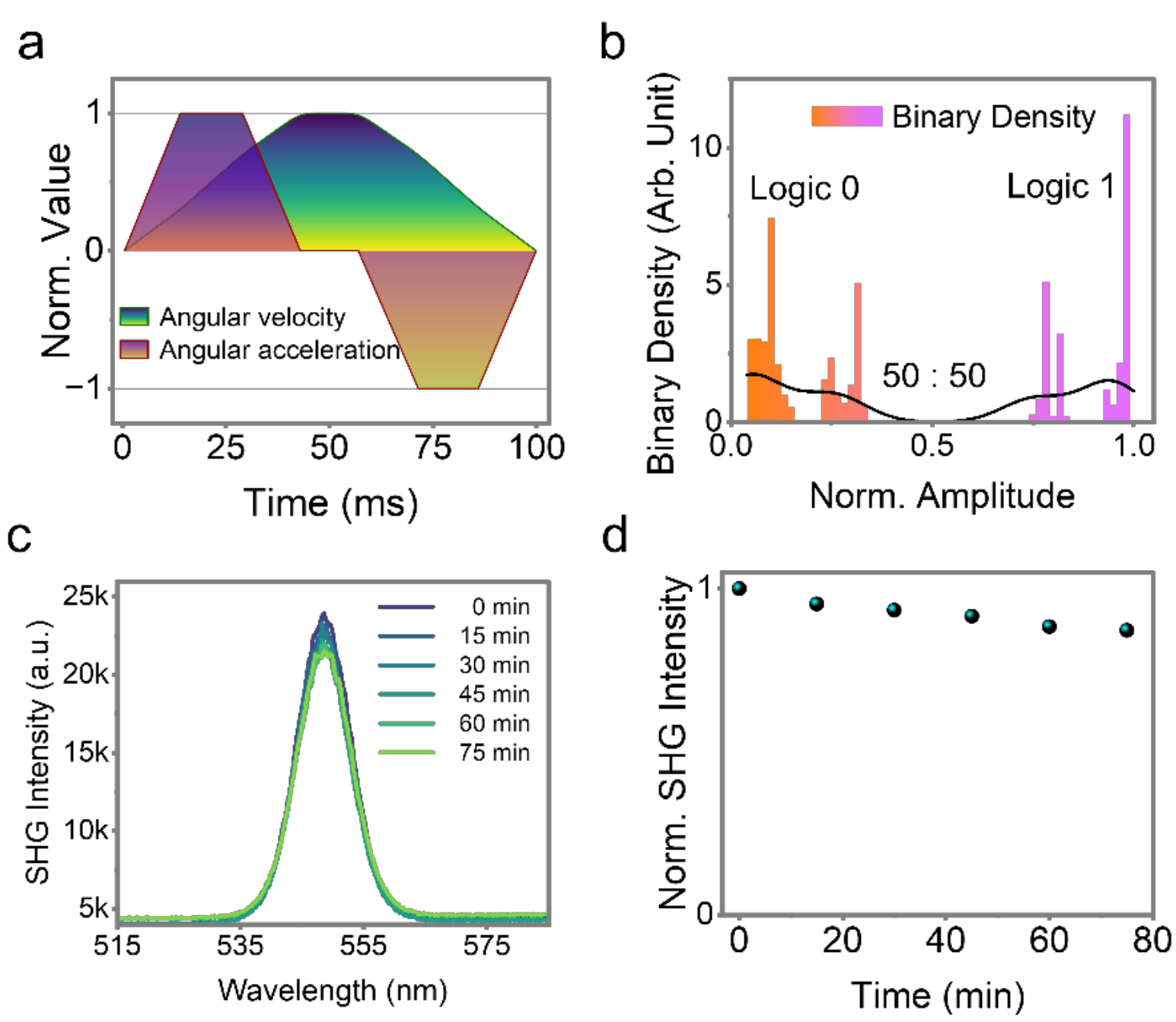

**Supplementary Figure 45. a,** Angular velocity and acceleration of the gate, normalized to its maximum value, plotted as a function of time, showing smooth acceleration and deceleration with zero velocity at the turning points. **b,** Histogram of intensity levels showing clear separation between '0' and '1' states, indicating a high extinction ratio of 50:50. **c, d,** Second-harmonic generation (SHG) intensity measured at a fixed pixel as a function of time, demonstrating stable nonlinear optical response over prolonged acquisition.

## 19. Supplementary Equations

**1. Photoisomerization conversion:**

a) Solution state

**forward process:**

$$\%Z = \frac{I_Z}{I_E+I_Z} \times 100 \text{ ......Supplementary Eq. (1)}$$

**Backward process:**

$$\%E = \frac{I_E}{I_E+I_Z} \times 100 \text{ ......Supplementary Eq. (2)}$$

Where, $I_E$ is the intensity (area) of the *E* and $I_Z$ is the intensity of the *Z*-peak.

rate $\propto \frac{\Delta I}{\Delta t}$

b) Solid State

**Emission-based isomerization:**

$$\text{Conversion (\%) } (\alpha_t) = \frac{\lambda_t-\lambda_E}{\lambda_Z-\lambda_E} \times 100 \text{......Supplementary Eq. (3)}$$

Where, $\lambda_t$: Emission maximum wavelength at time *t*. $\lambda_E$: Initial wavelength (0% *Z*). $\lambda_Z$**:** Final wavelength at full conversion (100% *Z*).

**2. Angular velocity and acceleration:**

a) Average angular velocity ($\omega_{avg}$).

$$\omega_{avg} = \frac{\Delta\theta}{\Delta t} \text{ ......Supplementary Eq. (4)}$$

$\omega_{avg}$: Average angular velocity, $\Delta\theta$: Change in angular displacement. $\Delta t$ Time interval taken for the displacement.

b) Average angular acceleration

$$\theta_{avg} = \frac{\Delta\omega}{\Delta t} \text{......Supplementary Eq. (5)}$$

$\alpha_{avg}$: Average angular velocity.

c) Maximum angular velocity ($\omega_{max}$).

$$\omega_{max} = \left\|\frac{\partial\theta}{\partial t}\right\|_{\infty} \text{......Supplementary Eq. (6)}$$

Where, $\omega_{max}$ : Maximum instantaneous angular velocity. $\frac{\partial\theta}{\partial t}$ : The first derivative of angular position with respect to time.

d) Maximum angular acceleration ($\alpha_{max}$).

$$\alpha_{max} = \left\|\frac{\partial^2\theta}{\partial t^2}\right\|_\infty \text{......Supplementary Eq. (7)}$$

Where, $\alpha_{max}$ : Maximum instantaneous angular velocity. $\frac{\partial^2\theta}{\partial t^2}$ : The second derivative of angular position.

**3. Binary density calculation**

a) Binary density

**Bin centre calculation($x_i$):**

$$x_i = \frac{e_i+e_{i+1}}{2} \text{......Supplementary Eq. (8)}$$

Where, $x_i$: The center of the $i^{th}$ bin, $e_i$, $e_{i+1}$: The lower and upper boundaries of the bin.

**Frequency count ($n_i$):**

$$n_i = \sum_{j=1}^{N} 1(e_i \leq X_j \leq e_{i+1} \text{......Supplementary Eq. (9)}$$

Where, $N$: Total number of data points, $Xj$: Individual intensity samples, 1: The indicator function (counts 1 if the condition is true, 0 otherwise).

**Probability Density Function ($P_i$):**

$$P(x_i) = \frac{n_i}{N*\Delta\omega} \text{......Supplementary Eq. (10)}$$

Where, $P(x_i)$: The probability density at bin center $x_i$, where $n_i$ represents the frequency count in the $i^{th}$ bin, $N$ is the total number of samples, and $w$ is the bin width.

b) Binary kernel density estimate (KDE)

**Signal Distribution Analysis:**

To analyse the distribution of the nonlinear optical (NLO) signal transmission, a Kernel Density Estimation (KDE) was performed on the raw intensity data.

The probability density function $P(I)$ was estimated using a Gaussian kernel:

$$P(I) = \frac{1}{Nh}\sum_{i=1}^{N} K\left(\frac{I-I_i}{h}\right) \text{......Supplementary Eq. (11)}$$

Where, $K(u) = \frac{1}{\sqrt{2\pi}}e^{-\frac{1}{2}u^2}$ is the standard normal Gaussian kernel. A narrow bandwidth factor (h = $0.05 \cdot \sigma$ was utilized to ensure high-resolution capture of the logic-level peaks (Logic 0 and Logic 1) while suppressing stochastic noise. All statistical computations were performed using the SciPy library in Python.[3]

## 20. Electronic Componants

| Components | Parameters |
|---|---|
| FS Laser | Pulse width: 80 fs, Rep. Rate: 1KHz, Peak Power: few Gega Watts |
| Si Photodiode | Spectral Range: 200 – 1100 nm (visible to NIR), Peak Responsivity: ≈0.6 A/W at 970 nm, Dark Current: < 1 nA at 10 V reverse bias, Rise Time ≈1 ns (small-area diodes) to ≈10 ns |
| Boxcar Amplifier | Bandwidth: 300 MHz, Sensitivity: 1 V/√Hz, Input Impedance: 50 Ω, Input Voltage Range ±200 mV, Output Voltage Range: ±1 V |
| Lock-in Amplifier | Freq: 1 MHz to 250 kHz, Sensitivity: 2 nV/√Hz, Input Impendence: 10 MΩ, Input Voltage Range ±1 nV to ±1 V |
| Arduino Nano | ATmega328 AVR operating at 5 V, 16 MHz, and 20 mA (I/O Pins)[4] |
| IRLZ44N MOSFET | VDSS = 55 V, RDS (on) = 17.5 mΩ, ID = 49 A |
| Power Supply Unit (PSU) | 3 V, 1 A |

**Supplementary Table 1:** Electronic components utilized and their parameters.

## 21. Notes

**1.** The spectral collection was done on OceanFX spectrometer, and the baseline correction was done in the Y-axis with the blank spectrum during the analysis.

**2. Femtosecond laser:**

Output from the topaz is set at 1100 nm, our system has a systematic error of about 8 to 10nm off. Corresponding SHG peak also shifted by 4 nm. One can meticulous tune the wavelength for each case when we are working with resonance conditions. This happened during the power dependent studies. The boxcar integrator synchronised with the femtosecond laser TTL trigger.[5]

**3. Quantum Chemical Calculations:**

Theoretical calculations have been performed to investigate molecular stability as well as to understand the origin of the polarity of investigated molecules. We have carried out the density functional theory (DFT) method using the Gaussian 16 program based on the single-crystal structure of E/Z-**BTDPA**. The molecule's single point energy was performed using the WB97XD level of theory and 6-311++g (2d,2p) basis set. The unit cell of the *E*-**BTDPA** isomer was found to possess a substantial dipole moment value of approximately 16.04 Debye. Conversely, the *Z*-**BTDPA** isomer exhibited an exceptionally small dipole moment value of 0.002 Debye. This calculation strongly supports the highly polar nature observed for the *E*-isomer when compared to the *Z*-isomer.

The SHG characteristics of the *E*-isomer was supported by its higher hyperpolarizability ($\beta$), which is a 3rd-rank tensor parameter. The calculated first hyperpolarizability for the *E*-isomer ($\beta_E$) is 1.2083 x $10^{-28}$ esu, which significantly exceeds the value of the reference molecule urea, 0.392 x $10^{-30}$ esu.[7] In contrast, the *Z*-isomer has a $\beta z$ value of 6.39 x $10^{-33}$ esu. Noncovalent interaction (NCI) surface analyses were carried out using Multiwfn and visualized with VMD.[8,9]

**4. Micro-Raman Spectroscopic Analysis of E-Z Photoisomerization:**

Micro-Raman spectroscopy was utilized to examine the conformational changes occurring during the photoisomerization process. The spectral analysis indicates that the E and Z isomers display distinct spectral differences, as illustrated in Figure 3d,e in the main manuscript. The peak spans from 508 to 637 $cm^{-1}$, which pertains to C-Br stretching. A band identified at 1191 $cm^{-1}$ was attributed to the C-S stretching of the *E* isomer, while the higher value at 1198 $cm^{-1}$, exhibiting decreased intensity, corresponds to the *Z* isomer due to increased steric effect with the -CN group near the C-S bond in the Z isomer, which makes the C-S bond slightly stiffer (stronger restoring force, less polarizable). The band observed at 1450 $cm^{-1}$ is attributed to the C-H bending of the *E* isomer compared to the broad bands at 1484 $cm^{-1}$ resulting from the intramolecular hydrogen bonding in the *Z* isomer. This hydrogen bonding alters the local electronic environment and increases the anisotropy of the polarizability tensor, broadening the Raman spectral features. The C=C stretching vibrations appear at 1572 $cm^{-1}$ and 1560 $cm^{-1}$ for the E and Z isomers, respectively. The lower Raman intensity observed for the *E* isomer's C=C stretch is ascribed to steric repulsion around the double bond, which diminishes π-electron delocalization and thus the amplitude of polarizability changes during vibration. Although C=C stretching vibrations generally manifest as moderate Raman signals between 1680-1640 $cm^{-1}$, the conjugation through the C(9)-C(10) double bond in both isomers stabilizes electron distribution, shifting signals to the lower wavenumbers. A weak band at 2220 $cm^{-1}$ in the *E* isomer is assigned to the C≡N stretching vibration. Its low intensity results from steric hindrance between the benzene rings ortho-hydrogen (*o*-H) and the nitrile group, reducing polarizability changes during vibration. The *Z* isomer exhibits a similar band at a slightly lower frequency of 2205 $cm^{-1}$, reflecting its enhanced conjugation and higher overall molecular polarizability. These spectroscopic observations align with single-crystal X-ray diffraction (SC-XRD) data, revealing a reduced torsional angle (8.42°) for the *Z* isomer versus 12.02° for the *E* isomer, indicative of a more planar, delocalized structure. Complementary Density Functional Theory (DFT) calculations of Raman spectra further substantiate the experimental Raman results, demonstrating excellent correlation in peak positions and intensity variations between the E and Z isomers.